\documentclass{article}

\usepackage{fortschritte}
\usepackage{epsf}
\usepackage{amssymb,amsmath}
\def\beq{\begin{equation}}                     %
\def\eeq{\end{equation}}                       %
\def\bea{\begin{eqnarray}}                     
\def\eea{\end{eqnarray}}                       
\def\barr{\begin{array}}
\def\earr{\end{array}}
\font\mybb=msbm10 at 12pt
\def\bb#1{\hbox{\mybb#1}}

\def\bT {\bb{T}}
\def\bM {\bb{M}}

\font\mybb=msbm10 at 10pt
\def\bb#1{\hbox{\mybb#1}}

\def\bsM {\bb{M}}

\def\spa#1{\phantom{\fbox{\rule[-#1cm]{0cm}{0cm}}}}
\input{tcilatex}
\begin {document}                 
\def\titleline{
%
%
%
Time--dependent Orbifolds and                 
\newtitleline                                 
String Cosmology                              
}
\def\email_speaker{
{\tt 
%
%
miguelc@fc.up.pt
}}
\def\authors{
%
%
%
%
%
L. Cornalba\1ad , M. S. Costa\2ad\sp                   %
}
\def\addresses{
%
%
%
%
\1ad                                        %
Instituut voor Theoretische Fysica,         %
Universiteit van Amsterdam\\                %
Valckenierstraat 65, 1018 XE Amsterdam,     
The Netherlands\\                           
\2ad                                        
Centro de F\'\i sica do Porto and           
Departamento de F\'\i sica,\\               
Faculdade de Ci\^encias,                    %
Universidade do Porto\\                     %
Rua do Campo Alegre 687,                    %
4169-007 Porto, Portugal\\                  %
}
\def\abstracttext{In these lectures, we review the physics of time--dependent 
orbifolds of string theory, with particular attention to orbifolds of 
three--dimensional Minkowski space. We discuss the propagation of free particles 
in the orbifold geometries, together with their interactions. We address the issue
of stability of these string vacua and the difficulties in defining a
consistent perturbation theory, pointing to possible solutions. In particular,
it is shown that resumming part of the perturbative expansion gives 
finite amplitudes. Finally we discuss the duality of some orbifold models with the 
physics of orientifold planes, and we describe cosmological models based on the 
dynamics of these orientifolds.}
\large
\makefront
\newpage
\section{Introduction}
A fundamental problem of theoretical physics is concerned with the nature of
the initial cosmological singularity. General relativity,
which describes the universe at large scales \cite{HawkingEllis}, predicts
that, under generic assumptions, the universe went through a phase of high
curvature, where quantum gravity should have been important. Hence, the
reasons behind the present evolution of the universe can only be answered by
understanding the Planck era. At the present stage of our knowledge, string
theory is the most developed \cite{GSW,Polchinski}, even though still far
from complete, description of gravitational quantum phenomena, and therefore
should provide the right tools to address such fundamental question.

To investigate quantum gravity effects at the cosmological singularity,
there has been, over the past two years, a considerable activity in the
development of time--dependent string orbifolds. In this review we shall
give an introduction to the subject. We shall not give a
detailed analysis of all the time--dependent orbifolds in the literature,
since we will mostly concentrate on the orbifolds of three--dimensional
Minkowski space, but we will provide a guide through the basic techniques in 
the subject. The subject is very young, far from being clearly understood, so
that the problems we address are still quite basic, starting from a
consistent definition of time--dependent string vacua. Recent developments
in the field have shown that perturbation theory breaks down in many
time--dependent orbifolds, leading to the belief that a strong coupling
problem arises, similar to the case of black hole curvature singularities.
We shall give some evidence that these divergences can actually be resolved,
and therefore that time--dependent string orbifolds are, in fact, a good
laboratory for studying the physics of the cosmological singularity.

Let us review a simplified version of the singularity theorems 
\cite{HawkingEllis}. Consider a four--dimensional FRW universe with metric 
\begin{equation}
ds^{2}=-dt^{2}+a^{2}(t)\,ds^{2}\left( \mathcal{M}\right) \ ,  \notag
\end{equation}
where $\mathcal{M}$ is a maximally symmetric space. From Einstein equations
one immediately concludes that 
\begin{equation*}
\dot{H} = -4\pi G\,\left( \rho +p\right) \ +\frac{k}{a^{2}}\ ,
\end{equation*}
where $H=\dot{a}/a$ is the Hubble parameter and where $\rho $ and $p$ are
the matter energy density and pressure, respectively. If $k=0,-1$ and matter 
satisfies the {\em null energy condition} $\rho +p\ge 0$, then the universe 
cannot reverse from a contracting phase ($H<0$) to an expanding phase ($H>0$). 
Moreover, from Einstein equations it also follows that 
\begin{equation*}
\frac{\ddot{a}}{a} = -\frac{4\pi G}{3}\,\left( \rho +3p\right) \,.
\end{equation*}
Thus, if the stronger {\em strong energy condition} $\rho +3p\ge 0$ holds, 
then for any curvature $k$, as we go back in time, we expect an
initial singularity \cite{HawkingEllis}. In this case, one immediately faces
the horizon problem, because today's observable universe consisted, at the
Planck era, of $(10^{30})^{3}$ causally disconnected regions. If, on the
other hand, only the null energy condition holds, then we can have a
non--singular behavior in the past, or just a coordinate singularity,
usually signalling the presence of cosmological horizons. Note that a scalar
field $\phi $ with potential energy $V(\phi )$ has energy density and
pressure given by 
\begin{equation}
\rho =\frac{1}{16\pi G}\left( \frac{1}{2}\,\dot{\phi}^{\,2}+V(\phi )\right)\ ,
\ \ \ \ \ \ \ \ \ \ \ p=\frac{1}{16\pi G}\left( \frac{1}{2}\,
\dot{\phi}^{\,2}-V(\phi )\right) \ ,  \notag
\end{equation}
and therefore generically satisfies the null energy condition, but not the
strong one. This fact is the basis of the solution of the horizon problem
based on inflation \cite{Guth}.

Currently there are two conventional ways of thinking about the cosmological
singularity problem. One possibility is to describe the singularity by a
quantum gravity initial state from which the universe inflated
\cite{HartleHawking}. Alternatively, the universe went through a bounce where
quantum gravity was relevant. Here various scenarios have been considered in the
literature. In the Veneziano pre big--bang model \cite{Veneziano:00} 
and in the ekpyrotic model \cite{ekpyrotic} initial conditions must
be set in the far past before the bang in order to solve the homogeneity and
flatness problems. This problem of initial conditions can be solved,
within the ekpyrotic setup, with the cyclic model \cite{cyclic}, where
the observed homogeneity, flatness and density perturbations are dynamically
generated by periods of dark energy domination before the bangs. Moreover,
the big--bang singularity of the ekpyrotic and cyclic models are given by a
specific time--dependent orbifold of M--theory \cite{Khoury}. The problem of defining 
transition amplitudes across the singularity has received considerable attention
in the literature \cite{Tolley}, however it is far from being settled in
quantum gravity. Therefore, this question ought to be addressed in string theory and,
in particular, time--dependent string
orbifolds are useful models, where one has computational power to
investigate the high curvature cosmological phase. Additionally, a new
possibility for the universe global structure, where the conventional
curvature singularity is replaced by a past cosmological horizon, will be
derived from a string theory orbifold construction.

This review is organized as follows. In section 2 we describe generalities
of time--dependent orbifolds \cite{HoroSteif} such as their classification,
geometry, single particle wave functions, free particle propagation and
linear backreaction. We shall work out in detail the time--dependent
orbifolds of three--dimensional flat space. Each example is reasonably
self--contained, so that the reader has at his/her disposal an independent
review of the basic facts about each orbifold.

Section 3 will be devoted to the important topic of particle
interactions. We shall start by reviewing the non--linear response of the
gravitational field when a particle is placed in the orbifold geometry. This
includes the argument for formation of large black holes put forward by
Horowitz and Polchinski \cite{HP}, and a particular exact solution of the
problem that uses the powerful techniques of two--dimensional dilaton
gravity \cite{Lawrence,CC2}. Then we consider tree level particle
interactions, deriving the divergences that appear in the four--point
amplitude at specific kinematical regimes, as found by Liu, Moore and
Seiberg \cite{LMS1,LMS2,BCKR}. It is shown that these divergences can be
cured by using the eikonal approximation which resums generalized ladder
graphs \cite{CC3}. Moreover, we shall see, with a specific example, that the
Horowitz--Polchinski non--linear gravitational instability and the breakdown
of perturbation theory are unrelated, contrary to claims in the literature.
Finally, we describe the present status of the one loop string amplitude
computations \cite{CC1,Nekrasov,LMS1}, and we analyze  the wave 
functions of on--shell winding states \cite{BerkoozBoris}.

In section 4 we review a new cosmological scenario in string theory,
which we call orientifold cosmology, where the presence of negative tension
branes generates a cosmological bounce \cite{CCK}. In this scenario, the
standard cosmological singularity is replaced be a past cosmological horizon 
\cite{KounnasLust,Q0,CC1}. Behind the horizon there is a time--like naked
singularity, interpreted as a negative tension brane \cite{CCK,Q1}. We shall
start by establishing a duality between a specific M--theory orbifold and a
type IIA orientifold $8$--plane \cite{CC2,CC3}. This duality is relevant for
describing the near--singularity limit of a two--dimensional toy cosmology
associated to the bounce of an $O8/\overline{O}8$ pair. Using a flux
compactification in supergravity, this construction is extended to the case
of a four--dimensional cosmology, which is shown to exhibit cyclic periods
of acceleration during the cosmological expansion \cite{CCV}.

We conclude in section five. For an extensive list of references on
time--dependent orbifolds of flat and curved spaces, together with related work,  
see references [27]--[59].

\section{Time--dependent orbifolds}

Given a conformal field theory (CFT) which is invariant under the action of
a discrete group $\Gamma $, there is a well known procedure to construct a
new CFT \cite{Dixon}: (1) Add a twisted sector to the theory satisfying 
\begin{equation}
\phi (\sigma ^{1}+2\pi )=h\,\phi (\sigma ^{1})\ ,\ \ \ \ \ \ \ \ \ \ \ \ \ \
h\in \Gamma \ ,  \notag
\end{equation}
where $\phi $ is a conformal field and $\sigma ^{1}$ the space--like
worldsheet coordinate; (2) Restrict the spectrum to $\Gamma $--invariant
states. The new theory is called an orbifold of the old CFT. In string
theory, the bosonic conformal fields are the target space fields $%
X^{a}(\sigma )$. Then, when the group $\Gamma $ is a discrete subgroup of
the target space isometries, the orbifold theory describes strings
propagating on the quotient space.

The simplest example of an orbifold is toroidal compactification. One breaks
the Lorentz group of $D$--dimensional Minkowski space to $SO(D-2,1)\times
U(1)$, by identifying points under a discrete translation by $2\pi R$ along
some direction. Then, winding strings are added, and the spectrum is
restricted by quantizing the momentum along the compact direction. Another
example, which is a close analogue of the orbifolds we shall be interested
in, is the $\mathbb{Z}_{N}$ orbifold, where the discrete subgroup $\Gamma
\sim \mathbb{Z}_{N}$ is generated by a rotation $r$ on a plane by an angle $%
2\pi /N$. The group $\Gamma $ consists of the elements of the form $r^{n}$,
with $n\in \{0,\dots ,N-1\}$. Moreover, the quotient space is a cone, which
has a delta function curvature singularity. It turns out that string theory
is well defined on this singular space. This is an intrinsically stringy
phenomenon, since one is forced to add twisted states, which wind around the
tip of the cone, to have a well defined and finite perturbation theory (for
example, a modular invariant partition function) \cite{Dixon}.

The fact that string theory can resolve space--time singularities (the
conifold being another example \cite{KachSilv,LNV}), led many authors to the
investigation of string orbifolds where the group action on the target space
generates time--dependent quotient spaces. In particular, one would hope
that possible singularities could be harmless, just like in the previous
example. Unfortunately, things are more complicated, but nevertheless one
still has cosmological space--times where the string coupling and the
curvature are under control. Moreover, we shall see that the situation for
some orbifold cosmological singularities is clearly better than the still
unsolved black hole singularities.

\subsection{Orbifold classification and generalities}

Consider a Killing vector field $\kappa $ on a manifold $\mathcal{M}$ with
isometry group $G$. Points along the orbits of $\kappa $ can be identified
according to 
\begin{equation}
P\sim e^{n\kappa }P\ ,\ \ \ \ \ \ \ \ \ \ \ \ \ \ n\in \mathbb{Z}\ , 
\notag
\end{equation}
where $e^{\kappa }$ generates a discrete subgroup $\Gamma \subset G$,
isomorphic to $\mathbb{Z}$ or $\mathbb{Z}_{N}$. Killing vectors related by
conjugation by $G$ 
\begin{equation}
\kappa \rightarrow h^{-1}\kappa \,h\ ,\ \ \ \ \ \ \ \ \ \ \ \ \ \ 
h\in G\ ,
\notag
\end{equation}
define the same orbifold. Here we shall analyze in detail the simplest cases
of time--dependent orbifolds, in particular, we shall consider the covering
space $\mathcal{M}$ to be the flat three--dimensional Minkowski space 
$\mathbb{M}^{3}$. The model can then be embedded in a critical string theory
adding extra spectator directions. 

To classify the orbifolds of $\mathbb{M}^{3}$ of the type described above,
we therefore simply have to analyze the Killing vectors of $\mathbb{M}^{3}$,
up to conjugation by $ISO\left( 2,1\right) $. Let us start by introducing
Minkowski coordinates $X^{0}$, $X^{1}$, $X^{2}$ and light--cone coordinates 
\begin{equation}
X^{\pm }=\frac{1}{\sqrt{2}}\left( X^{0}\pm X^{1}\right)\,.  
\notag
\end{equation}
A general Killing vector $\kappa $ is of the form 
\begin{equation*}
\kappa =2\pi i\left( \alpha ^{a}P_{a}+\beta ^{ab}J_{ab}\right)\,,
\end{equation*}
where 
\begin{eqnarray*}
iJ_{ab} &=&X_{a}\partial _{b}-X_{b}\partial _{a}\,, 
\\
iP_{a} &=&\partial _{a}\,,
\end{eqnarray*}
are the usual generators of the Poincar\'{e} algebra. In three dimensions the
situation is quite simple, since we can define the dual form to $\beta ^{ab}$ by
\begin{equation}
\beta ^{ab}=\epsilon ^{abc}\beta _{c}\,.  
\label{ss1}
\end{equation}
If we consider conjugating $\kappa $ with an element 
$h\in SO\left(2,1\right) \subset ISO\left( 2,1\right) $, then the vectors 
$\alpha ^{a}$ and $\beta _{a}$ transform by the corresponding (hyper)rotation. 
If, on the other hand, $h$ is an infinitesimal translation, then 
\begin{eqnarray}
\alpha ^{a} &\rightarrow &\alpha ^{a}+\beta ^{ab}\omega _{b}\,\,,
\label{ss2} \\
\beta ^{ab} &\rightarrow &\beta ^{ab}\,,  \notag
\end{eqnarray}
with $\omega _{a}$ infinitesimal. Therefore, using (\ref{ss1}), it is
simple to see that the two quantities 
\begin{equation*}
\alpha ^{a}\beta _{a}\,,
\ \ \ \ \ \ \ \ \ \ \ \ \ 
\beta ^{a}\beta_{a}
\end{equation*}
are invariant under conjugation. We shall assume that $\beta _{a}\neq 0$
(otherwise we have a pure translation orbifold). Then, depending on the sign
of $\beta ^{2}$, we have an elliptic ($\beta ^{2}<0$), hyperbolic 
($\beta^{2}>0$) or parabolic ($\beta ^{2}=0$) orbifold, and the two invariants
just described characterize completely the orbifold.

Let us start with the hyperbolic orbifolds, where we can choose, after a
Lorentz transformation, $\beta _{2}=\Delta $, $\beta_{\pm }=0$. Using 
(\ref{ss2}) we can eliminate $\alpha^{\pm }$, and we are left with 
$\alpha^{2}=R $. Therefore we have arrived at the general hyperbolic orbifold,
parametrized by a two--parameter family of inequivalent conjugacy classes,
given by 
\begin{equation}
\kappa =2\pi i\,\left( \Delta \,J_{+-}+R\,P_{2}\right) \,,
\label{sbk}
\end{equation}
and generated by a boost along one direction, say the $X^{1}$--direction,
and a translation along the transverse $X^{2}$--direction \cite{CC1}. In
this review, we shall call this orbifold the \emph{shifted--boost orbifold},
whenever $R\neq 0$. The particular case with $R=0$ gives the \emph{boost
orbifold} first studied in \cite{Khoury}, which is relevant for the
ekpyrotic universe.

Secondly, we can consider the parabolic orbifolds, with $\beta _{-}=\Delta $, 
and $\alpha ^{-}=R$. In this case, the inequivalent conjugacy classes are given 
uniquely by the invariant $\alpha \cdot \beta =\Delta R$, and are defined by the
Killing vector 
\begin{equation}
\kappa =2\pi i\,\left( \Delta \,J_{+2}+R\,P_{-}\right) \,.  
\label{Opk}
\end{equation}
We will denote the case $R\neq 0$ the \emph{$O$--plane orbifold} \cite{CC3}. 
The unconventional nomenclature will be justified in section 4.
Setting $R=0$ one obtains the \emph{null--boost orbifold} considered in 
\cite{Simon,LMS1}. Adding a translation along a fourth spatial direction
to the null boost generator, therefore considering an orbifold of 
$\mathbb{M}^{4}$, one obtains the so--called \emph{null--brane orbifold} , which has
been studied in the literature in \cite{FigueroaSimon,LMS2,Fabinger}. We
shall comment on this orbifold throughout the review, but will not give the
details, which are a simple generalization of the null--boost orbifold.

\begin{table}[tbp]
\caption{Time--dependent orbifolds of $\bsM^{3}$.}
\begin{center}
\begin{tabular}{|c|c|}
\hline
Orbifold & Generator \\ \hline\hline
Shifted--boost & $2\pi i\left(\Delta\,J_{+-} + R\,P_2\right)$ \\ \hline
Boost & $2\pi i\,\Delta\,J_{+-}$ \\ \hline
$O$--plane & $2\pi i\left(\Delta\,J_{+2} + R\,P_-\right)$ \\ \hline
Null--boost & $2\pi i\,\Delta\,J_{+2}$ \\ \hline
\end{tabular}
\end{center}
\end{table}

Thirdly, we briefly comment on the elliptic case, where 
$\beta _{0}=\Delta $, $\alpha ^{0}=R$ and where 
\begin{equation}
\kappa =2\pi i\,\left( \Delta \,J_{12}+R\,P_{0}\right) \,.  
\notag
\end{equation}
For $R=0$ and $\Delta =1/N$, the quotient space is the $\mathbb{Z}_{N}$ cone
briefly discussed in the previous section. We shall not consider it here
because it gives a time--independent quotient space. The case $R\neq 0$ has
never been studied, since it has a quite unconventional global space--time
structure, and since it is probably unphysical\footnote{The time--dependent
orbifold defined by $\kappa=2\pi i R P_0$ was considered in
\cite{Moore}.}. Table 1 shows the
generators of the time--dependent orbifolds of three--dimensional Minkowski
space which will be the subject of this review.

After having defined the orbifold identifications in the covering space, one
moves to the study of the quotient space geometry. This is done by changing
to the coordinate system where the Killing vector $\kappa $ has the trivial
form $\kappa \propto \partial _{z}$. Then, starting from the 
three--dimensional flat space--time, one can read, from the Ka\l u\.{z}a--Klein
ansatz 
\begin{equation}
ds_{3}^{\,2}=ds^{2}+\Phi ^{2}\left( dz+A\right) ^{2}\,,  
\notag
\end{equation}
the $2D$--metric, the scalar field $\Phi $ and the 1--form potential $A$. Of
course, one still has the freedom of using the scalar field $\Phi $ to
rescale the lower dimensional metric. This is important in string
compactifications, when one defines the string or Einstein frame. Of course,
in such compactifications extra spectator directions must be added.

Once the basic geometric aspects are understood one proceeds with the
investigation of quantum field theory and string theory on the orbifold,
whose starting point is the construction of single particle wave functions.
These functions will be important to understand single particle propagation
through the previous cosmological spaces, as well as particle interactions.
Moreover, the single particle wave functions are necessary to define the
string theory vertex operators. For simplicity we shall consider scalar
fields with three--dimensional mass $m$, obeying, on the covering space, the
Klein--Gordon equation 
\begin{equation}
\square \psi =m^{2}\psi \ .  \notag
\end{equation}
In order for $\psi $ to be invariant under $\Gamma $, it must also satisfy
the boundary conditions 
\begin{equation}
\psi (X)=\psi (e^{n\kappa }X)\ ,\ \ \ \ \ \ \ \ \ \ \ \ \ \ \ \ \ \ n\in 
\mathbb{Z}\ .  \label{orbbc}
\end{equation}
The quotient space inherits the continuous symmetry generated by $\kappa$,
which commutes with the d'Alembertian operator, and it is therefore convenient to
choose a basis of functions that satisfy
\begin{equation}
\kappa \,\psi _{n}= 2\pi i n\,\psi _{n}\ ,  
\label{ss3}
\end{equation}
where $n\in \mathbb{Z}$ is one of the quantum numbers of the different wave
functions, and must be integral in order to satisfy (\ref{orbbc}). In all
orbifolds discussed in these lectures, there is always a second Killing
vector which commutes with $\kappa $ and whose eigenvalues can be used to
classify the wave--functions completely. We shall see concrete examples
case--by--case. There is also another general way to construct invariant
wave--functions, which always works, even though it might not be the fastest
choice in a particular situation. This representation, though, will be
important when studying particle interactions, since it writes the wave
functions $\psi$ as linear combinations of the usual plane--waves on the
covering space. Start, in the covering space, with the plane wave 
\begin{equation*}
\phi _{p}\left( X\right) =e^{ip\cdot X}
\end{equation*}
and note that, under the action of the continuous isometry $e^{s\kappa }$
one has, in general, that 
\begin{equation*}
\phi _{p}\left( e^{s\kappa }X\right) 
=\phi _{e^{s\kappa }p}\left(X\right)\,e^{i\varphi \left( p,s\right)}\ ,
\end{equation*}
where $\varphi $ is independent of $X$, and $e^{s\kappa }p$ is the momentum $%
p$ transformed under the isometry $e^{s\kappa }$. Choose now $p^{2}+m^{2}=0$, 
and construct the function 
\begin{equation*}
\psi _{p}\left( X\right) =\sum_{n}\phi _{p}\left( e^{n\kappa}X\right)\,,
\end{equation*}
which is clearly invariant under the action of the orbifold group. Actually,
in order to obtain functions satisfying (\ref{ss3}), it is more convenient to
Fourier transform the sum over $n$ in the previous formula, and to consider
the following single--particle wave--functions 
\begin{eqnarray}
\psi _{p,n}\left( X\right) &=&\int ds\,\phi _{p}\left( e^{s\kappa }X\right)
\,e^{-2\pi ins}  \notag \\
&=&\int ds\,\phi _{e^{s\kappa }p}\left( X\right)
\,e^{i\varphi \left(p,s\right) -2\pi ins}\ .  
\label{ss5}
\end{eqnarray}
The last expression is the general expression for the integral
representation of the single particle wave functions, of which we shall see
concrete examples in the sections that will follow.

To conclude this introductory section, we shall consider three basic
problems related to the single particle wave functions. Firstly, in any
geometry with a contracting period followed by an expansion, there may be a
large backreaction of matter fields as they propagate through the bounce.
This fact is a simple consequence of particle acceleration during the
collapse. Within the linear approximation, the single particle wave functions
will tell us whether this problem is under control. Secondly, one question
that naturally arises in time--dependent orbifolds is whether there is
particle production, since the geometry is varying with time. The wave
functions will define asymptotic particle states and transition amplitudes
for free fields. Thirdly, using the covering space plane wave
representation, it is possible to derive $n$--point amplitudes from the
knowledge of such amplitudes in the covering space. This latter problem will
be considered only in section 3.

A note about notation. The $X^a$ coordinates will always be the Minkowski
coordinates on the flat covering space. To make the notation lighter we
refer to the $X^2$--direction as the $X$--direction. Given a vector, it will
be clear from the context when we refer to the vector itself 
$p=p^a\partial_a$, or to its component $p=p^2$.

\subsection{Shifted--boost orbifold}

\begin{figure}[t]
\begin{center}
\begin{tabular}{c}
\epsfysize=9cm\epsfbox{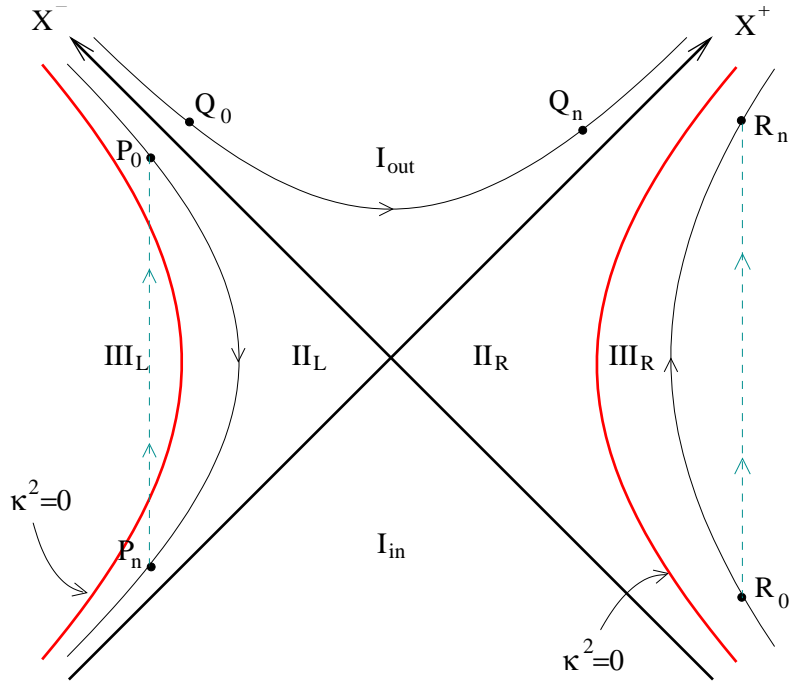}
\end{tabular}
\end{center}
\caption{The different space--time regions for the shifted--boost orbifold in
$X^\pm$--plane. Image points are displaced in the $X$--direction by 
$2\pi Rn$. All CTC's must cross region III and none goes into region I.
The dashed lines correspond to closed time--like geodesics.}
\label{fig1}
\end{figure}

The shifted--boost orbifold is defined by identifying points along the
orbits of the Killing vector \cite{CC1} 
\begin{equation}
\kappa =2\pi i\,\left( \Delta \,J_{+-}+R\,P_{2}\right) \ .  \notag
\end{equation}
Then, from the explicit representation of the Lorentz algebra, it is simple
to deduce the orbifold identifications 
\begin{eqnarray*}
X^{\pm } &\sim &e^{\pm 2\pi \Delta }X^{\pm }\,, \\
X &\sim &X+2\pi R\,.
\end{eqnarray*}
The norm of the Killing vector $\kappa $ becomes null on the surface 
\begin{equation}
2X^{+}X^{-}=-\frac{1}{E^{2}}\ ,\ \ \ \ \ \ \ \ \ \ \ \ \ \ \ \ \ \ (E=\Delta
/R)\ .  \notag
\end{equation}
It is then convenient to divide space--time in three different regions.
Referring to figure \ref{fig1} of the $X^{\pm}$--plane, we will call regions 
$\mathrm{I}_{in}$ and $\mathrm{I}_{out}$, respectively, the past and future
light--cones where $X^{+}X^{-}>0$, and regions $\mathrm{II}_{L}$ and 
$\mathrm{II}_{R}$ the regions, defined by $-E^{-2}<2X^{+}X^{-}<0$, between
the light--cones and the $\kappa ^{2}=0$ surface. In both regions $\mathrm{I}$ 
and $\mathrm{II}$ the Killing vector $\kappa $ is space--like. Finally, we
define the regions $\mathrm{III}_{L}$ and $\mathrm{III}_{R}$, where 
$2X^{+}X^{-}<-E^{-2}$ and where $\kappa $ is time--like.

To understand the causal structure in each region of space--time, consider
the geodesic distance square between a point with coordinates $X^{a}$ and
its $n$--th image, given by 
\begin{equation}
8\sinh ^{2}(n\pi \Delta )\,X^{+}X^{-}+(2\pi Rn)^{2}\ .  \notag
\end{equation}
Clearly, image points in region I are space--like separated. In region II,
provided $n$ is large enough, every point will have a time--like separated
image, as shown for points ${\rm P}_0$ and ${\rm P}_n$ in the figure. 
However, notice that the corresponding geodesic always crosses the
$\kappa ^{2}=0$ surface. In region III all images are time--like separated.
We conclude that there are closed time--like geodesics through regions II
and III, which always go through region III. In region I there are no closed
time--like geodesics. We shall see below that these results are, in fact,
more general, and apply to every causal
curve. Thus, if one excises region III from space--time, there will be no
closed time--like curves (CTC's). 

Particularly interesting points are those
which are light--like related to their $n$--th image. These points lie on 
the so--called polarization surfaces \cite{Thorne}
\begin{equation*}
2X^+X^- = -\frac{c_n}{E^2}\ ,\ \ \ \ \ \ \ \ \ \ \ \ \ \ \ \ \ \ 
c_n=\frac{(n\pi \Delta)^2}{\sinh^2{(n\pi \Delta)}}\ ,
\end{equation*}
which all lie on region II and get arbitrarily close to the horizon for
$n\rightarrow\infty$. These surfaces are potentially problematic when one considers
loop diagrams in perturbation theory. We shall come back to this issue in section 4.1.

The coordinate transformation such that the Killing vector $\kappa $ becomes
trivial is 
\begin{eqnarray*}
X^{\pm } &=&y^{\pm }e^{\pm Ez} \\
X &=&z\,.
\end{eqnarray*}
With this coordinate transformation $\kappa =\left( 2\pi R\right) \partial
_{z}$ and the coordinate $z$ has periodicity $2\pi R$. In order to follow
section 2.1, and to write the three--dimensional flat metric in the Ka\l
u\.{z}a--Klein form in terms of the two--dimensional metric, scalar field
and 1--form potential, it is convenient to move to Lorentzian polar
coordinates in the $y^{\pm }$--plane. In the Milne wedge, corresponding to
the regions I, we choose coordinates 
\begin{equation*}
y^{\pm}=\frac{t}{\sqrt{2}}\,e^{\pm Ey}
\end{equation*}
and the Ka\l u\.{z}a--Klein fields are given 
\beq
\barr{c}
\displaystyle{ds^{2} = -dt^{2}+\frac{(Et)^{2}}{\Phi ^{2}}\,dy^{2}\ ,}
\spa{0.4}\\
\displaystyle{\Phi ^{2} = 1+(Et)^{2}\ ,
\ \ \ \ \ \ \ \ A=\left( 1-\Phi^{-2}\right) \,dy\ .}
\earr
\label{sbgeomI} 
\eeq
We recall that the orbifold has a continuous $U\left( 1\right) $ symmetry
associated to the Killing vector $\kappa $. Moreover, since $J_{+-}$ and $%
P_{2}$ commute, one expects a $SO(1,1)$ and a $U(1)$ symmetry. The $SO(1,1)$
corresponds to translations along $y$, and the $U(1)$ to gauge
transformations of the 1--form potential.

\begin{figure}[t]
\begin{center}
\begin{tabular}{c}
\epsfysize=10cm\epsfbox{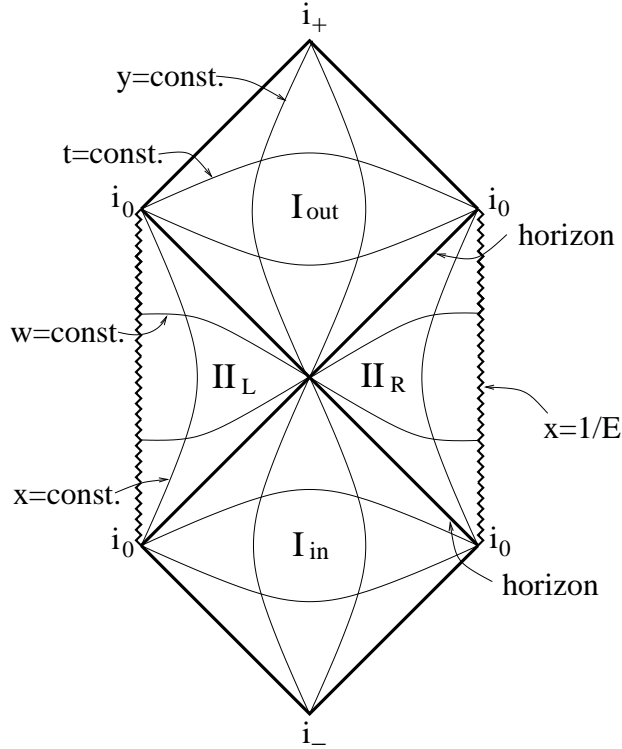}
\end{tabular}
\end{center}
\caption{CP diagram for the shifted--boost orbifold quotient space. The surface
where the Killing vector $\kappa$ is null becomes, in the compactified
geometry, a time--like singularity. The future horizon of the
contracting region ${\rm I}_{in}$ is a Cauchy Horizon.
Region III is excluded from the diagram.}
\label{fig2}
\end{figure}

For $(Et)\ll 1$, the above metric becomes the two--dimensional Milne metric,
and therefore ($t=0$, $y\in \left[ -\infty ,\infty \right] $) is a horizon.
For $t\rightarrow \pm \infty $ the geometry becomes flat and space--time
decompactifies. Region $\mathrm{I}_{in}$, where $t<0$, is contracting
towards a future cosmological horizon, while region $\mathrm{I}_{out}$,
where $t>0$, is expanding from a past cosmological horizon. It is natural to
ask what happens if one crosses the horizons. This can be done by defining
the coordinate transformation that covers regions II and III of the orbifold
space 
\begin{equation*}
y^{\pm }=\pm \frac{x}{\sqrt{2}}\,e^{\pm Ew}\,.
\end{equation*}
Then, the lower dimensional fields read 
\beq
\barr{c}
\displaystyle{ds^{2} = -\frac{(Ex)^{2}}{\Phi ^{2}}\,dw^{2}+dx^{2}\ ,}
\spa{0.4}\\
\displaystyle{\Phi ^{2} = 1-\left( Ex\right) ^{2}\ ,
\ \ \ \ \ \ \ \ A=\left( 1-{\Phi ^{-2}}\right)\,dw\ .}
\earr
\label{sbgeomII}
\eeq
Now the geometry is static and for $(Ex)\ll 1$ the metric is just the
Rindler metric. Hence, in regions II there is a horizon at ($x=0$, $w\in %
\left[ -\infty ,\infty \right] $), that looks just like a black hole
horizon. As one moves away from the horizon there is a curvature singularity
at $Ex=1$. This singularity corresponds to the surface where the
compactification Killing vector $\kappa $ becomes null and region II ends.
Behind the singularity the compactification scalar is imaginary because 
$\kappa$ becomes time--like. It is interesting to compare this with
the five--dimensional BMPV black--hole \cite{bmpv}. 
For this geometry there are also CTC's which are absent when uplifting
the geometry to ten dimensions, however this CTC's are not hidden
behind a singularity of the compactified space \cite{garrycarlos}.
The Carter--Penrose (CP) diagram for this cosmological geometry is 
shown in figure \ref{fig2}. Two--dimensional cosmological models with similar
global structure were also considered in \cite{KounnasLust,Q0}.
Immediately one could worry about the instability of the 
Cauchy horizon when fields propagate from the contracting region. We shall address
this delicate issue at the end of this section, but notice that, in
contrast with the boost and null--boost orbifolds reviewed below, the
compact direction does not shrink to zero size so that classical
backreaction may be under control.

It is now a simple exercise to show that all closed causal curves passing in
regions II must go through the singularity. The proof is identical to the
one for the BTZ black hole \cite{BTZ}. Suppose that such a curve exists and
has tangent vector 
\begin{equation}
l=l^{a}\frac{\partial \ }{\partial x^{a}}\ ,\ \ \ \ \ \ \ \ \ \ 
x^{a} \equiv (w,x,z)\ .  
\notag
\end{equation}
If the curve is closed and time--like in region II there will be a point
where $l^{w}=0$. Then the norm of the tangent at this point has to be
space--like, as can be seen by the form of the three--dimensional metric in
the $x^{a}$--coordinates, which is a contradiction. In order to close the
CTC's one needs to go to regions III, where $z$ becomes the time--like
direction. This brings us to an important point. If one excises region III
from space--time, the geometry has no CTC's. However, one needs to justify
this procedure and to provide boundary conditions at the naked
singularities. When embedded in string theory, we shall see that these
singularities behave like mirrors, and therefore the propagation of fields
through the geometry is well defined.

\subsubsection{Single particle wave functions}

Next, let us describe the single particle wave functions on the orbifold.
Consider the basis of wave functions that diagonalizes the operators 
$\square $, $J_{+-}$ and $\kappa $. In the $(t,y,z)$ coordinates these
operators have the form 
\begin{eqnarray*}
\square &=&-\partial_{t}^{2}-\frac{1}{t}\,\partial_{t}
+\frac{1}{(Et)^{2}}\,\partial_{y}^{2}
+\left( \partial_{z}-\partial_{y}\right)^{2}\,, 
\\
EJ_{+-} & = & -i\partial_{y}\,, 
\\
\frac{1}{2\pi i}\,\kappa &=&-iR\partial_{z}\,,
\end{eqnarray*}
with eigenvalues $m^{2}$, $p$ and $n$, respectively. Omitting the mass label,
we start by writing the wave functions as 
\begin{equation}
\psi _{p,n}=f(t)\,e^{i\left( py+\frac{n}{R}z\right) }\ .  \notag
\end{equation}
The Klein--Gordon equation then becomes 
\begin{equation}
\left[ t^{2}\,\frac{d^{2}\ }{dt^{2}}+t\,\frac{d\ }{dt}+(\omega t)^{2}-\nu
^{2}\right] \,f(t)=0\ ,  \notag
\end{equation}
where 
\begin{equation}
\omega ^{2}=m^{2}+\left( p-\frac{n}{R}\right) ^{2}\ ,
\ \ \ \ \ \ \ \ \ \ \ \ \ \ \ \ \ \ \ 
\nu =i\,\frac{p}{E}\ .
\label{omega}
\end{equation}
The function $f(t)$ is a Bessel function of imaginary order $\nu $. 
Hence, a complete basis for the wave functions in regions I of space--time is
given by 
\begin{equation}
\psi _{p,n}^{\pm }=J_{\pm \nu }(\omega |t|)\,e^{i\left( py+\frac{n}{R}%
z\right) }\ .  \label{sbwfI}
\end{equation}
A similar analysis can be done in regions II, where the wave functions have
the form 
\begin{equation}
\psi _{p,n}^{\pm }=J_{\pm \nu }(i\omega |x|)\,e^{i\left( pw+\frac{n}{R}%
z\right) }\ .  \label{sbwfII}
\end{equation}
These wave functions are defined in each region of space--time. They will be
particularly useful to analyze the propagation of fields near the
cosmological horizons.

It will be quite useful to express the wave functions as a superposition
of the covering space plane waves, and in order to do so we use the general
technique of section 2.1. Consider the on--shell plane wave 
\begin{equation*}
\exp i\left( \pm \frac{\omega }{\sqrt{2}}X^{+}\pm 
\frac{\omega }{\sqrt{2}}X^{-}+kX\right) ,
\end{equation*}
with 
\begin{equation*}
k=\frac{n}{R}-p
\end{equation*}
being the eigenvalue of $P_{2}$. Then it is immediate to use equation 
(\ref{ss5}) to obtain the representation 
\begin{eqnarray*}
&&e^{ikX}\int d\sigma\exp i\left( \pm \frac{\omega }{\sqrt{2}}\,e^{\sigma}X^{+}\pm 
\frac{\omega }{\sqrt{2}}\,e^{-\sigma}X^{-}-\frac{p}{E}\,\sigma\right) \\
&=&e^{i\frac{n}{R}z}\int d\sigma\exp i\left( \pm \frac{\omega }{\sqrt{2}}\,
e^{\sigma}y^{+}\pm \frac{\omega }{\sqrt{2}}\,e^{-\sigma}y^{-}-\frac{p}{E}\,\sigma\right)\,.
\end{eqnarray*}
Next, let us consider the above function in region I. It is given by 
\begin{equation*}
e^{i\left( \frac{n}{R}z+py\right) }\int d\sigma\exp \left( \pm i\omega t\cosh \sigma-i
\frac{p}{E}\,\sigma\right) \,.
\end{equation*}
To see that we have obtained the same result as before, we just need to
notice that the above integral over $\sigma$ is nothing but the integral
representation 
\begin{equation*}
H_{\nu }^{(1,2)}(x)=\pm \frac{1}{\pi i}\,e^{\mp \frac{i\pi \nu }{2}}\int
d\sigma\,\exp \,\left( \pm ix\cosh\sigma-\nu\sigma\right)
\end{equation*}
of the Hankel functions $H_{\nu }^{\left( 1,2\right) }$, which are given by
specific linear combinations of the Bessel functions.

From the above form of the wave functions, we can anticipate a problem
common to all the orbifolds here reviewed \cite{LMS1}. 
In fact, because these functions have a large UV support on the covering space 
single particle states, one expects an enhancement of the graviton exchange 
when they interact gravitationally, which may lead to divergences already at 
tree level.

\subsubsection{Thermal radiation}

Let us now move to the analysis of the cosmological particle production, due
to the time--dependence of the geometry \cite{CCK}. The only subtlety here is to define
uniquely the transition between particle states in the $\mathrm{I}_{in}$ and 
$\mathrm{I}_{out}$ vacua. To analyze the behavior of the wave functions $%
\psi ^{\pm }$ defined in (\ref{sbwfI}) and (\ref{sbwfII}) near the horizons,
we first recall the expansion of the Bessel functions 
\begin{equation}
J_{\nu }(z)=\left( \frac{z}{2}\right) ^{\nu }F_{\nu }\left( z^{2}\right) \ ,
\notag
\end{equation}
where $F_{\nu }$ is the entire function 
\begin{equation}
F_{\nu }(x)=\sum_{n=0}^{\infty }\,\frac{(-1)^{n}}{4^{n}\,n!\,\Gamma (n+1+\nu
)}\,x^{n}\ .  \notag
\end{equation}
The wave function $\psi _{p,n}^{+}$ in the contracting region 
\textrm{\ I}$_{in}$ becomes $(y^{\pm }<0)$ 
\begin{equation}
\psi _{p,n}^{+}=\left( \frac{\omega }{\sqrt{2}}\right) ^{\nu }
\left(-y^{+}\right) ^{\nu }F_{\nu }(2\omega ^{2}y^{+}y^{-})\,
e^{\,i\frac{n}{R}z}\ ,  
\notag
\end{equation}
which, near the horizon, behaves like a conformally coupled scalar. 
As is clear from the above representation,
these wave functions can be continued into the intermediate region $\mathrm{%
II}_{L}$, where $y^{+}<0$ and $y^{-}>0$. Now we come to the delicate issue
of boundary conditions at the singularity. We shall argue, in section 4,
that the singularity can be understood in string theory as an orientifold
plane, where fields obey either Newman or Dirichlet boundary conditions.
With this in mind, let us choose for simplicity the Dirichlet boundary
conditions, by requiring that the field vanishes at the singularity. This
means that we should add, in the region $\mathrm{II}_{L}$, and therefore
also in region ${\rm I}_{out}$, the function 
\begin{equation}
-C\,\psi _{p,n}^{-}=-C\left( \frac{\omega }{\sqrt{2}}\right)^{-\nu }
\left(y^{-}\right) ^{-\nu }F_{-\nu }(2\omega ^{2}y^{+}y^{-})\,
e^{\,i\frac{n}{R}z}\ ,  
\notag
\end{equation}
where $C$ is determined by the boundary condition at $2E^{2}y^{+}y^{-}=-1$
to be 
\begin{equation}
C=\left( \frac{\omega }{2E}\right) ^{2\nu }
\frac{F_{\nu }(-\omega^{2}/E^2)}{F_{-\nu}(-\omega^{2}/E^2)}=
\frac{J_{ip/E}\left( i\omega /E\right)}{J_{-ip/E}\left(-i\omega/E\right)}\ .  
\notag
\end{equation}
Note that $C$ is pure phase, i.e. $C\overline{C}=1$. Physically, the
functions $-C\psi ^{-}$ can be seen as the reflection of the incident waves 
$\psi ^{+}$ at the singularity, or that, in evolving from region I$_{in}$ to
I$_{out}$, one has 
\begin{equation}
\psi ^{+}\longrightarrow -C\,\psi ^{-}\,.  
\label{ref1}
\end{equation}
Similarly we have that 
\begin{equation}
\psi ^{-}\longrightarrow -\overline{C}\,\psi ^{+}\,.
\label{ref2}
\end{equation}

We are now ready to determine the Bogoliubov coefficients by considering the
full effect of the boundary condition on incoming plane waves in the far
past. The functions 
\beq
\barr{l}
\displaystyle{\mathcal{H}_{p}^{+}=
\sqrt{\frac{\pi}{2}}\,
\frac{e^{\frac{i\pi}{4}}}{\sinh\left(\pi p/E\right)}
\left(e^{\frac{\pi p}{2E}}\,\psi_{p,n}^{+}
-e^{-\frac{\pi p}{2E}}\,\psi_{p,n}^{-}\right)\,,}  
\spa{0.6}\\
\displaystyle{\mathcal{H}_{p}^{-}=
\sqrt{\frac{\pi}{2}}\,
\frac{e^{-\frac{i\pi}{4}}}{\sinh\left(\pi p/E\right)}
\left(-e^{-\frac{\pi p}{2E}}\,\psi_{p,n}^{+}
+e^{\frac{\pi p}{2E}}\,\psi_{p,n}^{-}\right) \,,}  
\earr
\label{hankel} 
\eeq
have the plane--wave asymptotic behavior, for $\left| t\right| \rightarrow
\infty $, given by 
\begin{equation*}
\mathcal{H}_{p}^{\pm }\simeq 
\frac{1}{\sqrt{\omega\left|t\right|}}\,
e^{\pm i\omega\left|t\right|+ipy}\,e^{\,i\frac{n}{R}z}\ ,
\end{equation*}
which follows from the large argument behavior of the Hankel functions.
We may then consider, in the far past $Et\ll -1$, the positive frequency
plane wave 
\begin{equation}
\mathcal{H}_{p}^{-}\simeq \frac{1}{\sqrt{-\omega t}}\,
e^{i\left( \omega t+py\right)}\,e^{\,i\frac{n}{R}z}\ .  
\label{vacuumpast}
\end{equation}
Using the reflection equations (\ref{ref1}) and (\ref{ref2}), together with
the defining relations (\ref{hankel}), the above plane wave will evolve in
the future to the following combination of positive and negative frequency
waves 
\begin{equation*}
\alpha \,\,\mathcal{H}_{p}^{+}+\beta \,\mathcal{H}_{p}^{-}\simeq 
\frac{1}{\sqrt{\omega t}}\left[ \alpha \,e^{i\left( pz+\omega t\right) }+\beta
\,e^{i\left( pz-\omega t\right) }\right] e^{\,i\frac{n}{R}z}\,,
\end{equation*}
where the Bogoliubov coefficients $\alpha $ and $\beta $ are given
explicitly by 
\begin{eqnarray*}
&&\alpha =\frac{1}{2i\sinh\left(\pi p/E\right)}
\left( e^{-\pi p/E}C-e^{\pi p/E}\,\overline{C}\,\right) \,, 
\\
&&\beta =\frac{1}{2\sinh \left( \pi p/E\right)}
\left(\, C-\overline{C}\,\right)\,.
\end{eqnarray*}
Using the fact that $C\overline{C}=1$, one can easily check that 
\begin{equation*}
\left| \alpha \right| ^{2}-\left| \beta \right| ^{2}=1\,.
\end{equation*}

\begin{figure}[t]
\begin{center}
\begin{tabular}{c}
\epsfysize=5cm\epsfbox{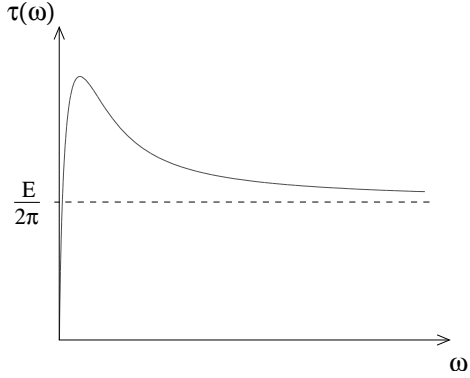}
\end{tabular}
\end{center}
\caption{Effective temperature plot for $m=0$. This curve is the cosmological
analogue of the Hawking radiation grey body factor for black holes. 
For massive particles the maximum of the curve is shifted.}
\label{fig3}
\end{figure}

The natural choice of the cosmological vacuum is the one defined in the far
past $Et\ll -1$ by the plane waves (\ref{vacuumpast})\footnote{Another choice
is the intermediate region II vacuum. This is the Hartle--Hawking
\cite{HartleHawking} vacuum, which gives a thermal spectrum in both past and
future cosmological regions.}. Hence, the
observer in the expanding universe will detect an average number 
$N\left(p\right) $ of particles of momentum $p$ given by the usual formula 
$N\left(p\right) =\left| \beta \right|^{2}$. Moreover, we can define an effective
dimensionless temperature 
\begin{equation}
\frac{1}{\tau (\omega )}=\frac{1}{\omega }\,\ln \left| \frac{\alpha }{\beta }
\right| ^{2}=\frac{2}{\omega }\,\ln \left|\,\frac{e^{-\pi p/E}C-e^{\pi p/E}\,
\overline{C}\,}{C-\overline{C}}\right|\ ,  
\notag
\end{equation}
which defines particle states with respect to the $\mathrm{I}_{out}$ vacuum.
The function $\tau (\omega )$ is plotted in figure \ref{fig3}. 
Notice that for large $\omega $ one has 
\begin{equation}
\tau (\omega )\simeq \frac{E}{2\pi }\ .  
\label{effT}
\end{equation}
Then the physical temperature measured by an observer that is comoving with the
expansion in the far future is given by 
\begin{equation}
T=\frac{\tau (\omega )}{\sqrt{-g_{tt}}}\ .  \notag
\end{equation}
For the compactification from three to two dimensions associated to the
geometry (\ref{sbgeomI}) this gives $T=\tau (\omega )$. More generically,
when there is an extra conformal factor $\Phi ^{2\alpha }$ in the
compactified metric, as it is the case in M--theory compactifications, the
temperature becomes $T=\tau (\omega )/a(t)$, because asymptotically the
scale factor $a(t)$ converges to $\Phi ^{\alpha }$. Moreover, since a
comoving cosmological observer will measure a red--shifted local energy 
\begin{equation}
\Omega =\frac{\omega }{\sqrt{-g_{tt}}}\ ,  \notag
\end{equation}
for fixed $\Omega $, the effective frequency $\omega $ becomes very large
and the asymptotic formula for the temperature is 
\begin{equation}
T=\frac{E}{2\pi a(t)}\ .  \notag
\end{equation}
This temperature can be interpreted as Hawking radiation due to the presence
of a cosmological horizon with non--vanishing surface gravity. In fact, the
horizon surface gravity with respect to the Killing vector defined by $%
\partial _{y}$ in region I and $\partial _{w}$ in region II is $E$. This
defines the effective temperature (\ref{effT}) for a state of momentum $p$,
which has frequency $\omega $ defined in (\ref{omega}). We shall use this
fact in section 4 to generalize the argument for particle production to
higher dimensions.

\subsubsection{Classical stability of Cauchy horizon}

Finally, let us consider the single particle backreaction within the linear
regime. The above wave functions are well behaved everywhere except at the
horizons $y^{\pm }=0$, where there is an infinite blue--shift of the
frequency. To see this, consider the leading behavior of the wave function 
$\psi _{p,n}^{+}$ as $y^{+}y^{-}\rightarrow 0$ 
\begin{equation*}
\psi _{p,n}^{+}\propto \left| y^{+}\right| ^{i\frac{p}{E}}\,
e^{\,i\frac{n}{R}z}\ .
\end{equation*}
Near the horizon $y^{-}=0$ the wave function is well behaved and can be
trivially continued through the horizon. Near $y^{+}=0$, on the other hand,
the wave function has a singularity which can be problematic. In fact, close to the
horizon, the derivative  
$\partial_{+}\psi_{j,n}^{+}\propto\left( y^{+}\right)^{ip/E-1}$ 
diverges as $y^{+}\rightarrow 0$, and this
signals an infinite energy density, since the metric near the horizon has
the regular form $ds^{2}\simeq -dy^{+}dy^{-}$. This fact was noted already
in \cite{HoroSteif,Q1}.

A natural way to cure the problem is to consider wave functions which are
given by linear superpositions of the above basic solutions with different
values of $p$. The problem is then to understand if general perturbations in
the far past $Et\ll -1$ will evolve into the future and create an infinite
energy density on the horizon, thus destabilizing the geometry. This problem
is well known in the physics of black holes where, generically, Cauchy
horizons are unstable to small perturbations of the geometry 
\cite{PenroseSimpson}. Following the work of Chandrasekhar and Hartle for black
holes \cite{CH}, this study was done for cosmologies with a Cauchy horizon
in \cite{Q2,CC2}. In particular, in \cite{CC2}, the following was shown.
Consider, for simplicity, a perturbation corresponding to an uncharged 
field of the form $\psi \left( t,y\right)$. At some
early time $t_0\ll -E^{-1}$, before the field is scattered by the potential 
induced by the curved geometry, the perturbation is given by a function 
$\psi\left( t_0,y\right)=f\left( y\right)$ which is localized in $y$ 
(for example of compact support or, at most, with a Fourier transform that does 
not have poles on the strip $|\mathrm{Im}\,p|<E$). Then we can follow the 
evolution of the field $\psi$ and one discovers that it is perfectly regular at 
the cosmological horizon. The interested reader can see the details of the
computation in \cite{CC2}. The result is quite different from the case of charged 
black holes, where the evolution of regular perturbations at the 
\textit{outer horizon} produces, quite generally, diverging perturbations at 
the inner horizon.

\subsection{Boost orbifold}

\begin{figure}[t]
\begin{center}
\begin{tabular}{c}
\epsfysize=9cm\epsfbox{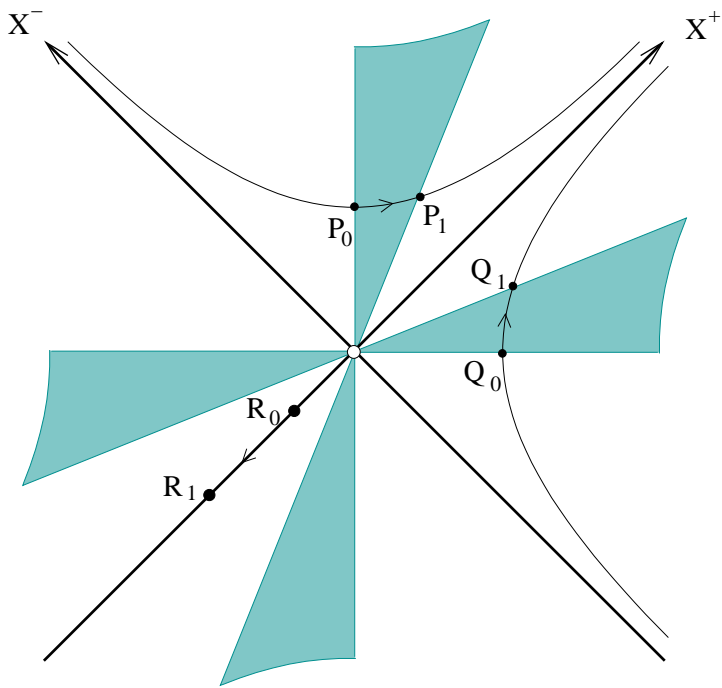}
\end{tabular}
\end{center}
\caption{The fundamental domain for the boost--orbifold. There are CTC's in the 
whiskers and light--cone points have images arbitrarily close to the origin.}
\label{fig4}
\end{figure}

The first time--dependent orbifold to be investigated when the subject was
revived, was the boost orbifold \cite{Khoury}. It is the $R\rightarrow 0$
limit of the shifted--boost orbifold; however, in this limit, the geometry
changes drastically. Space--time points are identified according to 
\begin{equation}
X^{\pm }\sim e^{\pm 2\pi \Delta }X^{\pm }\ ,  \notag
\end{equation}
and the spatial $X$--direction plays no role. Each quadrant in the $X^\pm$--plane 
is mapped onto itself, and the origin is a fixed point of
the orbifold action. Moreover, points on the light--cone have images arbitrarily
close to the origin and, consequently, space--time is not Hausdorff. In
figure \ref{fig4} the orbifold identifications along the orbits of $\kappa $ are
represented schematically. This orbifold describes the collision of branes
in the ekpyrotic scenario \cite{ekpyrotic}. There, one considers the boost
orbifold together with an additional $\mathbb{Z}_{2}$ projection. The
orbifold fixed lines are then identified with branes, which are extended
along three transverse non--compact space directions, as in the brane world
scenario \cite{RandSund}.

The geodesic distance square between images can be easily computed 
\begin{equation}
8\sinh ^{2}(n\pi \Delta )\,X^{+}X^{-}\ ,  \notag
\end{equation}
from which we immediately see that there are closed time--like curves
(CTC's) on both left and right quadrants, which are usually called 
\emph{the whiskers}.

\begin{figure}[t]
\begin{center}
\begin{tabular}{c}
\epsfysize=6cm\epsfbox{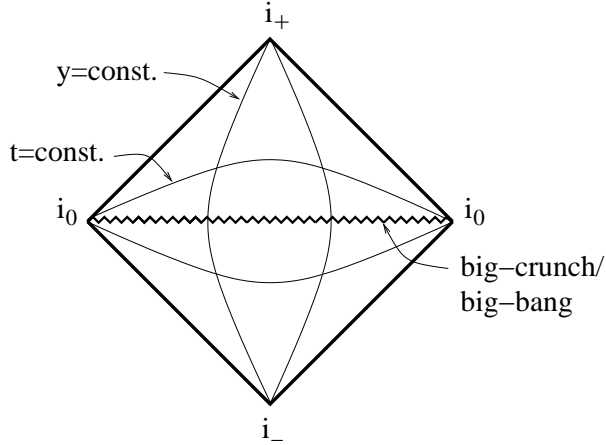}
\end{tabular}
\end{center}
\caption{CP diagram for the boost--orbifold quotient space. Here we consider the
covering space $\bsM^3$, but the generalization to higher dimensions is trivial.
Only the contracting and expanding regions are represented in this diagram. The
remaining regions are the whiskers, which end at the singularity.}
\label{fig5}
\end{figure}

The coordinate transformation 
\begin{eqnarray*}
X^{\pm } &=&\frac{t}{\sqrt{2}}\,e^{\pm \Delta z} \\
X &=&y
\end{eqnarray*}
brings the three--dimensional flat metric to the Ka\l u\.{z}a--Klein form 
\begin{equation}
ds_{3}^{\,2}=-dt^{2}+dy^{2}+(\Delta t)^{2}dz^{2}\ ,  \notag
\end{equation}
where the $z$--coordinate has periodicity $2\pi $ and the compactification
radius varies with time according to $R(t)=2\pi |\Delta t|$. From the
original Poincar\'{e} invariance on $\mathbb{M}^{2}$, the orbifold breaks
translation invariance, but preserves the continuous $SO(1,1)$ associated to
translations along the $z$--direction. The CP diagram for
the geometry is represented in figure \ref{fig5}.

The Ricci scalar has a delta function space--like singularity at $t=0$. The
initial hope was that, like for the Euclidean $\mathbb{Z}_{N}$ orbifold,
string winding states would resolve this singularity. However, as we shall
see in section 3.5, the 1--loop partition function for this orbifold has
divergences whose physical interpretation remains unclear \cite{CC1,Nekrasov}. 
This problem is yet to be understood, in particular, the role of the
winding states and its relation with the poles of the partition function
which originate the above divergence. For recent work on this problem see 
\cite{BerkoozBoris}. This issue is important because it should clarify what
is the role of the whiskers, which terminate at the singularity and are not
covered by the above coordinate transformation. To embed this construction
in M--theory consider the map between the $D=11$ and the type IIA
supergravity fields 
\begin{equation}
ds_{11}^{\,2}=e^{-\frac{2}{3}\phi }ds_{10}^{\,2}+e^{\frac{4}{3}\phi }\left(
dz+A\right) ^{2}\ .  \notag  \label{M-theory}
\end{equation}
Then, the orbifold of $\mathbb{M}^{11}$ by a boost gives the type IIA
background fields 
\begin{equation}
ds^{2}=|\Delta t|\,ds^{2}(\mathbb{M}^{10})\ ,\ \ \ \ \ \ \ \ \ \ \ 
e^{\phi}=(\Delta t)^{3/2}\ .  
\notag
\end{equation}
This geometry describes a universe with a contracting phase for $t<0$ and an
expanding phase for $t>0$. At the curvature singularity the string coupling
vanishes.

Next let us analyze the single particle wave functions. These can be deduced
from the results for the shifted--boost orbifold with little effort by
sending $R\rightarrow 0$. The integral representation \cite{Nekrasov} 
\begin{equation*}
e^{ikX}\int ds\exp i\left( \pm \frac{\omega }{\sqrt{2}}\,e^{s}X^{+}\pm 
\frac{\omega }{\sqrt{2}}\,e^{-s}X^{-}-\frac{n}{\Delta }\,s\right) ,
\end{equation*}
with $\omega^{2}=m^{2}+k^{2}$, defines invariant functions in the full
covering space. In particular, in the $X^{\pm}$--plane the functions are
nothing but Bessel functions of the radial coordinate with imaginary order 
$\nu=i\frac{n}{\Delta}$. In the Milne wedge we have the functions 
\begin{equation*}
J_{\pm i\frac{n}{\Delta }}\left( \omega \left| t\right| \right) 
\,e^{i\left(ky+\frac{n}{R}z\right)}\,.
\end{equation*}

We wish to consider the problem of particle production. We can follow the
same arguments of section 2.2.2, and extend the above functions to the
Rindler wedge. This time, though, we do not have a natural boundary where to
impose the boundary condition, and we must then impose that the field vanishes
at spacial infinity in the whiskers, thus picking the exponentially damped
solution \cite{Tolley}. We can then, following again section 2.2.2, define the
reflection constant $C$
\begin{equation*}
C=\lim_{\eta\rightarrow \infty}\frac{J_{in/\Delta}\left(i\eta\right)}
{J_{-in/\Delta}\left(-i\eta\right) }=1\,.
\end{equation*}
The corresponding Bogoliubov coefficients are given by 
\begin{equation*}
\alpha =i\,,\,\ \ \ \ \ \ \ \ \ \ \ \ \ \ \ \beta =0\,,
\end{equation*}
and we have no particle production. The temperature vanishes. Note that this
is not the limit $R\rightarrow 0$ of the results in section 2.2.2, which is, 
on the other hand 
\begin{equation*}
C=\lim_{E\rightarrow \infty }\frac{J_{i\left( n/\Delta -k/E\right)}
\left(i\sqrt{m^{2}+k^{2}}/E\right) }{J_{-i\left( n/\Delta -k/E\right)}
\left(-i\sqrt{m^{2}+k^{2}}/E\right)}\sim 
\lim_{\eta\rightarrow 0}
\frac{J_{in/\Delta}\left(i\eta\right)}{J_{-in/\Delta }\left(-i\eta\right)}\,.
\end{equation*}
The amusing fact is that the above formula still gives $C=1$ for $n=0$, which 
corresponds to the case considered in \cite{Tolley}, by requiring continuity 
of the wave functions on the covering space. However, for $n\neq 0$ the 
limit is ill--defined, thus signalling the fact that the $R\rightarrow 0$ limit 
of the shift--boost orbifold is far more complex than the $R\neq 0$ situation, 
if the prescription of section 2.2.2 (to be justified in section 4) is correct.

Finally, as for the shifted--boost orbifold, the above single particle wave
functions with $n\neq 0$ will induce a large backreaction at the singularity. 
A simple calculation shows that the corresponding energy density scales near 
the big crunch/big bang singularity as $t^{-2}$. In this case, however, one 
can not form a wave packet because the Hankel functions are of discrete order.
Physically this problem arises because the compact circle is shrinking to
zero size, so that any non--constant perturbation will necessarily induce a
large backreaction. In this sense, the addition of a shift to the boost
orbifold can be seen as a regulator of the singularity because there is no
fixed point.

\subsection{$O$--plane orbifold}

\begin{figure}[t]
\begin{center}
\begin{tabular}{c}
\epsfysize=9cm\epsfbox{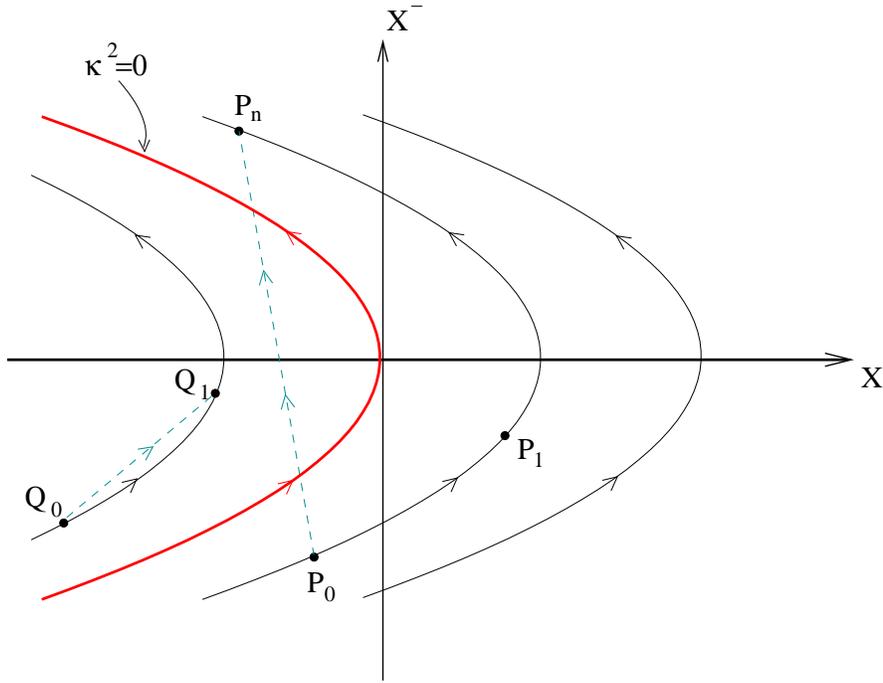}
\end{tabular}
\end{center}
\caption{The orbits of the Killing vector $\kappa$ for the $O$--plane orbifold 
represented in the $X^-X$--plane. Image points are displaced in the $X^+$--direction 
according to (\ref{Oplane}). All CTC's must cross the region with $\kappa^2<0$.
The dashed lines represent closed time--like geodesics.}
\label{fig6}
\end{figure}

In section 2.1 we have introduced the $O$--plane orbifold, defined by the
Killing vector 
\begin{equation}
\kappa =2\pi i\,\left( \Delta \,J_{+2}+R\,P_{-}\right)\,.  \notag
\end{equation}
One can check that, under the action of $e^{\kappa }$, space--time points
are identified according to \cite{CC3}
\begin{eqnarray}
X^{-} &\sim &X^{-}+2\pi R  \notag \\
X^{+} &\sim &X^{+}-(2\pi \Delta )X+\frac{1}{2}\,\left(2\pi \Delta\right)^{2}X^{-}
+\frac{1}{6}\,\left( 2\pi \right) ^{3}R\Delta ^{2}  \label{Oplane} \\
X &\sim &X-(2\pi \Delta )X^{-}-\frac{1}{2}\,\left( 2\pi \right) ^{2}R\Delta \ .
\notag
\end{eqnarray}
The Killing vector $\kappa $ has norm 
\begin{equation}
\kappa ^{2}=8\pi ^{2}\Delta R\left( X+\frac{1}{2}E\left( X^{-}\right)
^{2}\right) \ ,\ \ \ \ \ \ \ \ \ \ \ \ \ \ (E=\Delta /R)\ ,  \notag
\end{equation}
and therefore space--time is naturally divided in two regions, with $\kappa $
space--like or time--like, by the surface 
\begin{equation}
X+\frac{1}{2}E\left( X^{-}\right) ^{2}=0\ .  \notag
\end{equation}
The geodesic distance square between image points satisfies 
\begin{equation*}
2E(2\pi Rn)^{2}\left( X+\frac{1}{2}\,E\left( X^{-}\right) ^{2}
-\frac{1}{12}\,E (2\pi R n)^2\right)\,.
\end{equation*}
Hence, provided $n$ is large enough, points that are in
the region where $\kappa $ is space--like are connected to their $n$--th
image by a time--like geodesic. Notice, however, that this geodesic always
crosses the $\kappa ^{2}=0$ surface. More generally, any closed causal curve
must cross this surface. This is similar to what happens in regions II and
III of the shifted--boost orbifold. In fact this is not a coincidence,
since, as we shall see, the $O$--plane orbifold is the limit of the
shifted--boost orbifold near the surface where $\kappa $ is null. In figure
\ref{fig6} the identifications on the $X^{-}X$--plane are represented.

The orbifold breaks the Poincar\'{e} invariance of the covering space, and
preserves the symmetries generated by $\kappa $ and the translations $P_{+}$
along the $X^{+}$--direction. Also, when embedded in a supersymmetric
theory, this orbifold preserves some supersymmetry. Consider, as an example $%
N=2$, $D=10$ supersymmetry. For the spin structure with periodic boundary
conditions on the orbifold circle, supersymmetry transformations generated
by spinors satisfying the condition 
\begin{equation}
\Gamma ^{-}\epsilon =0\ ,  \notag
\end{equation}
are inherited. Thus, this orbifold preserves half of the $N=2$
supersymmetries.

\begin{figure}[t]
\begin{center}
\begin{tabular}{c}
\epsfysize=6cm\epsfbox{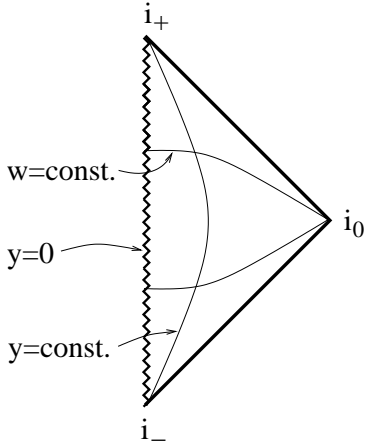}
\end{tabular}
\end{center}
\caption{The CP diagram for the compactified $O$--plane orbifold geometry. 
The time--like singularity corresponds to the surface $\kappa^2=0$ 
and only the region where $\kappa$ is space--like is represented.}
\label{fig7}
\end{figure}

To better study the orbifold geometry, it is very useful to consider the
following coordinate transformation 
\begin{eqnarray*}
X^{-} &=&y^{-} \\
X^{+} &=&y^{+}-Eyy^{-}+\frac{E^{2}}{6}\left( y^{-}\right) ^{3} \\
X &=&y-\frac{E}{2}\left( y^{-}\right) ^{2}\,.
\end{eqnarray*}
Then, the flat three--dimensional metric looks like a (trivial) plane wave 
\begin{equation}
ds_{3}^{\,2}=-2dy^{+}dy^{-}+2Ey\left( dy^{-}\right) ^{2}+dy^{2}\ ,  \notag
\end{equation}
where the $y^{-}$ direction has periodicity $2\pi R$. In terms of the
coordinate $y^{a}$, the norm of $\kappa $ is simply$\ 8\pi ^{2}\Delta Ry$,
so the surface $y=0$ is the locus where the Killing vector $\kappa $ is
null. For $y>0$, $\kappa $ is space--like, and for $y<0$, it is time--like.
Moreover, the polarization surfaces, where image points are light--like 
related, are given by $y=E(2\pi Rn)^2/12$.
If we rewrite the line element in the Ka\l u\.{z}a--Klein form 
\begin{equation}
ds_{3}^{\,2}=-\frac{\left( dy^{+}\right) ^{2}}{2Ey}+dy^{2}+2Ey\,
\left( dy^{-}-\frac{dy^{+}}{2Ey}\right) ^{2}\ ,  
\notag
\end{equation}
we can easily show that it corresponds precisely to the near--singularity
limit of the shifted--boost orbifold geometry (\ref{sbgeomII}), provided one
replaces $y^{+},y^{-}$ by $w,z$ and considers the limit  
$Ey=|Ex\mp 1|\ll 1$, for $x>0$ or $x<0$, respectively. 
Thus, the near--singularity limit of the shifted boost
orbifold is the $O$--plane orbifold. It then follows that there are CTC's
everywhere, but all these curves must cross the singularity at $y=0$. The CP
diagram for this geometry is shown in figure \ref{fig7}.

Finally, to find the single particle wave functions, let us choose a basis
that diagonalizes the following operators, expressed in the $(y^{\pm },y)$
coordinates, 
\begin{eqnarray*}
\square  &=&-2\partial _{+}\partial _{-}-2Ey\,\partial _{+}^{\,2}+\partial
_{y}^{\,2}\ , \\
P_{+} &=&-i\partial _{+}\,, \\
\frac{1}{2\pi i}\,\kappa  &=&-iR\partial _{-}\,.
\end{eqnarray*}
For a particle of mass $m\,$, the wave functions are labelled by $(p_{+},n)$%
, and we can use separation of variables to write them as 
\begin{equation}
\psi _{p_{+},n}=f(y)\,e^{i\left( p_{+}y^{+}+\frac{n}{R}y^{-}\right) }\ , 
\notag
\end{equation}
where $f(y)$ satisfies the differential equation 
\begin{equation}
\left[ 2\,\frac{n}{R}\,p_{+}+2Ey\,p_{+}^{\,2}+\frac{d^2\ }{dy^2}-m^2\right]\,
f(y)=0\ .  
\notag
\end{equation}
Defining the new variable 
\begin{equation}
\omega =-\left( 2Ep_{+}^{\,2}\right) ^{\frac{1}{3}}\left( y+\frac{n}{ERp_{+}}%
-\frac{m^{2}}{2Ep_{+}^{\,2}}\right) \ ,  \notag
\end{equation}
the above differential equation simplifies to 
\begin{equation}
\frac{d^2f}{d\omega^2}=\omega \,f\ ,  
\notag
\end{equation}
which describes, in quantum mechanics, a zero energy particle subject to a
linear potential. The solutions are the Airy functions $Ai(\omega )$ and $%
Bi(\omega )$, which are, respectively, exponentially damped or exponentially
growing in the $\omega >0$ region. This region corresponds mostly to
negative $y$, where the Killing vector $\kappa $ is time--like. Choosing the
normalizable solution, we have just shown that 
\begin{equation*}
\psi _{p_{+},n}\propto Ai(\omega )\,e^{i\left( p_{+}y^{+}+\frac{n}{R}%
y^{-}\right) }.
\end{equation*}
This choice has a clear physical interpretation. Consider a particle of mass 
$m$ and Ka\l u\.{z}a--Klein charge $n$. Since the Airy function $Ai(\omega )$
and its derivative are exponentially damped for $\omega <0$, the probability
of finding the particle in the region 
\begin{equation}
y<y_{c}=\frac{m^{2}}{2Ep_{+}^{\,2}}-\frac{n}{ERp_{+}}\ ,  \notag
\end{equation}
is negligible. This behavior is clear physically: in the covering space, all
time--like geodesics that go through the region $y<0$ remain there for a
finite proper time, explaining why the wave function is damped. Moreover,
for very large $p_{+}$, the particle gets arbitrarily close to the
singularity. Finally, the case of charged particles is particularly
interesting, since $y_{c}$ is linear with$\,n$. Particles with positive
charge are attracted towards the singularity, whereas negatively charged
particles are repelled.

We shall now obtain an integral representation of the function $\psi_{p_{+},n}$, 
as a superposition of standard plane waves in the covering space. 
We start from the integral representation of the Airy function 
\begin{equation}
Ai(\omega )=\frac{1}{2\pi }
\int dt\,\,e^{\,i\left( \omega t+\frac{t^{3}}{3}\right) }\ ,  
\notag
\end{equation}
which immediately yields 
\begin{equation*}
\psi _{p_{+},n}\propto \,e^{i\left( p_{+}y^{+}+\frac{n}{R}y^{-}\right) }\int
ds\,e^{i\left( y+\frac{n}{ERp_{+}}-\frac{m^{2}}{2Ep_{+}^{\,2}}\right) s-
\frac{i}{6}\frac{s^{3}}{Ep_{+}^{2}}}\,.
\end{equation*}
Changing coordinates to the original Minkowski coordinates $X^{a}$, and
defining the new integration variable $p=s+X^{-}p_{+}E$, one gets after
choosing a specific normalization 
\begin{equation}
\psi _{p_{+},n}=\frac{1}{\sqrt{\left| p_{+}\right| }}\int dp\,\,
\phi_{p_{+},p}\left( X\right) \,\,
\exp\frac{i}{E}\left( \frac{np}{Rp_{+}}
-\frac{1}{2}\frac{m^{2}p}{p_{+}^{\,2}}
-\frac{1}{6}\frac{p^{3}}{p_{+}^{\,2}}\right)\ ,  
\label{Owf}
\end{equation}
where 
\begin{equation}
\phi _{p_{+},p}\left( X\right) =\,e\,^{i\left(
p_{+}X^{+}+p_{-}X^{-}+pX\right) }\ ,\ \ \ \ \ \ \ \ \ \ \ \ 
p_{-}=\frac{m^{2}+p^{2}}{2p_{+}}\ ,  \notag
\end{equation}
is the usual on--shell flat space plane wave. The integral representation (\ref{Owf}) 
is nothing but the representation described in general in section 2.1, 
as it is possible to show starting from the identifications (\ref{Oplane}). 
We leave this check to the interested reader. It is a matter of
computation to show that the above single--particle functions satisfy the
orthogonality condition 
\begin{equation}
\langle m^{2},p_{+},n|m^{\prime 2},p_{+}^{\prime},n^{\prime}\rangle 
=32\pi^{4} ER\,|p_{+}| \,\delta \left( m^{\,2}-m^{\prime \,2}\right)
\delta \left( p_{+}-p_{+}^{\prime }\right) \delta _{n-n^{\prime}}\ ,
\end{equation}
where we have reinserted the mass label. 

Finally, since $\partial _{+}$ is
a globally defined null Killing vector these functions define the same
particle states in the $y_{+}\rightarrow \pm \infty $ regions. Consequently,
it is possible to define a global vacuum and there is no particle production.

\subsection{Null--boost orbifold}

\begin{figure}[t]
\begin{center}
\begin{tabular}{c}
\epsfysize=6cm\epsfbox{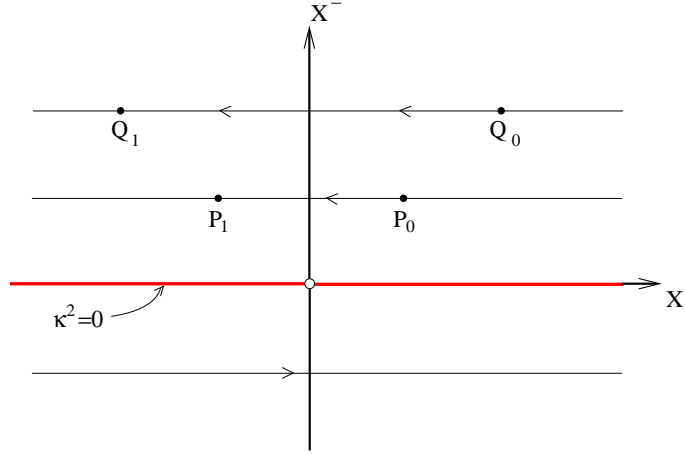}
\end{tabular}
\end{center}
\caption{The orbits of the Killing vector $\kappa$ for the null--boost orbifold 
represented in the $X^-X$--plane. Image points are displaced in the $X^+$--direction 
according to (\ref{nullboost}). There are CNC's on the surface $X^{-}=0$ and
the origin is a fixed point.}
\label{fig8}
\end{figure}

The null--boost orbifold was studied recently in \cite{Simon,LMS1}. To
obtain the identification of space--time points for this orbifold, we can
simply set $R=0$ in the analogous equation (\ref{Oplane}) for the $O$--plane
orbifold 
\begin{eqnarray*}
X^{-} &\sim &X^{-} 
\notag\\
X^{+} &\sim &X^{+}-(2\pi \Delta )X+\frac{1}{2}\,\left(2\pi \Delta\right)^{2}X^{-} 
\label{nullboost}\\
X &\sim &X-(2\pi \Delta )X^{-}\,.
\notag
\end{eqnarray*}
The Killing vector $\kappa =2\pi i\,\Delta \,J_{+2}$ is everywhere
space--like except at $X^{-}=0$, where it is null. Moreover, $\kappa $
vanishes on the $X^{+}$--axis ($X^{-}=X=0$), which is a fixed line of the
orbifold. The geodesic distance square between image points is 
\begin{equation*}
(2\pi \Delta nX^{-})^{2}\ ,
\end{equation*}
which vanishes on the surface $X^{-}=0$. Hence, there will be closed null
curves (CNC's) on this surface. The orbifold action is represented
schematically on the $X^{-}X$--plane in figure \ref{fig8}. This orbifold preserves
the symmetries generated by $J_{+2}$ and $P_{+}$, and also the same
supersymmetries of the $O$--plane orbifold.

\begin{figure}[t]
\begin{center}
\begin{tabular}{c}
\epsfysize=7cm\epsfbox{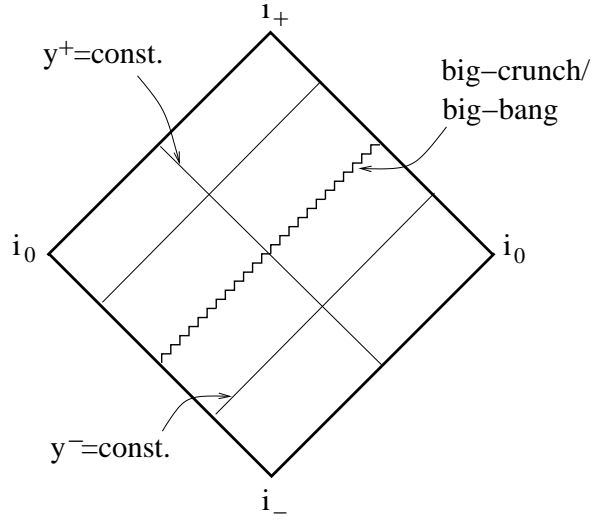}
\end{tabular}
\end{center}
\caption{The CP diagram for the compactified null--boost orbifold geometry. 
The null--like singularity corresponds to the surface where 
$\kappa$ is null and where the CNC's are.}
\label{fig9}
\end{figure}

The coordinate transformation 
\begin{eqnarray*}
X^{-} &=&y^{-} \\
X^{+} &=&y^{+}+\frac{\Delta ^{2}}{2}\,z^{2}y^{-} \\
X &=&\Delta zy^{-},
\end{eqnarray*}
brings the three--dimensional flat metric to the form ($z\sim z+2\pi $) 
\begin{equation}
ds_{3}^{2}=-2dy^{-}dy^{+}+(\Delta y^{-})^{2}dz^{2}\ ,  \notag
\end{equation}
so that the compact circle has radius $R(y^{-})=2\pi |\Delta y^{-}|$. This
metric describes a dilatonic wave that is singular at $y^{-}=0$. This is
where the orbifold action is fixed and where the energy density of infalling
matter will diverge, leading to a large backreaction. The CP diagram for
this geometry is shown in figure \ref{fig9}.

The wave functions for this orbifold can be obtained from (\ref{Owf}) by
setting $R=0$. This gives \cite{LMS1}
\begin{equation*}
\psi _{p_{+},n}=\frac{1}{\sqrt{\left| p_{+}\right| }}\int dp\,\ \phi
_{p_{+},p}\left( X\right) \,\ e^{\,i\frac{np}{\Delta p_{+}}}\ .
\end{equation*}
The $p$--integral is gaussian and can be explicitly done to obtain 
\begin{equation*}
\psi _{p_{+},n}=
\sqrt{\frac{2\pi i}{\left| X^{-}\right| }}\,\exp i\left[ \frac{m^{2}}{2p_{+}}
X^{-}+p_{+}X^{+}-\frac{p_{+}}{2X^{-}}\left( X+\frac{n}{\Delta p_{+}}\right)
^{2}\right] .
\end{equation*}
In the limit $X^{-}\rightarrow 0$ the wave functions become singular. More
precisely, 
\begin{equation}
\lim_{X^{-}\rightarrow 0}\psi _{p_{+},n}=2\pi\,\sqrt{\frac{i}{\left|p_{+}\right|}}
\,\delta \left( X+\frac{n}{\Delta p_{+}}\right)
\,e^{i\,p_{+}X^{+}}\ .  \notag
\end{equation}
Thus, the wave functions are focused on the lattice $X\in \left( \Delta
p_{+}\right) ^{-1}\mathbb{Z}$. This divergence is not integrable because $n$
is discrete and, consequently, these particle states create a large
backreaction on the geometry. Notice that $n/\Delta $ is the eigenvalue of $%
J_{2+}$, which is forced to be have discrete eigenvalues. 

For the null--brane \cite{LMS2,FigueroaSimon}, where the orbifold generator includes
a translation $P_{3}$ along an extra spatial direction, a similar focusing
occurs, at $X=-J_{2+}/p_{+}$. On the other hand, now, since only $\Delta
J_{2+}+RP_{3}\in \mathbb{Z}$, and since $P_{3}$ has continuous spectrum, so
does $J_{2+}$. Therefore we have a continuum of focusing points, and by
choosing wave--packets which are linear combinations with different values
of $J_{2+}$, we can construct regular wave--packets in both the covering 
and the quotient space. As we shall see in the next section, perturbation theory is
badly behaved in the case of both the null--boost and the $O$--plane
orbifold. In \cite{LMS2} the authors show that, on the other hand,
perturbation theory is well--behaved in the case of the null--brane, and
they claim that this is due to the possibility of constructing regular
single particle states, as we have just discussed. This claim is clearly not
correct, since the $O$--plane orbifold has perfectly regular wave--functions, 
but suffers from the same pathology of the null--boost case.

\section{Interactions}

So far we have reviewed in detail the geometry and the single particle wave
functions for the time--dependent orbifolds of three--dimensional flat
space. The next natural step is to consider interactions. In fact, the same
phenomenon that leads to the blue--shift of single particle states during a
cosmological contracting phase, could give rise to instabilities due to
particle interactions. Physically, the acceleration induces a stronger
coupling to the graviton, enhancing the exchange of this particle. One way
to see this is given by the argument, put forward by Horowitz and Polchinski 
\cite{HP}, for the formation of large black holes. We shall review this
argument below and comment on its regime of validity and limitations. A more
precise analysis can be done, in three dimensions, using the powerful
techniques of two--dimensional dilaton gravity, which permits an exact study
of conformal matter propagating in the quotient space of the previously described 
orbifolds \cite{Lawrence,CC2}. Another way to study particle interactions, which does
not always gives the same result regarding stability, is by direct
computation of tree level amplitudes \cite{LMS1,LMS2,BCKR}. We shall review
how divergences are found in four--point amplitudes, and how these
amplitudes can be made finite by resumming generalized ladder graphs in the
eikonal approximation \cite{CC3}. One--loop amplitudes will also be reviewed
\cite{CC1,Nekrasov,LMS1}, together with on--shell winding states wave functions 
\cite{BerkoozBoris}.

\subsection{Formation of large black--holes}

It has been argued, in \cite{HP}, that a large class of time--dependent
orbifolds are unstable to small perturbations, due to a large backreaction
of the geometry. These results do not rely on string theory arguments, and
are obtained within the framework of classical General Relativity. The
argument is quite simple and starts by consider a particle in the orbifold
geometry, which corresponds to an infinite collection of particles in the
covering space. If the interaction between image particles produces a black
hole in the covering space, then this signals that a black hole is formed in
the orbifold quotient space. The condition for black hole formation in the
covering space is that, given a particle and its $n$--th image, their impact
parameter $b$ should be smaller than the Schwarzchild radius associated to
the center of mass energy $\mathcal{E}$, 
\begin{equation}
G\mathcal{E}>b^{D-3}\ ,  \notag
\end{equation}
where $D$ is the dimension of space--time. In practice, one is interested in
interactions with large boosted images, so that one can consider, without
loss of generality, particles moving along null geodesics. The condition for
black hole formation can then be made quite precise, because it corresponds
to the existence of a trapped surface in space--time when two shock--waves,
described by the Aichelberg--Sexl metrics \cite{AichSexl}, collide \cite
{Death}.

Let us now be more quantitative and consider a null geodesic with
world--line 
\begin{equation}
X^{a}(\lambda )=p^{a}\lambda +C^{a}\ ,  \notag
\end{equation}
where $p^{a}$ is the momentum and $C^{a}$ a point along the geodesic. 
The $n$--th image geodesic has world--line 
\begin{equation}
X_{n}^{\,a}(\lambda )=p_{n}^{\,a}\lambda +C_{n}^{\,a}\ ,  \notag
\end{equation}
where $p^a_{n}$ and $C^a_{n}$ are the images, under the orbifold action $%
e^{n\kappa }$, of the momentum $p^{a}$ and of the point $C^{a}$,
respectively. Simple kinematics shows that the impact parameter $b$ and
center of mass energy $\mathcal{E}$ are given by 
\begin{equation}
b^{\,2}=Y^{\,2}-\frac{2(p\cdot Y)(p_{n}\cdot Y)}{p\cdot p_{n}}\ ,\ \ \ \ \ \
\ \ \ \mathcal{E}^{2}=-2p\cdot p_{n}\ ,  \notag
\end{equation}
where $Y=C-C_{n}$.

It is now a mater of computation to determine which orbifolds are stable or
unstable according to these criteria. Consider, as an example, the 
$O$--plane orbifold. Let $p^{a}$ and $C^{a}$ be given by 
\begin{equation}
p=\left( 
\begin{array}{c}
p^{+} \\ 
p^{-} \\ 
p \\ 
\vec{p}_{\perp }
\end{array}
\right) \ ,\ \ \ \ \ \ \ \ \ C=\left( 
\begin{array}{c}
C^{+} \\ 
0 \\ 
C \\ 
\vec{C}_{\perp }
\end{array}
\right) \ ,  \notag
\end{equation}
where we allowed for possible extra spectator directions, and where we
parametrize the null geodesic so as to set $C^{-}=0$ (the case when $p^{-}=0$
implies that $p^{a}$ and $p_{n}^{a}$ are collinear, with $\mathcal{E}=0$ and
no black--hole formation). Then the momentum of the $n$--th image particle
reads 
\begin{equation}
p_{n}=\left( 
\begin{array}{c}
\displaystyle{p^{+}-\beta p+\frac{\beta^2}{2}\,p^{-}} \\ 
p^{-} \\ 
p-\beta p^{-} \\ 
\vec{p}_{\perp }
\end{array}
\right) \ ,  \notag
\end{equation}
where $\beta =2\pi n\Delta $ and the constant $Y$ satisfies 
\begin{equation}
EY=\left( 
\begin{array}{c}
\displaystyle{-\beta E C+\frac{1}{6}\,\beta^3}\spa{0.2} \\ 
\displaystyle{\beta}\spa{0.2} \\ 
\displaystyle{-\frac{1}{2}\,\beta^2}\spa{0.3} \\ 
\vec{0}
\end{array}
\right) \ .  \notag
\end{equation}
Finally, for large $n$, the impact parameter and the center of mass energy are 
\begin{equation}
b\simeq \frac{2}{3}\,R\Delta (\pi n)^{2}\ ,\ \ \ \ \ \ \ \ \ \ \ \ \ 
\mathcal{E}\simeq |p^{-}|2\pi \Delta n\ ,  \notag
\end{equation}
and we conclude that the $O$--plane orbifold is stable, according to this
criteria.

For the null--boost orbifold, start by setting $R=0$ (or $E\rightarrow
\infty $) in the above expressions for the momenta $p_{n}$ and constant $Y$.
Then one obtains that the large $n$ behavior for the center of mass energy
remains unchanged, while the impact parameter becomes $b\simeq 2C$. Hence,
the null--boost orbifold is unstable. In the case of the null--brane, where
one adds a translation to the null--boost orbifold action in a direction
orthogonal to the $\mathbb{M}^{3}$, again $\mathcal{E}$ is unchanged but $%
b\simeq 2\pi Rn$. In this case, provided $D\geq 5$, black holes do not form.
The cases of the boost and shifted--boost orbifolds can be analyzed in a
similarly way. One obtains that $b$ is polynomial in $n$, while $\mathcal{E}$
grows exponentially, with the result that both are unstable. In table 2 we
give a summary of the results.

\begin{table}[tbp]
\caption{Horowitz--Polchinski analysis for the time--dependent orbifolds of $
\bsM^3$.}
\begin{center}
\begin{tabular}{|c|c|c|c|}
\hline
Orbifold & $b$ & $\mathcal{E}$ & Result \\ \hline\hline
Boost & $\spa{0.4}\displaystyle{\sqrt{\frac{2p^-}{p^+}}\,|C^+|}\spa{-0.7}$ & $\sqrt{2p^+p^-}
e^{\pi\Delta n}$ & Unstable \\ \hline
Shifted--boost & $\spa{0.1}2\pi Rn\spa{-0.3}$ & $\sqrt{2p^+p^-} e^{\pi\Delta n}$ & Unstable \\ 
\hline
Null--boost & $\spa{0.1}2C\spa{-0.3}$ & $|p^-|2\pi\Delta n$ & Unstable \\ \hline
$O$--plane & $\spa{0.3}\displaystyle{\frac{2}{3}\,(\pi n)^2 \Delta R}\spa{-0.5}$ & $%
|p^-|2\pi\Delta n$ & Stable \\ \hline
\end{tabular}
\end{center}
\end{table}

This stability argument should be taken with some criticism. In fact, it
seems unlikely that a correct guess on the final qualitative features of the
scattering problem can be obtained by looking at the interaction between two
(or, for that matter, a finite number of) light--rays. This fact is already
true if we just consider the \textit{linear reaction }of the gravitational
field to the image geodesics. Then, very much like in electromagnetism, it
is incorrect to guess the qualitative features of fields by looking at just
a finite subset of the charges (matter in this case), whenever the charge
distribution is infinite (this infinity is really not an approximation in
this case, since it comes from the infinite copies of the particle in the
covering space). In particular, it was shown in \cite{KST}
that, for the boost orbifold with four extra non--compact directions, the
linear gravitational field produced by all the images is pure gauge. 
Therefore, to decide if the problem exists, much more work
is required, already in the linear regime of gravity, but most importantly
in the full non--linear setting. Also, we saw that, according to this argument, 
the $O$--plane orbifold is stable. We shall see, on the other hand, that it 
suffers from the same infinities in the two--particles scattering amplitudes 
found in \cite{LMS1}, questioning the agreement between the two approaches.

Finally, notice that \textit{the only case in which the HP argument is fully correct is
exactly in dimension} $D=3$, \textit{where the gravitational interaction is
topological and when, therefore, the interaction of an infinite number of
charges can be consistently analyzed by breaking it down into finite subsets}. 
This indeed is what we shall find in the next section.

\subsection{Backreaction in three--dimensions}

As we have described in the last section, it is quite important to
understand the full non--linear response of the orbifold geometry due to
small perturbations. Fortunately, at least in three dimensions, the problem
can be solved exactly for a specific type of matter fields. The reason is
that the orbifold geometry is described by two--dimensional dilaton gravity.
Then, for conformally coupled matter, one can derive the full non--linear
solution, including the backreaction of the conformal field. We shall
consider the null--boost and the shifted--boost orbifolds in some detail,
and we shall ask if conformal matter gives rise to a space--like
singularity, changing abruptly the space--time global structure. Notice that
the analysis of the shifted--boost orbifold includes the $O$--plane orbifold
if one takes the near--singularity limit.

Recall the general form for the dimensional reduction of the
three-dimensional metric 
\begin{equation}
ds_{3}^{\,2}=ds_{2}^{\,2}+\Phi ^{2}\left( dz+A\right) ^{2}\ ,  \notag
\end{equation}
where $\partial _{z}$ is a Killing direction. The three--dimensional Hilbert
action is proportional to 
\begin{equation}
\int d^{2}x\,\sqrt{-g}\left( \Phi R-\frac{1}{2}\,\Phi^{3}F^{2}\right) \ . 
\notag
\end{equation}
The equation of motion for the gauge field implies that the scalar $\Phi
^{3}\star F$ is constant. By rescaling $z$, $A$ and $\Phi ^{-1}$ we can fix
the constant to any desired value (provided it does not vanish) so that $%
\star F=2/\Phi ^{3}$. This will be possible for the $O$--plane and for the
shift--boost orbifold. On the other hand, $F=0$ for the boost and the
null--boost orbifolds. Focusing, for now, on the case $F\neq 0$, the
equations of motion for the scalar $\Phi $ and the metric can be derived
from the action 
\begin{equation}
\int d^{2}x\left( \Phi R-V(\Phi )\right) \ ,\ \ \ \ \ \ \ \ \ V(\Phi )=
\frac{2}{\Phi ^{3}}\ .  \notag
\end{equation}
We conclude that the problem of finding the geometry for the orbifolds of $%
\mathbb{M}^{3}$ can be rephrased in the language of two--dimensional
gravity. Note that, in this theory, the metric and the scalar $\Phi $ should
be considered together as the gravitational sector.

Next we wish to add the matter sector, which results in an action of the
general form 
\begin{equation}
S_{2D}(g,\Phi )+S_{M}(g,\Phi ,\mathrm{Matter})\ .  \notag
\end{equation}
The corresponding equations of motion are easily derived to be 
\begin{equation}
\begin{array}{rcl}
\displaystyle{2\nabla _{a}\nabla _{b}\Phi } & = & \displaystyle{g_{ab}\left(
2\,\Box \Phi +V\right) -\tau _{ab}\,}\ ,\phantom{\fbox{%
\rule[-0.4cm]{0cm}{0cm}}} \\ 
R & = & \displaystyle{\frac{dV}{d\Phi }+\rho }\ ,
\end{array}
\notag  \label{metric-eq}
\end{equation}
where $\tau_{ab}$ and $\rho $ are 
\begin{equation*}
\tau _{ab}=-\frac{2}{\sqrt{-g}}\frac{\delta S_{\mathrm{M}}}{\delta g^{ab}}\ ,
\ \ \ \ \ \ \ \ \ \ \ \ \ \ \ \ \ \ \ \ \ \ \ 
\rho =-\frac{1}{\sqrt{-g}}\frac{\delta S_{\mathrm{M}}}{\delta \Phi}\ .
\end{equation*}
Moreover, the conservation of the stress energy tensor $\tau _{ab}$ is
modified by the dilaton current $\rho $ to 
\begin{equation*}
\nabla ^{a}\tau _{ab}+\rho \nabla _{b}\Phi =0\ .
\end{equation*}

The inherent simplicity of the dilaton gravity model lies in the following
observations \cite{Kunstatter, 2Dgravity}. Define $J\left( \Phi \right) $ by 
\begin{equation*}
J=\int Vd\Phi
\end{equation*}
and consider the function 
\begin{equation}
C=\left( \nabla \Phi \right) ^{2}+J\left( \Phi \right)  \label{Cfunct}
\end{equation}
and the vector field 
\begin{equation*}
\kappa ^{a}=\frac{2}{\sqrt{g}}\,\epsilon ^{ab}\,\nabla _{b}\Phi \ .
\end{equation*}
Then, for any vacuum solution $\tau _{ab}=\rho =0$, the function $C$ is
constant and $\kappa $ is a Killing vector. The first fact follows from the
equations of motion, which imply that 
\begin{equation}
\nabla _{a}C=-\tau _{ab}\nabla ^{b}\Phi +\nabla _{a}\Phi \,(\tau
_{bc}g^{bc})\ .  \label{Cvariation}
\end{equation}
The second fact is proved most easily in conformal coordinates $z^{\pm }$,
with metric 
\begin{equation*}
-dz^{+}dz^{-}e^{\Omega }\ .
\end{equation*}
Then $\kappa _{\pm }=\mp \nabla _{\pm }\Phi$, and the non--trivial Killing
equations become $\nabla _{+}\nabla _{+}\Phi =\nabla _{-}\nabla _{-}\Phi =0$%
, which hold whenever $\tau_{ab}=0$. Finally note that these equations are
equivalent to 
\begin{equation*}
\partial _{-}\kappa ^{+}=\partial _{+}\kappa ^{-}=0\ .
\end{equation*}

\begin{figure}[t]
\begin{center}
\begin{tabular}{c}
\epsfysize=6cm\epsfbox{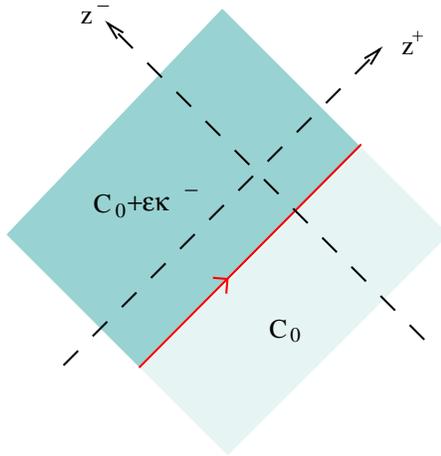}
\end{tabular}
\end{center}
\caption{Shock wave solution in two--dimensional dilaton gravity.}
\label{fig10}
\end{figure}

Let us now analyze the geometry in the presence of matter. For our purposes,
we are going to consider only matter Lagrangians which do not depend on the
dilaton, and which are conformal. This implies that 
\begin{equation}
\begin{array}{c}
\tau _{+-} =\rho =0\ , 
\\ 
\partial _{-}\tau _{++} =\partial _{+}\tau _{--}=0\ .
\end{array}
\notag
\end{equation}
The simplest example is clearly a conformally coupled scalar $\eta $ with $%
S_{M}=-\int \left( \nabla \eta \right) ^{2}$. The effect of this type of
matter is best described by considering a shock wave \cite{CGHS},
which is represented in conformal coordinates by a stress energy tensor of
the form 
\begin{equation*}
\tau _{--}\left( z^{-}\right) =\epsilon \,\delta
\left(z^{-}-z_{0}^{-}\right)\ ,\ \ \ \ \ \ \ \ \ \ \ \ \ \ \ \ \ \ \ \ \ \ \
\ \left( \epsilon >0\right)\,.
\end{equation*}
The positivity of $\epsilon $ can be understood by looking at the
conformally coupled scalar, for which $\tau _{--}=2\left( \nabla _{-}\eta
\right) ^{2}>0$. Recalling from (\ref{Cvariation}) that 
\begin{equation}
\nabla _{-}C=2\tau _{--}\nabla _{+}\Phi \,e^{-\Omega }=\tau _{--}\kappa
^{-}\,,  \label{Cjump}
\end{equation}
we conclude that the shock front interpolates, as we move along $z^{-}$,
between the vacuum solution with $C=C_{0}$ and the vacuum solution with 
$C=C_{0}+\epsilon\kappa^{-}(z_{0}^{-}) $ (see figure \ref{fig10}). 
As a consistency check note that, since in the vacuum $\tau _{--}$ and 
$\kappa ^{-}$ are functions only of $z^{-}$, equation (\ref{Cjump}) defines a
jump in the function $C$ which is independent of the position $z^{+}$ along
the shock wave.

We are now in a position to study the coupling of conformal matter to the
orbifolds geometry, including non--linear effects. Consider first the case
of the shifted--boost orbifold \cite{CC2}. It is easy to verify that the
corresponding geometry corresponds to a constant $C$ given by 
\begin{equation}
C=-E\ .  \notag
\end{equation}
For example, in the static regions II one has 
\begin{equation}
\begin{array}{rcl}
ds_{2}^{\,2} & = & \displaystyle{-dt^{2}\left( \frac{E^{2}x^{2}}{1-E^{2}x^{2}%
}\right) +dx^{2}}\ ,\phantom{\fbox{\rule[-0.4cm]{0cm}{0cm}}} \\ 
\sqrt{E}\,\Phi & = & \displaystyle\sqrt{1-E^{2}x^{2}}\ .\label{vacuumsol}
\end{array}
\end{equation}
Note that we have rescaled the field $\Phi $ from section 2.2 in order to
have a canonically normalized potential $2\Phi ^{-3}$. Given the parameter $%
E $, one has no freedom in the solution, which is unique. This shows that
there is no fine tuning in the choice of initial conditions for the metric
and the dilaton, in order to obtain a bounce cosmological solution with past
and future cosmological horizons.

\begin{figure}[t]
\begin{center}
\begin{tabular}{c}
\epsfysize=7cm\epsfbox{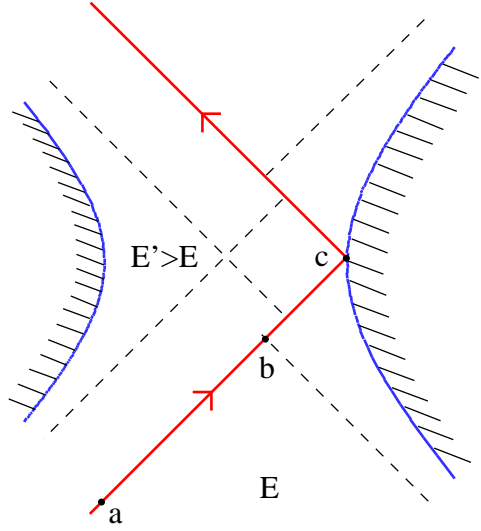}
\end{tabular}
\end{center}
\caption{Shock wave solution in the cosmological geometry associated to the
shifted--boost orbifold.}
\label{fig11}
\end{figure}

Next we add matter by considering shock wave solutions. Given the above
discussion, it is immediate to see that, after the wave, one has again a
vacuum solution, but with a different constant 
\begin{equation}
E^{\prime }=E-\epsilon \kappa ^{-}\ ,  \notag
\end{equation}
where $\epsilon >0$ and $\kappa ^{-}=2e^{-\Omega }\nabla _{+}\Phi $ must be
computed along the wave. In figure \ref{fig11} the new geometry is
represented. As one moves in this figure from point $a$ to points $b$ and $c$
along the shock wave, in the direction of increasing $z^{+}$, the value of $%
\Phi $ decreases to $0$ at $c$ on the singularity. Therefore $\nabla
_{+}\Phi <0$ and one has that 
\begin{equation}
E^{\prime }>E\ .  \notag
\end{equation}
Moreover, in any vacuum solution with $C=-E$, the value of the dilaton on
the horizons is $1/\sqrt{E}$, as can be seen from equation (\ref{vacuumsol})
at $x=0$. Therefore, since the value of $\Phi $ is continuous across the
shock wave, the horizon to the left of the wave, where the dilaton has value 
$\Phi =1/\sqrt{E^{\prime }}$, must intersect the wave between\ the points $b$
and $c$, as drawn. Let us briefly explain why the horizon at a constant
value of $z^{+}$ shifts as one passes the shock wave. It is easy to see that
the horizon in question is given by the curve $\kappa ^{+}=0$. In the
vacuum, $\kappa ^{+}$ is a function of $z^{+}$ alone, but in the presence of
matter one has that 
\begin{equation*}
\partial _{-}\kappa ^{+}=e^{-\Omega }\tau _{--}\,.
\end{equation*}
Then, since $\Omega $ is constant along the horizons (and therefore along
the shock wave) and since $\tau _{--}$ has a delta singularity, the function 
$\kappa ^{+}$ just jumps by a finite constant across the wave, thus
explaining the shift in the position of the horizon.

\begin{figure}[t]
\begin{center}
\begin{tabular}{c}
\epsfysize=7cm\epsfbox{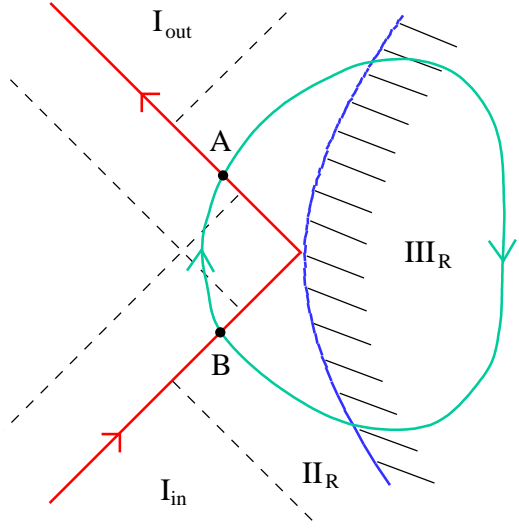}
\end{tabular}
\end{center}
\caption{A closed time--like curve in the three--dimensional geometry induced
by the matter surface, which is the oxidization of the shock front.}
\label{fig12}
\end{figure}

In conclusion, for the shifted--boost orbifold, the addition of matter does
not change the global structure of space--time, because the solution
interpolates between two non--BPS vacua with the same global structure. In
particular, there is no space--like singularity leading to a catastrophic
big--crunch. Is this result in contradiction with the argument reviewed in
the previous section? The answer is no. To see this consider the uplift to
three dimensions of the shock front geometry. It is given by two pieces of
flat space separated by a surface of matter. This distribution of matter is
nothing but the continuous image of a light ray generated by the action of
the orbifold Killing vector. Then a simple generalization of the argument
for the formation of large--black holes to continuous surfaces leads to the
same instability as before. However, in three--dimensional gravity there are
no black holes, a fact that follows simply because $GM$ is dimensionless. In
this case, the instability analogous to the formation of black holes is the
appearance of CTC's in the covering space \cite{Gott,Deser,Kabat,Gott2}. In
fact, given a two particle scattering process, the instability condition in
three dimensions 
\begin{equation}
G\mathcal{E}>1\ ,  \notag
\end{equation}
becomes the condition for the formation of CTC's. A careful analysis of the
covering space geometry corresponding to the shock front geometry, indeed
shows that such CTC's do appear, and there is no contradiction. However, all
these covering space CTC's cross the surface that corresponds to the
time--like singularities of regions II. Such a closed time--like curve is
represented in the compactified space in figure \ref{fig12}. Provided one
interprets the singularities as boundaries of space--time, and accordingly
excises from the geometry the region behind it, one concludes that the final
geometry is free of CTC's and the above instability is cured.

\begin{figure}[t]
\begin{center}
\begin{tabular}{c}
\epsfysize=6cm\epsfbox{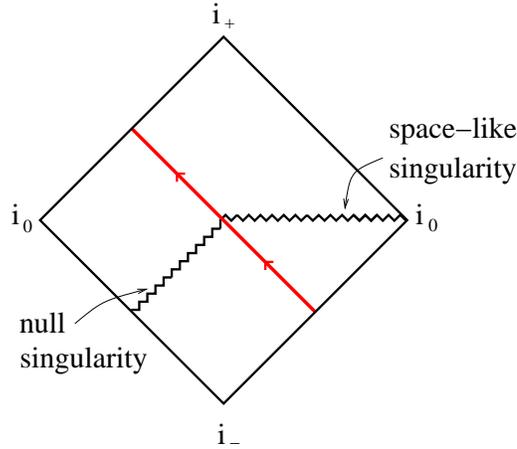}
\end{tabular}
\end{center}
\caption{CP diagram for the shock wave solution in the null--boost geometry.}
\label{fig13}
\end{figure}

Let us now consider the case of the null--boost orbifold. Lawrence analyzed
the reaction of the geometry when conformally coupled matter strikes the
null singularity \cite{Lawrence}. In this case, however, an instability is
found, which indicates a behavior already expected from the limiting form of
the wave functions at the big--crunch singularity. The problem can again be
rephrased in the language of two--dimensional dilaton gravity, but now in a
theory with a vanishing dilaton potential. The vacuum null--boost geometry
is the $C=0$ solution, which is supersymmetric. When one introduces a shock
wave heading towards the singularity, the constant $C$ will become negative
and one connects, after the shock wave, to the non--supersymmetric pure
boost space--time. The latter geometry has a totally different global
structure, as we saw in section 2, with a space--like singularity
corresponding to the big crunch. In figure \ref{fig13} the CP diagram
representing the gluing of both geometries across the shock wave is shown.

Finally, if one considers perturbing the pure boost geometry with a shock
wave, one does not expect drastic changes of space--time structure, since
one interpolates, across the shock, between two non--BPS geometries with
equal global structure.

\subsection{Tree--level Amplitudes}

In this section we concentrate on the computation of tree--level amplitudes
of field theory and string theory on the orbifolds 
$(\mathbb{M}^{3}/e^{\kappa})\mathbb{\times T}^{D-3}$, 
where $D$ is the space--time dimension.
The basic tool used to compute these amplitudes is the inheritance
principle, which states that we may use directly the amplitudes of the
parent theory on $\mathbb{M}^{3}\mathbb{\times T}^{D-3}$, as long as we
restrict our attention to external states which are invariant under the
orbifold action. This principle is certainly valid in field theory, and is
also the correct prescription for string states which do not carry winding
charge. For concreteness of exposition, we shall consider only the 
$O$--plane orbifold \cite{CC3}, but notice that similar 
techniques have been used for other orbifolds discussed in these lectures:
the amplitudes for the null--boost and null--brane, which can be 
derived easily from the computation here presented, were considered 
in \cite{LMS1,LMS2} and for the boost--orbifold in \cite{BCKR}.

Let us discuss first, in general, the $n$--point amplitude and then
restrict our attention to $n=3,4$. Let the parent amplitude be given by 
\begin{equation*}
\delta ^{3}\left( \tsum\nolimits_{i}\vec{p}_{i}\right) \mathcal{A}\left( 
\vec{p}_{1},\cdots ,\vec{p}_{n}\right) \,,
\end{equation*}
where the momenta $\vec{p}_{i}$ refer to the momenta in the $\mathbb{M}^{3}$
directions . We will consider as given, once and for all, the discrete
momenta $\vec{p}_{i\perp }$ in the torus directions $\mathbb{T}^{D-3}$, with the
only obvious requirement that $\sum_{i}\vec{p}_{i\perp }=0$.

As we saw in section 2.3, the external states are characterized by their
mass $m_{i}$, together with the conserved quantum numbers $(p_{i+},n_{i})$.
Moreover, the mass is clearly related to the $D$--dimensional mass $M_{i}^{\,2}$ 
by 
\begin{equation*}
m_{i}^{\,2}=M_{i}^{\,2}+(\vec{p}_{i\perp })^2\,.
\end{equation*}
Using the basic external states (\ref{Owf}), we may directly apply the inheritance
principle to obtain the expression 
\beq
\frac{1}{\sqrt{\left| \prod_{i}p_{i+}\right| }}\int dp_{1}\cdots dp_{n}\,
\delta\left( \tsum\nolimits_{i}p_{i+}\right) 
\delta\left(\tsum\nolimits_{i}p_{i}\right) 
\delta\left( \tsum\nolimits_{i}p_{i-}\right)
e^{i\varphi \left( p_{i}\right) }\,\mathcal{A}\left( \vec{p}_{i}\right)\ .
\label{s1}
\eeq
As we just mentioned, the momenta $p_{i+}$ are fixed. On the other hand,
the momenta $p_{i}$, which are momenta in the $X$--direction, are integrated  
and the momenta $p_{i-}$ are given by the quadratic on--shell condition 
\begin{equation*}
p_{-}=\frac{p^{2}+m^{2}}{2p_{+}}\,.
\end{equation*}
Therefore, of the three delta functions, the one related to the 
$X^+$--direction factors out of the integral, whereas the ones related to the
directions $X$ and $X^-$ give, respectively, a linear and a quadratic
constraint on the integration variables $p_{i}$. Finally, the phase 
$\varphi\left( p_{i}\right) $ is given by 
\begin{equation*}
\varphi \left( p_{i}\right) =\frac{1}{E}
\sum_{i}\left( \frac{p_{i}n_{i}}{Rp_{i+}}
-\frac{p_{i}m_{i}^{\,2}}{2p_{i+}^{2}}
-\frac{p_{i}^{\,3}}{6p_{i+}^{\,2}}\right)\,.
\end{equation*}

In the above expression for the amplitude, we have actually over--counted
the final answer, due to the invariance of the full expression under 
the isometries generated by the Killing vector $\kappa$. 
To understand this fact, consider the
action of the isometry generated by $\kappa$ on the plane--wave external
momenta, by defining the transformed momenta $\vec{p}_{i}\,'$ 
\begin{eqnarray*}
p_{i+}' & = & p_{i+} \\
p_{i}' & = & p_{i}+\beta p_{i+} \\
p_{i-}' & = & p_{i-}+\beta p_{i}+\frac{1}{2}\,\beta^{2}p_{i+}\ ,
\end{eqnarray*}
where $\beta \in \mathbb{R}$ parametrizes the action of the isometry. Note
that, due to the conservation $\sum_{i}\vec{p}_{i}=0$, we can show that 
\begin{equation*}
\varphi \left( p_{i}^{\prime }\right) =\varphi \left( p_{i}\right) +\frac{%
\beta }{ER}\tsum\nolimits_{i}n_{i}\,.
\end{equation*}
Thus, if the charge $n_{i}$ is conserved, the phase $\varphi \left(
p_{i}\right) $ is invariant. Moreover, due to Lorentz invariance, the
amplitude $\mathcal{A}$ does not change under the isometry $\kappa $.
Therefore, in order to undo the over--counting, we follow the standard
Faddeev--Popov procedure. First we must choose a gauge--fixing, which
depends on convenience of computation. The simplest possible gauge choice is
a linear constraint $\tsum\nolimits_{i}c_{i}p_{i}^{\prime }=0$, where the
constants $c_{i}$ are chosen case--by--case to simplify the expressions. We
then insert, in the integral (\ref{s1}), the identity ``$1$'' 
\begin{equation*}
\left| \tsum\nolimits_{i}c_{i}p_{i+}\right| \int d\beta \,\delta \left(
\tsum\nolimits_{i}c_{i}p_{i}^{\prime }\right) ,
\end{equation*}
where we are implicitly assuming that $\tsum\nolimits_{i}c_{i}p_{i+}\neq 0$.
Changing variables to the primed momenta $p_{i}^{\prime }$, using the
invariance of the phase $\varphi $ and of the amplitude $\mathcal{A}$, and
dropping the primes, we are left with the integral (\ref{s1}) with the extra linear
delta function 
\begin{equation*}
\delta \left( \tsum\nolimits_{i}c_{i}p_{i}\right) \,,
\end{equation*}
together with the normalization 
\begin{equation*}
\left| \tsum\nolimits_{i}c_{i}p_{i+}\right| \int d\beta 
\,e^{i\frac{\beta }{ER} \tsum\nolimits_{i}n_{i}}
\longrightarrow 
2\pi ER\left|\tsum\nolimits_{i}c_{i}p_{i+}\right| 
\delta_{_{\sum_{i}n_i}}\ .
\end{equation*}
Note that we have eliminated the over--counting by restricting the
integration over $\beta $ to a single action of the orbifold generator, from 
$0$ to $2\pi ER$, thus replacing the Dirac delta function with the Kronecker
symbol. We are then left with the final expression 
\beq
\barr{rcl}
\mathcal{A}\left(p_{i+},n_{i}\right) & = & 
\displaystyle{\left( 2\pi ER\right)\,\, 
\delta_{_{\sum_{i}n_{i}}}\,\,
\delta \left(\tsum\nolimits_{i}p_{i+}\right)\, 
\frac{\left| \tsum\nolimits_{i}c_{i}p_{i+}\right|}
{\sqrt{\left|\prod_{i}p_{i+}\right| }}}
\spa{0.6}\\&&
\displaystyle{\int dp_{1}\cdots dp_{n}\,
\delta\left( \tsum\nolimits_{i}p_{i}\right) \,
\delta\left(\tsum\nolimits_{i}p_{i-}\right) \,
\delta\left(\tsum\nolimits_{i}c_{i}p_{i}\right)\,
e^{i\varphi \left( p_{i}\right) }\,
\mathcal{A}\left( \vec{p}_{i}\right)}\ .
\notag
\earr
\eeq
The three $\delta $ functions inside the integral reduce the $n$
integrations to $n-3$. Now we move to the concrete examples of the three-- and
four--point functions. In what follows we shall omit the overall factor 
$\left( 2\pi ER\right)\,\delta_{_{\sum_{i}n_{i}}}\,
\delta\left(\tsum\nolimits_{i}p_{i+}\right)$, 
which we leave as understood.

\subsubsection{The three--point amplitude}

Let us choose, for concreteness, particles $1$, $2$ to be incoming and
particle $3$ to be outgoing, so we have $p_{1+},$ $p_{2+}>0$ and $p_{3+}<0$.
We also assume, for simplicity, that the parent amplitude is just a constant 
$\mathcal{A}=1$. We choose the gauge $p_{3}=0$, so that the amplitude reads 
\beq
\sqrt{\left|\frac{p_{3+}}{p_{1+}p_{2+}}\right|}
\int dp_{1}dp_{2}dp_{3}\,
\delta\left(\tsum\nolimits_{i}p_{i}\right)\,
\delta\left(\tsum\nolimits_{i}p_{i-}\right)\, 
\delta\left(p_{3}\right)\,
e^{i\varphi\left( p_{i}\right)}\ .
\notag
\eeq 
Choosing as integration variable $p_{1}=-p_{2}$, with $p_{3}=0$, we obtain 
\beq
2\,\sqrt{\frac{\left|p_{3+}\right|}{\left|p_{1+}p_{2+}\right|}}
\int dp_{1}\,
\delta\left(4\alpha +p_{1}^{\,2}(\mu_{12})^{-1}\right)\,
e^{i\varphi \left( p_{i}\right)}\ ,
\notag
\eeq
where we have defined 
\begin{equation*}
\mu _{12}=\frac{p_{1+}p_{2+}}{p_{1+}+p_{2+}}\ ,
\ \ \ \ \ \ \ \ \ \ \ \ \ \ \ \ 
\alpha =\sum_{i}\,\frac{m_{i}^{\,2}}{4p_{i+}}\ .
\end{equation*}
Therefore the amplitude vanishes if $\alpha >0$. The result can, in general,
be written in terms of $\varphi(p_1,p_2,p_3)$ as
\begin{equation*}
2\,\frac{\sqrt{\mu _{12}}}{\bar{p}}\,
\theta\left(-\alpha\right)\,
\cos \varphi\left(\bar{p},-\bar{p},0\right) ,
\end{equation*}
where 
\begin{equation*}
\bar{p}=\sqrt{-4\alpha \mu _{12}}\ .
\end{equation*}

\subsubsection{The four--point amplitude}

We consider the scattering of incoming particles $1,2$ into outgoing
particles $3,4$, so that we have $p_{1+},$ $p_{2+}>0$ and $p_{3+},$ $p_{4+}<0
$. A natural gauge choice is $p_{1}+p_{2}=0$, so that the expression for the
amplitude reads 
\beq
\frac{p_{1+}+p_{2+}}{\sqrt{p_{1+}p_{2+}p_{3+}p_{4+}}}
\int dp_{1}\cdots dp_{n}\,
\delta\left(\tsum\nolimits_{i}p_{i}\right)\, 
\delta\left(\tsum\nolimits_{i}p_{i-}\right)\,
\delta\left( p_{1}+p_{2}\right)\,
e^{i\varphi \left( p_{i}\right)}\,
\mathcal{A}\left(s,t\right)\ ,
\notag
\eeq
where we have used the Lorentz invariance of $\mathcal{A}$ to replace the
momenta $\vec{p}_{i}$ with the Mandelstam variables 
\begin{eqnarray*}
&&s=-\left(\vec{p}_{1}+\vec{p}_{2}\right)^{2}+s_{\perp }\,, \\
&&t=-\left(\vec{p}_{1}+\vec{p}_{3}\right)^{2}+t_{\perp }\,,
\end{eqnarray*}
with $s_{\perp }=-\left( \vec{p}_{1\perp }+\vec{p}_{2\perp }\right)^{2}$, 
$t_{\perp}=-\left( \vec{p}_{1\perp }+\vec{p}_{3\perp }\right) ^{2}$. In order to solve the
quadratic $p_{-}$ constraint, it is convenient to introduce, as for the
three--point function, the positive constants 
\begin{equation*}
\mu _{12}=\frac{p_{1+}p_{2+}}{p_{1+}+p_{2+}}\ ,
\ \ \ \ \ \ \ \ \ \ \ \ \ \ \ \ 
\mu _{34}=-\frac{p_{3+}p_{4+}}{p_{3+}+p_{4+}}\ ,
\end{equation*}
together with 
\begin{equation*}
\alpha =\sum_{i}\frac{m_{i}^{\,2}}{4p_{i+}}\ .
\end{equation*}
It is then relatively straightforward to show that the amplitude reduces to
the following expression 
\begin{equation}
\int dq d\widetilde{q}\,\,
\delta\left(q\widetilde{q}-\alpha\right)\,
e^{i\varphi }\mathcal{A}\ ,  
\label{s2}
\end{equation}
where the momenta $p_{i}$ are defined in terms of the integration variables
by 
\begin{eqnarray*}
&&p_{1} = -p_{2}=\sqrt{\mu _{12}}\left( q-\widetilde{q}\right) \,, \\
&&p_{3} = -p_{4}=-\sqrt{\mu _{34}}\left( q+\widetilde{q}\right) \,.
\end{eqnarray*}

For generic kinematics the amplitude is well defined and, in fact, can be
approximated by doing a saddle point computation for small $E$. Let us then 
discuss the basic problem in the amplitude (\ref{s2}), which is common to the 
time--dependent orbifolds here considered, and was first analyzed in \cite{LMS1}. 
Consider the specific kinematical regime
\begin{equation}
n_{1}+n_{3}=p_{1+}+p_{3+}=0\text{\thinspace },  \label{s3}
\end{equation}
i.e. vanishing $t$--channel exchange in the conserved 
$(\mathbb{M}^{3}/e^{\kappa})$--charges. We also assume, for simplicity, that the masses 
$m_{i}=m$ are all equal. In this case we have that 
\begin{equation*}
\alpha =0\,,
\ \ \ \ \ \ \ \ \ \ \ \ \ \ \ \ \ \ \ \ \ \ \ \ \ 
\mu _{12}=\mu _{34}\,.
\end{equation*}
The integral (\ref{s2}) splits into two branches, with $q=0$ and 
$\widetilde{q}=0$, respectively. Let us focus on the $\widetilde{q}=0$ branch, 
where we have 
\begin{equation*}
p_{1}=-p_{2}=-p_{3}=p_{4}=\sqrt{\mu_{12}}\,q
\end{equation*}
and therefore the $t$--exchange $\vec{p}_{1}+\vec{p}_{3}=0$ vanishes
throughout the integral for all values of $q$. On this branch, the phase $%
\varphi$ also vanishes. Finally, the Mandelstam variables $s$, $t$ are
given by 
\begin{eqnarray*}
&&s\left( q\right) = 
s_{\perp }+\left( m^{2}(\mu _{12})^{-1}+q^{2}\right)
\left(p_{1+}+p_{2+}\right)\,,\\
&&t\left( q\right) = t_{\perp}\,.
\end{eqnarray*}
Putting everything together we arrive at the expression 
\begin{equation*}
\int \frac{dq}{\left|q\right|}\,\,
\mathcal{A}\left(s\left(q\right),t_{\perp }\right)\,.
\end{equation*}
As $\left|q\right|\rightarrow\infty$, the center of mass energy $s$ goes
to infinity as $q^{2}$, while the $t$--exchange is fixed at $t_{\perp }$.
Therefore we are in the small--angle Regge regime of the amplitude, where we
expect a similar behavior for the parent amplitude $\mathcal{A}$ in string
theory and in field theory, a behavior of the form 
\begin{equation*}
\mathcal{A}\sim G\,\frac{s^{J}}{-t}\ ,
\end{equation*}
where $G$ is the coupling and where $J$ is the spin of the exchanged
massless minimally coupled particle. For a field--theoretic graviton
exchange, $J=2$, whereas in string theory, which exhibits Regge behavior, 
$J=2+\frac{1}{2}\alpha ^{\prime }t$. In both cases, we should interpret 
$Gs^{J}$ as the effective coupling, which diverges in the 
$q\rightarrow\infty $ limit, rendering the integral ill--defined, and signaling the
breakdown of perturbation theory. Note that, since $t$ is fixed, the
divergence is present in string theory whenever 
$-\alpha ^{\prime }t_{\perp}\geq 4$, which is the basic result of \cite{LMS1}. 
Let us note that the $O$--plane orbifold discussed here is stable to 
formation of large black--holes, following the analysis in section 3.1. 
Therefore the above computation contradicts the claim, often found in the 
literature, that the Horowitz--Polchinski instability is responsible, 
indirectly, for the breakdown of perturbation theory. Moreover, note that 
the $O$--plane invariant external states (\ref{Owf}) are perfectly regular functions 
in the covering space, and do not exhibit any focusing with a diverging
wave--function. This is also not the cause of the breakdown of perturbation
theory.

\subsection{Eikonal Resummation}

We have seen in the previous section that, for vanishing $t$--exchange, the
amplitude diverges, signaling a breakdown of perturbation theory. We now
wish to better understand the structure of the divergence, by considering
the amplitude as $p_{1+}+p_{3+}\rightarrow 0$. In order to keep notation to
a minimum, and to be able to focus on the essential point, let us specialize
to the massless case with 
\begin{equation*}
\vec{p}_{i\perp }=M^{2}=m^{2}=0\text{\thinspace }.
\end{equation*}
The reader can think, for instance, at the case of scattering, in
superstring theory, of four dilatons which have no momentum in the
transverse compact directions. Let us start by relaxing the
condition (\ref{s3}) by defining 
\begin{equation*}
\delta =\frac{1}{2\sqrt{p_{1+}p_{2+}}}\left( p_{1+}+p_{3+}\right)\,, 
\end{equation*}
so that we shall study the amplitude as a function of $\delta\ll 1$. A
simple computation shows that the Mandelstam variables in string units are
now given, to leading order in $\delta $, by 
\begin{eqnarray*}
&&\alpha^{\prime}s = 
\alpha^{\prime}q^{2}\left(p_{1+}+p_{2+}\right)=\lambda^{2}\,, \\
&&\alpha ^{\prime }t \simeq 
-\alpha^{\prime}s\,\delta^{2}=-\lambda^{2}\delta^{2}\,,
\end{eqnarray*}
where we have defined the dimensionless integration variable 
\begin{equation*}
\lambda=q\sqrt{\alpha ^{\prime }\left( p_{1+}+p_{2+}\right) }\ .
\end{equation*}
Moreover, the phase $\varphi $ is given, again to leading order in 
$\delta $, by the expression 
\begin{equation}
\varphi (\lambda)\simeq \frac{\delta}{E\alpha^{\prime 3/2}p_{1+}p_{2+}}
\left[ -\frac{n}{R}\left( p_{1+}+p_{2+}\right) 
\alpha ^{\prime }\lambda+\frac{1}{6}\,\lambda^{3}\right]\,.  
\notag
\end{equation}
We see that the ratio 
\begin{equation*}
\frac{-t}{s}\simeq \delta ^{2}
\end{equation*}
is fixed for fixed $\delta $, and the large $\lambda$ region of the integration is
therefore dominated by fixed angle scattering. As it is well known in string
theory, at fixed angles, the amplitude is exponentially damped whenever 
$\alpha^{\prime }s,\alpha ^{\prime }t\gg 1$, due to the finite size of
the string, or, equivalently, to the presence of the infinite tower of
massive modes. Therefore, the integral defining the amplitude is effectively
cut at 
\begin{equation*}
\lambda_{t}=\frac{1}{\delta}\,.
\end{equation*}
Let us assume, to estimate the integral defining the amplitude, that the
parent amplitude is dominated,  up to $\lambda_{t}$, by the graviton exchange
\begin{equation*}
\mathcal{A}\sim G\,\frac{s^{2}}{-t}\sim \frac{G}{\alpha'}\,
\frac{\lambda^{2}}{\delta^{2}}\ .
\end{equation*}
We are omitting the correction due to the higher massive modes, 
which modify this formula and give the Regge behavior. 
Therefore we see that the integral (\ref{s2}) is given by 
\begin{equation*}
2\,\frac{G}{\alpha'}\frac{1}{\delta ^{2}}
\int_{0}^{1/\delta}d\lambda\,\lambda\,
e^{i\varphi\left(\lambda\right)}\,.
\end{equation*}
Neglecting the phase $\varphi $ in the $\delta \rightarrow 0$ limit, we see
that the orbifold amplitude goes as 
\begin{equation}
\frac{G}{\alpha'}\,\frac{1}{\delta^{4}}\,,
\label{s4}
\end{equation}
a highly non--integrable singularity in $\delta$ (note that one may consider 
building small wave packets and integrate the above result over $\delta $ 
to alleviate the divergence).

\begin{figure}[t]
\begin{center}
\begin{tabular}{c}
\epsfysize=9cm\epsfbox{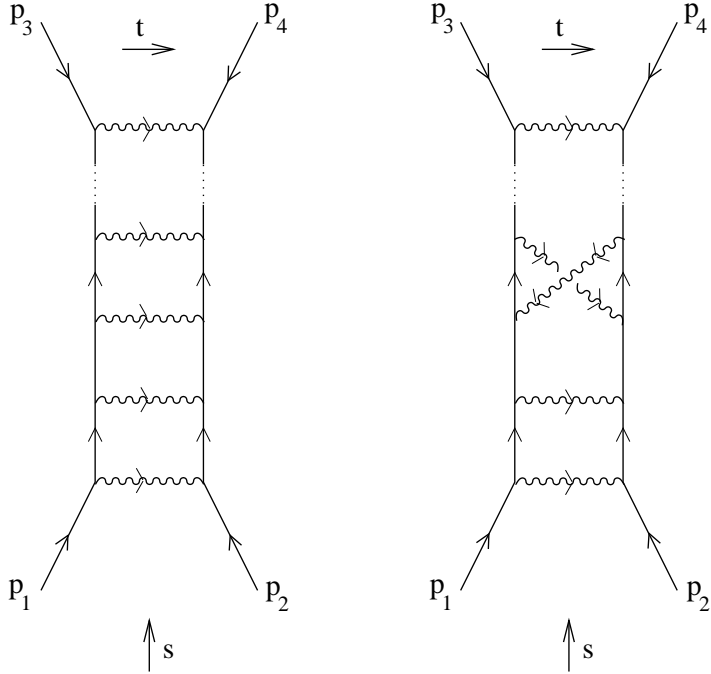}
\end{tabular}
\end{center}
\caption{Planar and non--planar ladder graphs for the four--point amplitude. In the
eikonal approximation one assumes that the momentum along the vertical lines is much
larger than the exchanged graviton momentum.}
\label{fig14}
\end{figure}

We have seen that the major divergence comes from the region of large $s$,
with $t$ bounded. This is the regime of high--energy small--angle scattering
in which, to estimate the amplitude it is necessary to go beyond tree level, 
and often one uses the standard eikonal approximation. This approximation resums part 
of the generalized ladder graphs represented in figure \ref{fig14}, in which the intermediate
gravitons are soft, and where the external scattered particles are
considered essentially as classical particles. In order to use the eikonal
approximation, though we will have to make the following assumptions:
\begin{itemize}
\item[$\bullet$] 
We will (naively) apply the inheritance principle to a loop amplitude
of the parent theory. This is certainly part of the full result in the
orbifold theory, but we are leaving out all graphs where the orbifold group
acts non--trivially in the internal loops.
\item  We are going to assume that the eikonal scheme is a good
approximation to the problem of high--energy scattering. For early 
references on the subject see \cite{Cheng,Abarbanel,Levy}.
\item  We assume, for convenience, that the problem is
essentially three--dimensional. In order to achieve this, it is simplest to take 
the compactification scale to be of the order of the string scale. Then, in
the scattering process, before the amplitude is damped exponentially by
string effects, the $t$--exchange is smaller than the compactification scale
and the compactified momenta are not appreciably excited.
\end{itemize}

The eikonal approximation in string theory has been considered in \cite{Muzinich,Veneziano}, 
and the result is analogous to the field theory results in \cite{tHooft}, where the
scattering is dominated by the eikonal graviton exchange. The graviton
exchange can be resummed, with a resulting expression depending on the
number of non--compact dimensions. The result in dimension three is given in
\cite{DMS} and reads 
\begin{equation*}
\mathcal{A}\sim-G\,\frac{s^{2}}{t+\left(2\pi Gs\right)^{2}}\ .
\end{equation*}
We therefore see that, for 
\begin{eqnarray*}
\left( 2\pi Gs\right) ^{2} &\gg &-t\,, \\
\lambda &\gg &\lambda_{e}\,,
\end{eqnarray*}
where we defined
\begin{equation*}
\lambda_{e}=\frac{\sqrt{\alpha ^{\prime }}}{2\pi G}\,\delta \,,
\end{equation*}
the amplitude $\mathcal{A}$ goes to a constant 
\begin{equation*}
\mathcal{A}\sim -\frac{1}{(2\pi)^2}\,\frac{1}{G}\,.
\end{equation*}
Again neglecting the phase, we conclude that a corrected version of the amplitude 
for the orbifold theory is given by
\beq
2\,\frac{G}{\alpha'}\frac{1}{\delta ^{2}}\int_{0}^{\lambda_{e}}d\lambda\,\lambda
-\frac{1}{2\pi^2}\,\frac{1}{G}\int_{\lambda_{e}}^{\lambda_{t}}\frac{d\lambda}{\lambda}
\,\,\sim\,\,
\frac{1}{(2\pi)^2}\,\frac{1}{G}
\left[1+2\ln\left(\frac{\sqrt{\alpha^{\prime}}\,\delta^2}{2\pi G}\right)\right]\,.
\eeq
The singularity is clearly much milder in $\delta $ then equation (\ref{s4}), 
and it is now perfectly integrable. As explained above, even if we are
not including all the orbifold graphs due to the internal loops, the graphs
here consider already cure the divergence. This is the first hint that, although
much needs to be understood in these orbifold models, gravity, or more
precisely string theory, might possibly be a valid description. We need
though more control over scattering at trans--planckian energies, a
notoriously difficult subject.

\subsection{One--loop Amplitudes}

In this section we discuss the computation of the partition function in
(bosonic) string theory. This is the simplest possible exact one--loop
computation in string theory. Even though these computations are formally
possible, their physical interpretation is still not clear. They generically
present divergences which are not understood and might again signal a
problem in perturbation theory, or alternatively, are related to the
quantization of the coupling constant to be discussed in section 4.1. As an
example of these kind of computations, we shall concentrate in this section
on the shifted boost orbifold \cite{CC1,Nekrasov}, with identifications given by 
\begin{equation}
X^{\pm }\sim e^{\pm 2\pi \Delta }X^{\pm }\ ,\ \ \ \ \ \ \ \ \ \ \ \ \ \ \
X\sim X+2\pi R\ .  \label{s100}
\end{equation}
These computations can be carried out in all the orbifolds discussed in
these lectures. For the null--boost and null--brane case see \cite{LMS1,LMS2}.

We will use units such that $\alpha ^{\prime }=2$. We concentrate on the
sector with winding number $w$. The mode expansion of the field 
$X\left( z,\overline{z}\right) $ is the usual one of a compact boson (where, as usual, 
$z$ is the complex coordinate on the Euclidean string world--sheet). The only
difference with the standard $S^{1}$ compactification is given by a modified
constraint on the total momentum $P$, which must be compatible with the
identification (\ref{s100}) and must therefore satisfy $e^{2\pi i\left(
RP+\Delta J\right) }=1$, or 
\begin{equation*}
P=\frac{1}{R}\left( n-\Delta J\right) \ ,
\end{equation*}
where $n$ is an integer and $J$ is the boost operator. The left and right
momenta for $X$ are then given, as usual, by 
\begin{equation*}
p_{L,R}=P\pm \frac{wR}{2}\,.
\end{equation*}
On the other hand the mode expansions of the fields 
$X^{\pm }\left( z,\overline{z}\right)$ are modified and are given explicitly by 
\begin{equation*}
X^{\pm }\left( z,\overline{z}\right) =i\sum_{n}\left( \frac{1}{n\pm i\nu }
\frac{a_{n}^{\pm }}{z^{n\pm i\nu }}+\frac{1}{n\mp i\nu }\frac{\widetilde{a}
_{n}^{\pm }}{\overline{z}^{n\mp i\nu }}\right) ,
\end{equation*}
where $\nu =w\Delta $ and where the oscillators satisfy the commutation
relations 
\begin{equation*}
\left[ a_{m}^{\pm },a_{n}^{\mp }\right] =-\left( m\pm i\nu \right) \delta
_{m+n}\ ,\ \ \ \ \ \ \ \ \ \ \ \ \ \ \ \left[ \widetilde{a}_{m}^{\pm },
\widetilde{a}_{n}^{\mp }\right] =-\left( m\mp i\nu \right) \delta _{m+n}\ ,
\end{equation*}
and the hermitianity conditions $\left( a_{m}^{\pm }\right) ^{\dagger
}=a_{-m}^{\pm }$, $\left( \widetilde{a}_{m}^{\pm }\right) ^{\dagger }=
\widetilde{a}_{-m}^{\pm }$.

Let us focus on the zero--mode sector, with oscillators satisfying the
relations 
\begin{equation*}
\left[ a_{0}^{\pm },a_{0}^{\mp }\right] =\mp i\nu \ ,\ \ \ \ \ \ \ \ \ \ \ \
\ \ \ \left[ \widetilde{a}_{0}^{\pm },\widetilde{a}_{0}^{\mp }\right] =\pm
i\nu \ .
\end{equation*}
The correct way \cite{BerkoozBoris} to quantize the above commutators is to
start from the usual position and momentum operators $x^{\pm }$ and $P^{\pm }
$, and to construct the combinations 
\begin{equation*}
a_{0}^{\pm }=P^{\pm }\pm \frac{\nu }{2}x^{\pm }\,,\,\ \ \ \ \ \ \ \ \ \ \ \
\ \ \ \ \ \ \ \ \ \ \widetilde{a}_{0}^{\pm }=P^{\pm }\mp \frac{\nu }{2}
x^{\pm }.
\end{equation*}
When $\Delta =0$ we recover the usual relation between the zero--modes and
the momenta. We see that the contribution of the winding is to make the
zero--modes $a_{0}^{\pm },\widetilde{a}_{0}^{\pm }$ non--commuting
coordinates on the Minkowskian two--plane $X^{\pm }$. This representation
for the zero--modes is very convenient if one wants to analyze the wave
functions associated with on shell winding states, as we shall discuss at
the end of this section. To compute the partition function, on the other
hand, it is technically more convenient to use, instead of the above
representation, the more naive representation used in \cite{CC1}, which
treats $a_{0}^{-}$ and $\widetilde{a}_{0}^{+}$ as creation operators. As
discussed in \cite{BerkoozBoris}, the two prescriptions give the same
result. More precisely, let us define the occupation number operators
$N_{n}^{\pm }=-\left( n\mp i\nu \right) ^{-1}a_{-n}^{\pm }a_{n}^{\mp }$ and 
$\widetilde{N}_{n}^{\pm }=-\left( n\pm i\nu \right)^{-1}\widetilde{a}_{-n}^{\pm }
\widetilde{a}_{n}^{\mp }$ , which we assume to have integral
eigenvalues. Start by defining the left and right parts of the boost operator 
$J=J_{L}+J_{R}$, according to $i\left[J_{L,R},X^{\pm}\right]=\pm X^{\pm}$, 
and given explicitly by 
\begin{equation*}
J_{L}=-i\sum_{n\geq 1}N_{n}^{+}+i\sum_{n\geq 0}N_{n}^{-}\ ,\ \ \ \ \ \ \ \ \
\ \ \ \ \ J_{R}=-i\sum_{n\geq 0}\widetilde{N}_{n}^{+}+i\sum_{n\geq 1}
\widetilde{N}_{n}^{-}\ .
\end{equation*}
The contribution to the Virasoro generators from the fields $X^{\pm }$ has
been computed in \cite{Bachas,TseytlinRusso} and is given by ($\cdots $
denotes contributions from other fields) 
\begin{eqnarray*}
&&L_{0} = \cdots +\ \frac{1}{2}i\nu \left( \ 1-i\nu \right) \ -\sum_{n\geq
1}a_{-n}^{+}a_{n}^{-}-\ \sum_{n\geq 0}a_{-n}^{-}a_{n}^{+}\,, 
\\
&&\widetilde{L}_{0} = \cdots +\ \frac{1}{2}i\nu \left( \ 1-i\nu \right) -\
\sum_{n\geq 0}\widetilde{a}_{-n}^{+}\widetilde{a}_{n}^{-}-\ \sum_{n\geq 1}
\widetilde{a}_{-n}^{-}\widetilde{a}_{n}^{+}\,.
\end{eqnarray*}
It is then clear that one can rewrite the \textit{total} Virasoro generators
for the three bosons $X^{\pm }$, $X$ in terms of the usual \textit{integral }
level numbers $\mathbb{L},\widetilde{\mathbb{L}}$ and the boost operator as 
\begin{eqnarray*}
&&L_{0}=\frac{1}{2}\,i\nu \left(1-i\nu \right) +\nu J_{L}
+\frac{1}{2}\,p_{L}^{\,2}+\mathbb{L}\ , 
\\
&&\widetilde{L}_{0}=\frac{1}{2}\,i\nu \left(1-i\nu \right) -\nu
J_{R}+\frac{1}{2}\,p_{R}^{\,2}+\widetilde{\mathbb{L}}\ .
\end{eqnarray*}
We are now ready to compute the partition function $\;Z\,\ $for the three
bosons $X^{\pm }$, $X$. We have 
\begin{eqnarray*}
Z &=&\left(q\overline{q}\right)^{-1/8}\sum_{w,n}\mathrm{Tr}\,q^{\,\mathbb{L}}
\overline{q}^{\,\widetilde{\mathbb{L}}}
\left(\frac{q}{\overline{q}}\right)^{\left( 1/2\right) nw}
\\
&&\times\, \left( q\overline{q}\right) ^{\left( 1/2\right) \left[ \left(
wR/2\right) ^{2}+\left( n-\Delta J/R\right) ^{2}\right] } \\
&&\times\, \left( q\overline{q}\right) ^{\left( 1/2\right) \nu \left(
J_{L}-J_{R}\right) }\left( q\overline{q}\right) ^{\left( 1/2\right) i\nu
\left( 1-i\nu \right) }\,,
\end{eqnarray*}
where $q=e^{2\pi i\tau }$, and where $\tau =\tau _{1}+i\tau _{2}$ is the
torus modular parameter. Performing the usual Poisson resummation on $n$
brings the above expression to the simpler form
\begin{eqnarray*}
Z &=&\left(q\overline{q}\right)^{-1/8}\frac{R}{\sqrt{2\tau_{2}}}\,
\sum_{w,w^{\prime }}
\exp\left[-\frac{\pi R^{2}}{2\tau_{2}}\,T\overline{T}-2\pi\tau_{2}\Delta^{2}w^{2}\right]  
\\
&&\times\,q^{\,\left(1/2\right)i\nu}\,\mathrm{Tr}_{L}
\left( e^{2\pi iT\Delta J_{L}}q^{\,\mathbb{L}}_{\phantom{\frac{1}{1}}}\right)  
\\
&&\times\,\overline{q}^{\,\left(1/2\right)i\nu}\,\mathrm{Tr}_{R}
\left(e^{2\pi i\overline{T}\Delta J_{R}}
\overline{q}^{\,\widetilde{\mathbb{L}}}\right)\,,
\end{eqnarray*}
\bigskip with 
\begin{equation*}
T=w\tau -w^{\prime }\,.
\end{equation*}
As usual, in the above sum, the term with $w=w^{\prime }=0$ is by itself
modular invariant, and gives the partition function of the uncompactified
theory (the one obtained by the naive application of the inheritance
principle). We therefore focus on the other terms in the sum, denoting the
restricted sum with $\sum^{\prime }$. If we define the constant $c$ by 
\begin{equation*}
c=e^{2\pi i\left( i\Delta T\right) }=q^{i\nu }e^{2\pi w^{\prime }\Delta },
\end{equation*}
the traces $\mathrm{Tr}_{L}$ and $\mathrm{Tr}_{R}$ are easy to compute, and
are given by 
\begin{eqnarray*}
&&\mathrm{Tr}_{L}\left( e^{2\pi iT\Delta J_{L}}q^{\,\mathbb{L}}\right) =
\mathrm{Tr}_{L}\left( c^{-iJ_{L}}q^{\,\mathbb{L}}\right) = \\
&=&\frac{1}{1-c}\prod_{n\geq 1}\frac{1}{\left( 1-q^{n}\right) \left(
1-q^{n}c\right) \left( 1-q^{n}c^{-1}\right) } \\
&=&iq^{1/8}c^{-1/2}\frac{1}{\theta_{1}\left(i\Delta T|\tau \right) }\,,
\end{eqnarray*}
and by $\mathrm{Tr}_{R}=\overline{c}\,\overline{\mathrm{Tr}_{L}}$. Therefore
the partition function $Z$ is given by the final expression 
(reinserting $\alpha^{\prime}$)
\begin{equation}
Z = \frac{R}{\sqrt{\alpha ^{\prime }\tau _{2}}}
\sum_{w,w^{\prime}}{}^{{}^{\prime}}e^{-\left( \pi R^{2}/\alpha ^{\prime }\right) \left( T
\overline{T}/\tau _{2}\right) -2\pi \tau _{2}\Delta ^{2}w^{2}}
\,\left| \theta_{1}^{\phantom 2}\left( i\Delta T|\tau \right) \right| ^{-2}\,.
\end{equation}

Let us comment briefly on the above result. First of all, we note the strong
similarity with the expression for the partition function of the Euclidean
BTZ black hole found in \cite{OoguriMalda2}. This is to be expected since
the BTZ black holes are nothing but orbifolds of $AdS_{3}$ space, and we are
therefore considering a special limit, with the radius of $AdS$ sent to
infinity \cite{CC1}. Secondly, and more problematically, the above partition function
exhibits poles at the zeros of $\theta _{1}\left( i\Delta T|\tau \right) $,
which are located at 
\begin{equation*}
i\Delta T=a\tau -b\ ,
\end{equation*}
for $a,b\in \mathbb{Z}$. These poles where interpreted in \cite{OoguriMalda2}
as coming from the contribution of long strings in the partition function 
of the Euclidean thermal BTZ black hole. In the present setting though, the
Euclidean interpretation is unclear, as is the presence and contribution of the 
long strings. Another possibility, is that these infinities have to do with the basic
problem of defining perturbation theory order by order in these models, as discussed 
previously. In fact, as is well known, the derivative of the partition function
with respect to $\alpha ^{\prime }$ is nothing but the one--loop tadpole for
the dilaton. Recall that we are discussing a space--time with closed time--like
curves, and that the space--time has surfaces of polarization, where points
are light--like related to their $n$--th image. If we compute the string
two--point function using the method of images, and evaluate it at equal
points, it will diverge at the polarization surfaces, thereby implying a
possible divergence of the full dilaton tadpole. We shall come back to this
important point more thoroughly in section 4.1.

Let us conclude this section by discussing the issue of the free spectrum of
on--shell winding strings in the shift--boost orbifold, by briefly
discussing their wave functions. This was done for the boost--orbifold in
\cite{BerkoozBoris}. For simplicity, we shall assume that we are
in the groundstate of all the non--zero oscillators $a_{n}^{\pm }$,
$\widetilde{a}_{n}^{\pm }$,  ($n\neq 0$), so that, in the $X^{\pm }$ plane,
we only consider the zero modes. The generators $L_{0}$, $\widetilde{L}_0$
can be written, using the $x$--$P$ representation of 
$a_{0}^{\pm }$, $\widetilde{a}_{0}^{\pm }$, as
\begin{equation*}
\begin{array}{l}
\displaystyle{L_{0} = \frac{\nu ^{2}}{2}-P_{+}P_{-}+\frac{\nu ^{2}}{4}\,x^{+}x^{-}
+\frac{\nu}{2}\,\mathcal{J}+\frac{1}{2}\,P_{L}^{\,2}+\ell_{0}\,,}
\spa{0.4} \\
\displaystyle{\widetilde{L}_{0} = \frac{\nu ^{2}}{2}-P_{+}P_{-}
+\frac{\nu^{2}}{4}\,x^{+}x^{-}
-\frac{\nu}{2}\,\mathcal{J}+\frac{1}{2}\,P_{R}^{\,2}+\widetilde{\ell}_{0}\,,}
\end{array}
\end{equation*}
where
\begin{equation*}
\mathcal{J}=-\left( x^{+}P^{-}-x^{-}P^{+}\right) =-i\left( x^{+}\partial
_{+}-x^{-}\partial _{-}\right) 
\end{equation*}
is the zero--mode part of the boost operator, and where $\ell _{0}$ and 
$\widetilde{\ell}_{0}$ are constants which come from the oscillator part of
the boson $X$, together with the conformal weight relative to the CFT of the
spectator directions. Note that the complex term $i\nu/2$ has
dropped from the expressions for the Virasoro generators. The level matching
condition reads $L_{0}-\widetilde{L}_{0}=wn+\ell _{0}-\widetilde{\ell }_{0}=0
$. The on--shell condition $L_{0}+\widetilde{L}_{0}-2=0$, on the other hand,
leads to the differential equation
\begin{equation}
2\partial_{+}\partial_{-}+M^{2}+P^{2}
+\frac{w^{2}\Delta^{2}}{2}\,x^{+}x^{-}=0\,,  
\label{PDEwinding}
\end{equation}
where we have defined the constant mass
\begin{equation*}
M^{2}=\frac{w^{2}R^{2}}{4}+w^{2}\Delta ^{2}+\ell _{0}+\widetilde{\ell }
_{0}-2\,\,,
\end{equation*}
and we recall that $P=\left( n-\Delta \mathcal{J}\right) /R$. Now equation 
(\ref{PDEwinding}) is a \textit{real} PDE, with classical solutions corresponding
to on--shell winding states, whose existence has been questioned in the
literature \cite{Nekrasov}. To solve (\ref{PDEwinding}), we look for
solutions, in region I, of the form
\begin{equation}
G\left(\omega t\right) e^{\,i\left( E\mathcal{J}y+\frac{n}{R}z\right)}\,,
\label{solwind}
\end{equation}
where $\omega ^{2}=P^{2}+M^{2}$, and where $G$ satisfies
\begin{equation*}
\left[ \frac{d^2\ }{d\sigma^2}
+\frac{1}{\sigma}\,\frac{d\ }{d\sigma}
+\frac{\mathcal{J}^{2}}{\sigma^{2}}+1+A^{2}\sigma^{2}\right] 
G\left(\sigma\right) =0\,.
\end{equation*}
The constant $A$ is given, reinserting $\alpha ^{\prime }$, by
\begin{equation*}
A=\frac{w\Delta }{\alpha ^{\prime }\omega ^{2}}\ .
\end{equation*}
Performing the change of variables $z=iA\sigma ^{2}$ and 
$F=e^{\frac{iA}{2}\sigma ^{2}}\sigma ^{-i\mathcal{J}}\,G$ we obtain the differential 
equation
\begin{equation}
zF^{\prime \prime }\left( z\right) +\left( \beta -z\right) F^{\prime }
\left(z\right) -\alpha F\left( z\right) =0\,,  
\label{confhypo}
\end{equation}
with $\alpha $, $\beta $ given by
\begin{equation*}
\alpha = \frac{1}{2}\left( 1+i\mathcal{J}\right) +\frac{i}{4A}\ ,
\ \ \ \ \ \ \ \ \ \ \ \ \ \ \ \ 
\beta = 1+i\mathcal{J}\,.
\end{equation*}
The independent solutions to (\ref{confhypo}) are
given by $\mathcal{F}\left( z,\alpha ,\beta \right)$ and
$z^{1-\beta }\mathcal{F}\left( z,\alpha-\beta +1,2-\beta \right)$, where
$\mathcal{F}$ is the confluent hypergeometric function, defined in the whole complex plane by
\begin{equation*}
\mathcal{F}\left( z,\alpha ,\beta \right) =1+\frac{\alpha}{\beta}\,z+
\frac{\alpha \left(\alpha+1\right)}{\beta\left(\beta +1\right)}\,\frac{z^{2}}{2!}
+\cdots .
\end{equation*}
We conclude that the two solutions for the winding modes wave function are given by
\begin{equation*}
\left| X^{\pm }\right| ^{\pm i\mathcal{J}}e^{-\frac{i\rho }{2}}\,
\mathcal{F}\left( i\rho ,\frac{1}{2}\left( 1\pm i\mathcal{J}\right) 
+\frac{i}{4A},1\pm i\mathcal{J}\right) e^{iPX}\,,
\end{equation*}
with
\begin{equation*}
\rho =\frac{2w\Delta }{\alpha ^{\prime }}\,X^{+}X^{-}\,.
\end{equation*}
Finally, let us discuss the asymptotics of the solutions, which are easily
deduced from the large $\rho $ asymptotic formula
\begin{equation*}
e^{-\frac{i\rho }{2}}\,\mathcal{F}\left( i\rho ,\alpha ,\beta \right) \sim 
\frac{\Gamma \left( \beta \right) }{\Gamma \left( \alpha \right)}\,
e^{\frac{i\rho }{2}}\left( i\rho \right) ^{\alpha -\beta }
+\frac{\Gamma \left( \beta\right) }{\Gamma \left( \beta -\alpha \right)}\,
e^{-\frac{i\rho }{2}}\left(i\rho \right) ^{-\alpha}\,.
\end{equation*}
We then easily see that the winding solutions are localized around the
cosmological horizons, since they decay, in modulus, as 
$(X^+X^-)^{-\frac{1}{2}}$,
both in region I and in regions II, III. In particular, in region I, the
wave functions go as $t^{-1}$, as opposed to the non--winding states whose
wave function decays only as $t^{-\frac{1}{2}}$.

\section{Orientifold cosmology}

Throughout this review we have referred to the orbifold of $\mathbb{M}^{3}$
by a null boost and a null translation as the $O$--plane orbifold. In several
occasions we used the fact that this orbifold's singularity should be
interpreted as a string theory orientifold plane, excising the region behind
it. In particular, we have imposed specific boundary conditions on the
fields at the singularity. This fact was used in the context of the
shifted--boost orbifold, since near the singularities it reduces to the 
$O$--plane orbifold. In the following we shall justify these assumptions by
arguing that the $O$--plane orbifold is dual to a type IIA orientifold
8--plane\cite{CC3}. 
Then we interpret the M--theory shifted--boost orbifold as an 
$O8/\overline{O}8$  system. The corresponding geometry is a two--dimensional
cosmological toy model. This construction is then generalized at the level
of supergravity to a four--dimensional model arising from a string theory
flux compactification. The late time evolution of this cosmological model
exhibits a cyclic acceleration \cite{CCV}.

\subsection{$O$--plane orbifold revisited}

Consider M--theory, with Planck length $l_{P}$, on the space 
\begin{equation}
\left( \mathbb{M}^{3}/e^{\kappa }\right) \times \mathbb{T}^{\,7}\times
S^{1}\ ,  \notag
\end{equation}
where $\kappa $ is the $O$--plane orbifold generator (\ref{Opk}). In the $%
(y^{\pm },y)$ coordinates introduced in section 2.4 the
eleven--dimensional supergravity metric has the form 
\begin{equation}
ds_{11}^{\,2}=-2dy^{+}dy^{-}+2Ey\left( dy^{-}\right) ^{2}+dy^{2}+ds^{2}\left( 
\mathbb{T}^{7}\times S^{1}\right) \,,  \notag
\end{equation}
with $y^{-}\sim y^{-}+2\pi R$.

Let us first look at the M--theory compactification on the $S^{1}$ circle of
radius $R_{11}$. This defines the type IIA $O$--plane orbifold 
\begin{equation}
\left( \mathbb{M}^{3}/e^{\kappa }\right) \times \mathbb{T}^{7}\ ,  \notag
\end{equation}
with string coupling and string length 
\begin{equation}
g_{s}=(R_{11}/l_{P})^{3/2}\,,\ \ \ \ \ \ \ \ \ \ \ \ \ \ \ \ \ \ \ \ \ \
l_{s}^{\,2}=l_{p}^{\,3}/R_{11}\,.  
\label{rel1}
\end{equation}
This orbifold is defined by three parameters $(R,\Delta ,V_{7})$, where $V_7$ is the
volume of the 7--torus (and we ignore the other torus moduli).

On the other hand, one can consider the M--theory compactification on the
orbifold circle, \textit{i.e.} along the $y^{-}$--direction. Then one
obtains type IIA with string length $l_{s}^{\prime }$ and with background
fields 
\begin{equation*}
\begin{array}{c}
\displaystyle{ds_{10}^{\ 2} = -H^{-1/2}\left( dy^{+}\right) ^{2}+H^{1/2}
\left(dy^{2}+ds^{2}\left( \mathbb{T}^{7}\times S^{1}\right) \right)}\ ,  
\spa{0.3}\\
\displaystyle{e^{\phi } = g_{s}^{\prime }\,H^{3/4}\ ,\ \ \ \ \ \ \ \ \ \ \ \ \ \ \ \
A=-H^{-1}dy^{+}}\,,
\end{array}
\end{equation*}
where $H=2Ey$ and where 
\begin{equation}
g_{s}^{\prime }=(R/l_{P})^{3/2}\,,\ \ \ \ \ \ \ \ \ \ \ \ \ \ \ \ \ \ \ \ \
\ l_{s}^{\prime }{}^{2}=l_{p}^{\,3}/R\,.  
\label{rel2}
\end{equation}
Notice that this geometry is only defined for $y>0$. Finally we T--dualize
along the 8--torus $\mathbb{T}^{7}\times S^{1}$ to obtain a solution of
the massive supergravity theory \cite{Romans,Green} with background fields 
\begin{equation}
\begin{array}{c}
\displaystyle{ds_{10}^{\ 2} = H^{-1/2}\left( -\left(dy^{+}\right)^2
+ds^{2}\left(\mathbb{T}^{7}\times S^{1}\right) \right)+H^{1/2}dy^{2}}\ , 
\spa{0.3}\\
\displaystyle{e^{\phi} = \hat{g}_{s}\,H^{-5/4}\ ,
\ \ \ \ \ \ \ \ \ \ \ \ \ \ \ \ \ 
\star F=2E}\ ,
\end{array}
 \label{O8geom} 
\end{equation}
where $F$ is a 10--form. The string coupling and string length of the dual theory 
are 
\begin{equation}
\hat{g}_{s}=g_{s}^{\prime }\,\frac{(2\pi)^7\,l_s^{\prime }{}^{8}}{R_{11}V_{7}}\ ,\ \
\ \ \ \ \ \ \ \ \ \ \ \ \ \ \ \ \ \ \hat{l}_{s}=l_{s}^{\prime }\ .
\label{rel3}
\end{equation}
This background preserves one half of the type IIA supersymmetries and it
has the standard form of RR charged objects in string theory. In particular,
the function $H$ is harmonic on the transverse $y$--direction. As one
approaches $y=0$ the curvature and the string coupling diverge. For $y<0$,
the background fields are not well defined, since the dilaton field would be
complex.

How are we supposed to interpret this singularity? The above background
fields solve the supergravity equations of motion with a localized source at 
$y=0$. This source is extended along the 8--torus and couples to the
graviton, dilaton field and 9--form gauge potential, with the action 
\begin{equation}
S=\tau \left( -\int d^{9}x\,e^{-\phi }\sqrt{-\hat{g}}\pm \int A_{9}\right) \,,
\notag
\end{equation}
where $\hat{g}$ is the induced metric and the $\pm $ signs correspond to
positive or negative charge. Placing this source at the singularity, the
geometry can be extended to the $y<0$ region by setting 
\begin{equation}
H=2E|y|\ .  \notag
\end{equation}
The singularity of $\bigtriangleup _{y}H=4E\delta (y)$ is then related to
the tension $\tau $ of the source by the equations of motion according to 
\begin{equation}
\frac{1}{\left( 2\pi \right) ^{7}\hat{l}_{s}^{\,8}\hat{g}_{s}^{\,2}}
\,\bigtriangleup _{y}H=-\tau \,\delta (y)\ ,  \notag
\end{equation}
and therefore the source 8--brane has negative tension. This object is
called an orientifold 8--plane in string theory and the geometry 
(\ref{O8geom}) is the same as that found in \cite{PolchWitten}. At zero coupling, 
an $O8$--plane is a $Z_{2}$ orbifold of the IIA theory with group element 
$g=I\Omega $, where $I$ is the reflection along the $y$--direction and where 
$\Omega $ is the world--sheet parity operator. The $O8$--plane tadpole has
opposite sign to the $D8$--brane tadpole, so it has negative tension and
charge. As one turns on the coupling, the closed strings react giving the
above geometry, which has a large dilaton field near the singularity.
However, supersymmetry suggests that quantum corrections are under control.
Unlike for D--branes, there is no freedom to have an arbitrary number of 
$O8$--planes, so that the tension $\tau $ is fixed to be \cite{Polch,Sagnotti} 
\begin{equation}
\tau =-\frac{N}{(2\pi )^{8}\hat{g}_{s}\hat{l}_{s}^{\,9}}\ ,
\ \ \ \ \ \ \ \ \ \ \ \ (N=16)\,,  
\notag
\end{equation}
where we have written $\tau$ in units of the tension of the $D8$--brane.
This condition fixes the parameter $E=\Delta /R$ of the orbifold 
to\footnote{There is a difference by a factor of 2 with respect to \cite{CC2} 
because there the function $H$ was taken to vanish for $y<0$, and a factor of 
$2\pi$ from the definition of $\hat{l}_{s}$.} 
\begin{equation}
8\pi \hat{l}_{s}E=N\hat{g}_{s}\ .  \label{cqc}
\end{equation}

The above duality chain leads to the conjecture that the type IIA $O$--plane
orbifold is dual to an $O8$--plane of the type IIA theory. Moreover, the
relations (\ref{rel1}), (\ref{rel2}), (\ref{rel3}) between the couplings and
string lengths can be used to write the orientifold charge quantization
condition (\ref{cqc}) in terms of the original parameters of the orbifold.
Quite surprisingly, $O8$--plane charge quantization becomes, in the dual
theory, quantization of the coupling constant 
\begin{equation}
g_{s}^{\,2}=\frac{4}{(2\pi)^6\,N} \,\frac{\Delta RV_{7}}{l_{s}^{\,8}}\ .
\label{cconst}
\end{equation}
Let us comment on the above result. First it depends on $R$ and $\Delta $ only
through the invariant combination $\Delta R$, which parametrizes the inequivalent
conjugacy classes of $O$--plane orbifolds. Secondly, it is $S$-- and 
$T$--duality invariant, since it can be written explicitly in terms of the $10$
and $3$ dimensional Newton constants as 
\begin{equation*}
\begin{array}{rcl}
G_{10}&=&\displaystyle{\frac{1}{2N}\,\Delta RV_{7}}\,, 
\spa{0.3}\\
G_{3}&=&\displaystyle{\frac{1}{2N}\,\Delta R}\,.
\end{array}
\end{equation*}
Therefore we might start just as well with a IIB orbifold.

How should we interpret the above result? In order to answer this question,
let us summarize the basic known facts about the $O$--plane orbifold. First
of all, and most importantly, it consists of string theory on a space with
CTC's. Therefore, although we can formally write down a perturbation theory
in $g_{s}$, it is clear that it will not define a unitary theory order by
order in the coupling (the conventional check of perturbative unitarity
fails due to loops which wind around the CTC's). On the other hand, the duality
just described suggests that the region where $\kappa ^{2}<0$ should act as
a wall for the propagation of string fields, which should bounce and be
reflected with \textit{unit probability}. The wall in the orbifold theory is
replaced by the whole region $\kappa ^{2}<0$. A first hint of this fact
comes from the analysis of the single particle wave functions at zero
coupling. Recall that, in the region $\kappa ^{2}<0$, the particles face a
linearly increasing potential, and their wave function is exponentially
damped. Therefore, without interactions, the picture is consistent. As one
turns on interactions, one usually looses unitarity. The natural conjecture
is then that \textit{unitarity is restored just at a specific value of the
coupling constant, given by (\ref{cconst}). Therefore charge quantization on
one side of the correspondence becomes, on the other side, unitarity in the
presence of CTC's}. Chronology protection is therefore restored, but with a 
mechanism quite different from the one advocated by Hawking \cite{Hawk2}, which
excludes CTC's from the start. Note that, if this conjecture is correct,
perturbative unitarity looses its significance, since we are no longer free
to choose the coupling at will. One example where perturbative computations fail
is the one--loop quantum stress--energy tensor, which diverges generically
at the polarization surfaces. This is also related to violations of 
causality and should therefore be solved by higher order corrections.
A more thorough discussion of these subtle points, 
as well as more evidence, can be found in \cite{CC3}.

\subsection{$O/\bar{O}$ system}

The near singularity limit of the shifted--boost orbifold is the $O$--plane orbifold. 
Clearly, the same duality arguments of the previous section lead to the conjecture that the 
type IIA shifted--boost orbifold is dual to a system with two orientifolds.

Consider again M--theory on the space
\beq
\left( \bM^3/e^\kappa\right) \times \bT^{\,7} \times S^1\ ,
\notag
\eeq
but now let $\kappa$ be the shifted--boost orbifold generator (\ref{sbk}). 
If $S^1$ is the circle along the eleventh
direction, then we have the type IIA shifted--boost orbifold. On the other hand, one can take the 
orbifold circle to  be the eleventh compact direction. The corresponding background fields are
similar to those of (\ref{sbgeomI}) and (\ref{sbgeomII}), 
appropriately embedded in M--theory. Then, a T--duality
transformation on the 8--torus $T^7\times S^1$ gives a type IIA background, with the following 
metric in regions ${\rm I}$ and ${\rm II}$ of space--time
\beq
\barr{rl}
{\rm I}: & 
\displaystyle{ds^2=\Phi^{-1}\left[ (Et)^2 dy^2 
+ ds^2\left(T^7\times S^1\right)\right] - \Phi\, dt^2}\ ,
\spa{0.4}\\
{\rm II}: &
\displaystyle{ds^2=\Phi^{-1}\left[ - (Ex)^2 dw^2 
+ ds^2\left(T^7\times S^1\right)\right] 
+ \Phi\, dx^2}\ ,
\earr
\label{O/Obar}
\eeq
where $\Phi^2=1+(Et)^2$ or $\Phi^2=1-(Ex)^2$, respectively. The dilaton field and the massive IIA
cosmological constant are
\beq
e^\phi=\hat{g}_s\,\Phi^{-5/2}\ ,\ \ \ \ \ \ \ \ \ \ \ 
\star F = 2E\ ,
\notag
\eeq
where $\hat{g}_s$ is the string coupling at the horizons. The string coupling and the curvature
diverge at the singularities, and vanish at late times in region I. Moreover, the
volume of the 8--torus is also converging to zero at late times, so that this geometry describes
a two--dimensional cosmology.

To analyze the geometry near the singularities, consider the coordinate transformations
$Ey=1\mp Ex\ll 1$ in regions ${\rm II}_R$ and ${\rm II}_L$, respectively. The geometry becomes
precisely that of the $O8$-- or $\overline{O}8$--planes of the previous section, so that
the background (\ref{O/Obar}) solves the massive supergravity equations of motion with two 
localized sources. Also, the near--singularity geometries preserve opposite halfs of the 
supersymmetries, just like D--branes of opposite charge. Using two--dimensional gravity 
techniques, we saw that this geometry is completely determined by a constant of motion. 
This fact shows that the $O8/\overline{O}8$ boundary conditions at the singularities 
determine the geometry uniquely. In other words, there is no fine tunning of boundary
conditions at the singularities to obtain the cosmological bounce. These arguments justify the
conjectured duality between the IIA shifted--boost orbifold and the $O8/\overline{O}8$ system.
Note that the relation between the couplings and string lengths $(g_s,l_s)$ and
$(\hat{g}_s,\hat{l}_s)$ is unchanged from the previous section, and therefore the coupling
quantization (\ref{cconst}) still holds.

At zero coupling the $O8/\overline{O}8$ system is a $Z_2\times Z_2$ 
orbifold of the type IIA theory \cite{ADSagn,Kachru}, where the 
first and second $Z_2$ group elements are, respectively,
\beq
g_1=I\Omega\ ,\ \ \ \ \ \ \ \ \ \ \ \ \ 
g_2=I\Omega(-1)^F \delta\ ,
\notag
\eeq
with $F$ space--time fermion number and $\delta$ a translation by $2L$. The orientifold is at the 
fixed point $y=0$ of the group element $g_1$, while the anti--orientifold is
at the fixed point $y=L$ of $g_2$. The distance $L$ between the 
$O$--planes is a free modulus. When the coupling is
turned on, the reaction of the closed strings changes drastically the space--time global
structure, even if one still has strong coupling near the $O$--plane and weak coupling far away. 
A slice of constant time in regions II of the  $O8/\overline{O}8$ geometry can then be used
to define the distance between the $O$--planes, with the result
\beq
L=\,\frac{2}{E}\,\int_0^1\left(1-u^2\right)^{1/4}\,du 
\,\propto\, \frac{\hat{l}_s}{\hat{g}_s}\ ,
\notag
\eeq
so that for $\hat{g}_s\rightarrow 0$ the pair is far apart. Since the geometry breaks 
supersymmetry, how reliable is this supergravity prediction for the moduli fixing in the
$O8/\overline{O}8$ system? If  $\hat{g}_s$ is very small, the coupling $e^\phi$ will be small 
everywhere except for $x\sim\pm 1/E$, however, this is precisely where the geometry
becomes approximately that of the supersymmetric orientifolds. For that reason it is 
reasonable that quantum corrections are small. Also, in the limit of small coupling, 
curvature corrections are small everywhere except near the singularities.

The previous analysis shows a sharp distinction between small coupling and 
strictly vanishing coupling. A different way to look at the problem is to start with
the $O8/\overline{O}8$ system and place sixteen $D8$-- and $\overline{D}8$--branes on top 
of the orientifolds to cancel the tadpole, obtaining a flat space--time. However,
since supersymmetry is broken, there will be a one--loop potential between the D--branes,
which will attract each other and annihilate. The end point of this process, described by the
condensation of the open string tachyon, is the $O8/\overline{O}8$ vacuum.

To understand better the dynamics behind the $O8/\overline{O}8$ system, let us revise some 
basic facts about domain wall physics using the linearized theory of gravity, even though 
the complete results must be derived in the full non--linear setting. The physics of domain
walls in gravity is rather non--intuitive, and it was first explored in some detail in
\cite{Vilenkin}. One of the interesting results is that, in pure gravity, positive tension 
domain walls repel, so we should be careful with our physical intuition. Consider 
ten--dimensional space--time and a gravitational brane source localized on an 
eight--dimensional hypersurface. If we denote by $y$ the transverse direction to the brane,
then the linearized equations for the Einstein metric perturbations are
\beq
-\bigtriangleup_y h_{ab} = \tau_{ab}\, \delta(y)\ ,
\notag
\eeq
where $\tau_{ab}$ is related to the stress tensor of the brane by
\beq
\tau_{ab}=T_{ab} - \frac{T}{8}\,\eta_{ab}\ ,\ \ \ \ \ \ \ \ \ \ \ \ \ \ \ 
\left(T=T^a_{\ a}\right)\ .
\notag
\eeq
For a BPS brane of tension $\tau$, we have $T_{00}=-T_{ii}=-\tau$, so that the effective 
gravitational mass is $\tau_0=-\tau/8$, which is negative for positive tension branes
and vice--versa. However, in type II strings, all brane--like sources, as well as massive 
probes, have non--trivial couplings to other supergravity fields. For a BPS 8--brane, the 
action, written in the Einstein frame, is
\beq
S = \tau \left( - \int d^9x\, e^{\frac{5}{4}\phi}\sqrt{-\hat{g}} \pm \int A_9\,\right)
\ ,
\notag
\eeq
where the tension $\tau$ can be either positive or negative. Then the linear equations 
for the dilaton and gauge potential are
\beq
\barr{rcl}
-\bigtriangleup_y \phi &=&
\displaystyle{ - \frac{5}{4}\,\tau\,\delta (y)}\ ,
\spa{0.3}\\
-\bigtriangleup_y A_9 &=&
\displaystyle{ \pm\,\tau\,\delta (y)}\ .
\earr
\notag
\eeq

Let us now consider the fields created in the linear regime by a $O$8-- or 
a $\overline{O}$8--plane source, which have negative tension $\tau$.
We can compute the potentials seen by a $D$8--brane probe due to graviton, dilaton
and $RR$ exchange with a $O$8-- or a $\overline{O}$8--plane, which are respectively
given by
\beq
\barr{rcl}
V_g + V_{\phi} & = & \tau\,y \ ,
\\
V_A & = & \mp\,\tau\,y\ .
\earr
\notag
\eeq
As expected, a $D$8-brane feels no force in the presence of a $O$8--plane and it is repelled by
a $\overline{O}8$--plane. One could, naively, consider a $O8$--plane probe in the presence
of the linear fields created by a $\overline{O}8$--plane. Then the resulting potential is 
attractive, predicting that such system is unstable. This analysis is, however, very naive because
$O8$--planes have no degrees of freedom and therefore cannot be analyzed with a dynamical 
probe action. Also, non--linear effects in the fields are ignored in this approximation.

\begin{figure}[t]
\begin{center}
\begin{tabular}{c}
\epsfysize=5cm\epsfbox{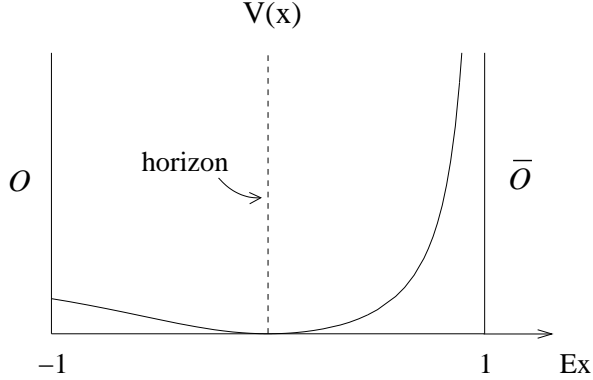}
\end{tabular}
\end{center}
\caption{The static potential for a $D8$--brane probe in the $O8/\overline{O}8$ geometry. 
The $D$--brane is repelled away from both orientifolds. The vertical dashed line represents
the horizon.}
\label{fig15}
\end{figure}

To investigate non--linear effects in the $O8/\overline{O}8$ geometry, recall that, within
the linear theory, a $D$8--brane probe placed in between the orientifolds will feel a 
repulsive force from the $\overline{O}8$--plane, which will drive it towards the $O8$--plane.
However, in this case we can compute the static potential for the $D$--brane in the full
$O8/\overline{O}8$ geometry. Recalling the form of the fields in region II (\ref{O/Obar}),
a simple calculation shows that a $D$8--brane placed at constant $x$, with vanishing velocity,
has a potential
\beq
V(x)=\frac{(Ex)^2}{1-(Ex)^2\,{\rm sign}(x)}\ .
\notag
\eeq
The full potential is plotted in figure \ref{fig15}, where one sees that now the
$D$8--brane feels a repulsive force in region ${\rm II}_L$, driving it away from the
$O$8--plane. What is the reason for this force? Clearly, it cannot be the repulsion from 
the $\overline{O}8$--plane, since the space--time global structure is such that the
$\overline{O}8$--plane is causally 
disconnected from any probe in region ${\rm II}_L$. In fact, as shown in the
figure, $x=0$ is at the horizon where the coordinate system becomes singular. This force is
due to the backreaction of the $O8/\overline{O}8$ system on the geometry. In other words,
the $O8/\overline{O}8$ vacuum has an extra energy density (or curvature) on its core, when
compared to the fields created by both orientifolds in the linear regime. This extra energy 
density will gravitationally attract the $D$--brane to the core of the geometry, explaining
the shape of the potential in region ${\rm II}_L$.

The above analysis is an example that the non--linear gravity effects due to the
orientifolds change drastically the naive expectation that the $O8/\overline{O}8$ system
is unstable under collapse and annihilation. In fact, in the $O8/\overline{O}8$ geometry
the orientifolds are not even in causal contact, in sharp contrast to the zero coupling situation.

\subsection{Four--dimensional cosmology}

The time--dependent geometries reviewed in these lectures are far from being realistic
cosmological models. In this section we shall construct a four--dimensional FRW
cosmology from a string compactification, using a generalization to higher
dimensions of the previous $O/\overline{O}$ system. The construction can be done in arbitrary 
dimensions but we shall concentrate on the four--dimensional case for obvious reasons. 

Let us start quite generically by considering four--dimensional
gravity coupled to a scalar field $\psi $
\begin{equation}
S=\frac{1}{2\kappa^2{\cal E}^{2}}\int d^4x\,\sqrt{-g}
\left[R-\frac{\beta}{2}\left(\nabla\psi\right)^{2}
-V\left(\psi\right)\right]\ , 
\label{eq100}
\end{equation}
where ${\cal E}$ is an energy scale. 
In this equation, $V\left(\psi\right)$ is the potential for the scalar
field, $\beta$ is a dimensionless constant and the coordinates are
dimensionless in units of $1/{\cal E}$. Following the previous two--dimensional toy model, 
we are interested in cosmological solutions to the equations of motion of the above action
with a contracting and an expanding phase, which we call, respectively,
regions ${\rm I}_{in}$ and ${\rm I}_{out}$, together with an intermediate region 
denoted by region ${\rm II}$. The expanding phase is the standard
FRW geometry for an open universe, described by the fields 
\begin{equation}
\begin{array}{rcl}
\displaystyle{ds^{\,2}_4} 
&=&
\displaystyle{-dt^{2}+a_{\rm I}^{\,2}
\left(t\right)ds^{2}\left(H_3\right)}\ ,
\spa{0.2}\\
\psi &=&
\displaystyle{\psi_{\rm I}\left(t\right)}\ ,  
\end{array}
\label{eq10}
\end{equation}
where $H_3$ is the three--dimensional hyperbolic space with unit radius.
Therefore, the dynamics of the system is described by the scalar and
Friedman equations 
\begin{equation}
\begin{array}{rcl}
\displaystyle{\ddot{\psi}_{I}+3\,\dot{\psi}_{\rm I}\,\frac{\dot{a}_{\rm I}}{a_{\rm I}}} 
&=&
\displaystyle{-\frac{1}{\beta}\,\frac{\partial V}{\partial \psi_{\rm I}}}\ ,
\spa{0.5}\\
\displaystyle{\left(\frac{\dot{a}_{\rm I}}{a_{\rm I}}\right)^{2}
-\frac{1}{a_{\rm I}^{\ 2}}} 
&=&
\displaystyle{\frac{1}{6}
\left[\,\frac{\beta}{2}\,\dot{\psi}_{\rm I}^{2}
+V\left(\psi_{\rm I}\right)\,\right]}\ , 
\end{array}
\label{eq20} 
\end{equation}
where dots denote derivatives with respect to the cosmological time $t$.
The contracting phase is nothing but the time--reversed solution,
where we replace in (\ref{eq10}) $t$ by $-t$.

\begin{figure}[t]
\begin{center}
\begin{tabular}{c}
\epsfysize=7cm\epsfbox{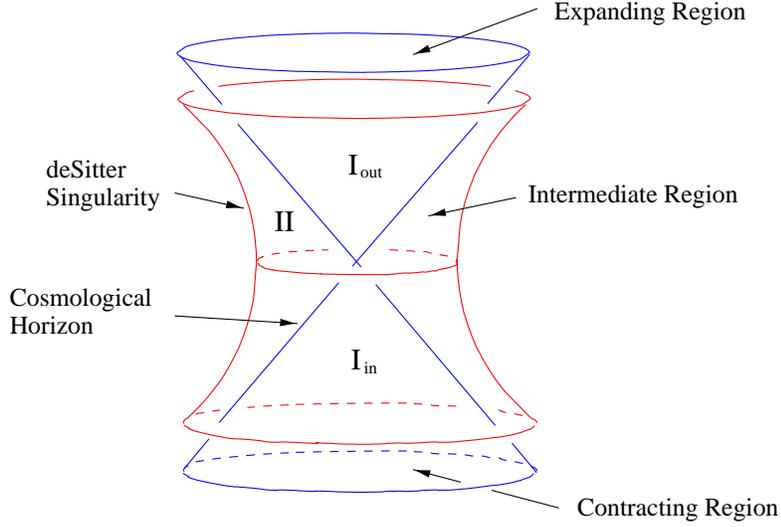}
\end{tabular}
\end{center}
\caption{The space--time global structure. We shall see that 
the geometry develops a time--like singularity in region $II$ interpreted
as a deSitter negative tension brane at the boundary of space--time. This geometry 
describes a transition from a contracting to an expanding cosmological
phase. Clearly, with this space--time global structure the standard 
cosmological horizon problem, associated with the conventional big--bang 
space--like singularity, does not arise.}
\label{fig16}
\end{figure}

In order to have an intermediate region II, we are interested in
solutions of (\ref{eq20}) where $a_{\rm I}(t)$ and $\psi_{\rm I}(t)$ are,
respectively, odd and even functions of $t$, with initial conditions 
\begin{equation}
a_{\rm I}\left(t\right)=t+{\cal{O}}\left(t^{3}\right)\ ,
\ \ \ \ \ \ \ \ \ \ \ \ \ \ \ \ \ 
\psi _{\rm I}\left(t\right)=\psi_{0}+{\cal{O}}\left(t^{2}\right)\ ,  
\label{eq50}
\end{equation}
for small cosmological time $t$. This implies that the $t=0$ surface does
not correspond to a big--bang singularity, but represents a null cosmological 
horizon \cite{CC1}. In this case, the space--time can be extended across the 
horizon to an intermediate region, where the solution has the form 
\beq
\begin{array}{rcl}
\displaystyle{ds^{\,2}_4} 
&=&
\displaystyle{a_{\rm II}^{\,2}\left(x\right)\,ds^2\left(dS_3\right)+dx^{2}}\ , 
\spa{0.2}\\
\psi &=&\displaystyle{\psi_{\rm II}\left(x\right)}\ ,
\end{array}
\label{regionII}
\eeq
with $dS_3$ the three--dimensional de Sitter space. $a_{\rm II}$ and
$\psi _{\rm II}$ are determined again by the
equations (\ref{eq20}), with the potential $V$ replaced by $-V$, and
are given by the analytic continuation
\begin{equation*}
a_{\rm II}(x)=-ia_{\rm I}(ix)\ ,\ \ \ \ \ \ \ \ \ \ \ \ \psi_{\rm II}(x)=\psi_{\rm I}(ix)\ .
\end{equation*}
Given the boundary conditions (\ref{eq50}), regions $I$
and $II$ can be connected along the null cosmological horizon, just as
a Milne universe can be glued to a Rindler wedge to form flat Minkowski
space, explaining the choice of an open universe. 
Moreover, it is clear that the solution possesses a global 
$SO\left(3,1\right)$ symmetry, which acts both on $H_3$ and on
$dS_3$ (see figure \ref{fig16}). 

The existence of an intermediate region with a generic metric of the form 
(\ref{regionII}) implies that any singularity that may develop in this region
is time--like, with a deSitter worldvolume. As we shall see bellow, this fact 
introduces the possibility of having a brane--like interpretation of the 
singularity. Moreover, the existence of this region demands generically
a field theory effective potential whose form is highly restrictive due to the
unnatural (fine--tuned) boundary conditions at the horizon. As we shall also see,
string effective supergravity theories provide a Liouville--Toda like potential
which naturally respects the boundary conditions at the horizon.

\subsubsection{Embedding in String Theory}

Let us now consider a particular case of the construction of the previous
section which can be embedded in Type IIA string theory (or M--theory).  
We consider a background with a non--trivial $RR$ 9--form potential. The
corresponding ten--dimensional Type II effective theory is given by the
massive supergravity theory, and in this case the relevant action is 
$$
S=\frac{1}{(2\pi)^7\,\hat{l}_s^{\,8}}\,
\left[\int d^{10}x\,\sqrt{-g}\,e^{-2\phi}
\left(R+4\left(\nabla\phi\right)^{2}\right)
-\frac{1}{2}\int F\wedge\widetilde{F}\,\right]\ ,
$$
where $F$ is a $RR$ 10--form field strength and 
$\widetilde{F}=\star F$ is related to the cosmological constant.
As mentioned before, we are considering a particular case for simplicity,
but the analysis can be extended to any dimensionality and degree of the 
form potential \cite{CCK}. Also, we consider a solution of massive SUGRA to 
make contact with the $O8$--plane interpretation of the previous sections.
A family of solutions, parameterized by an arbitrary 
constant $\hat{g}_s$, can be constructed by considering the following ansatz
\beq
\begin{array}{c}
\displaystyle{{\cal E}^2\,ds^{2}=
\Lambda^{-1}\,ds_{4}^{\,2}
+\Lambda^{-1/2}\,ds^{2}(\bT^{6})}\ ,
\spa{0.3}\\
\displaystyle{e^{\phi}=\hat{g}_s\,\Lambda^{-5/4}\ ,
\ \ \ \ \ \ \ \ \ 
\star F={\cal E}}\ ,
\end{array}
\label{eq1000}
\eeq
where the line element $ds_4$ is that of the previous section
and we conveniently define the scalar field $\psi$ by
\beq
\Lambda=e^{\,2\psi/7}\ .
\notag
\eeq
The  constants $\cal E$ and $\hat{g}_s$ define the electric field
and the string coupling at the horizon, respectively. It is now a mater 
of computation to show that the equations of motion for the effective 
four--dimensional theory can be derived from the action (\ref{eq100}) with
\beq
V=\frac{1}{2}\,e^{-\psi}\ ,\ \ \ \ \ \ \ \ \ \ 
\beta =\frac{1}{7}\ .
\label{potential}
\eeq

Let us start by considering the expanding region ${\rm I}_{out}$.
Solving in powers of the dimensionless coordinate $t$ 
around the cosmological horizon, a straightforward calculation gives
the first terms in this expansion
\beq
\barr{c}
\displaystyle{a_{\rm I}(t) = 
t\,\left( 1 + \frac{e^{-\psi_0}}{18}\,t^2+\cdots\right)}\ ,
\spa{0.4}\\
\displaystyle{\psi(t)=\psi_0 + \frac{7e^{-\psi_0}}{4}\,t^2+\cdots}\ .
\earr
\label{solHI}
\eeq 
At late times the universe becomes curvature dominated and the
solution has the asymptotic behavior
\beq
a(t)=\sqrt{\frac{7}{6}}\,t\ ,\ \ \ \ \ \ \ \ \ \ 
\psi(t)=\log\left(\,\frac{7}{8}\, t^2\,\right)\ ,
\label{larget}
\eeq
so that the scalar field is rolling down the potential. The solution in the 
contracting region ${\rm I}_{in}$ is the time reversal of the solution in the 
expanding region. 

\begin{figure}[t]
\begin{center}
\begin{tabular}{c}
\epsfysize=7cm\epsfbox{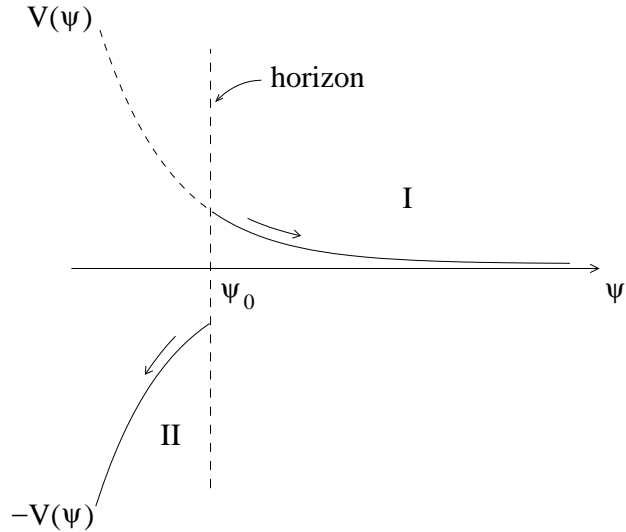}
\end{tabular}
\end{center}
\caption{Behavior of the scalar field along the potential in regions I and II.
Starting from the horizon, where $\psi=\psi_0$, the scalar field rolls down the potential
along region ${\rm I}_{out}$ and has the time reversed behavior in region ${\rm I}_{in}$.
In region II the equations of motion are equivalent to those with an inverted potential potential,
and therefore the field roles to $-\infty$ at the singularity.}
\label{fig17}
\end{figure}

\begin{figure}[t]
\begin{center}
\begin{tabular}{c}
\epsfysize=10cm\epsfbox{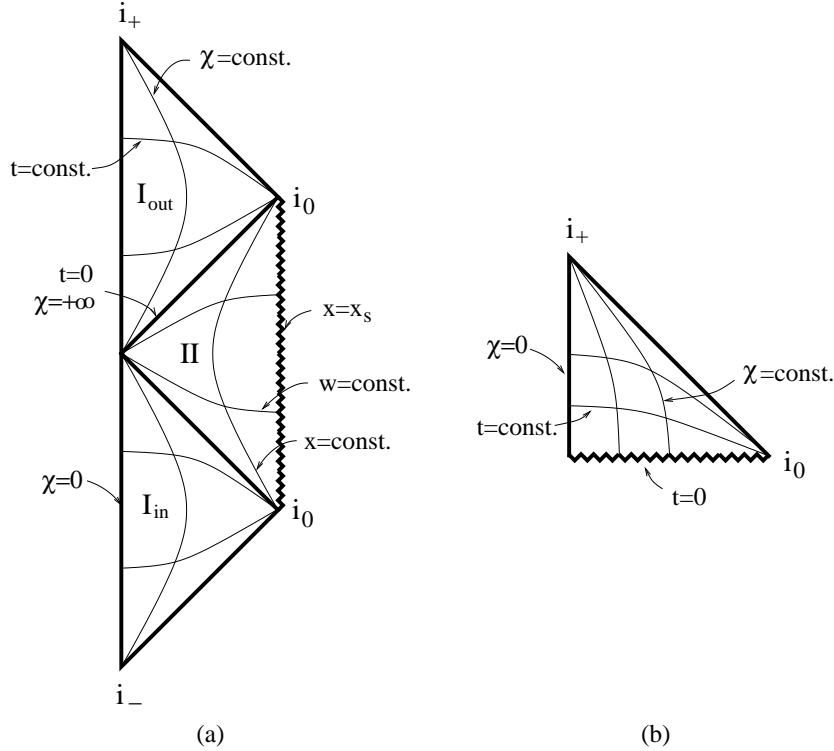}
\end{tabular}
\end{center}
\caption{CP diagrams  for open Universe cosmologies. In both diagrams each point
represents a two--sphere and $\chi=0$ is a coordinate singularity. The standard 
diagram, with a space--like singularity, is presented for comparison with the
orientifold cosmology diagram.}
\label{fig18}
\end{figure}

Next consider the intermediate region ${\rm II}$. The form of the solution
near the horizon can be obtained simply from the analytic continuation 
\beq
\barr{c}
\displaystyle{a_{\rm II}(x)=x\,\left( 1 - \frac{1}{18}\,e^{-\psi_0}\,x^2+\cdots\right)}\ ,
\spa{0.4}\\
\displaystyle{\psi_{\rm II}(x)=\psi_0 - \frac{7}{4}\,e^{-\psi_0}\,x^2+\cdots}\ .
\earr
\label{solHII}
\eeq 
One can then integrate the differential equations away from the horizon, to see that the 
geometry develops a time--like singularity, where
the scale factor and the scalar field have the behavior
\begin{equation}
\begin{array}{c}
\displaystyle{a_{\rm II}\left(x\right) =
a_s\left(x_s-x\right)^{1/7}}\ ,
\spa{0.3}\\
\displaystyle{\psi_{\rm II}\left(x\right) =2\,
\log{\left(\frac{7}{4}\,\left(x_s-x\right)\right)}}\ .
\label{eq30}
\end{array}
\end{equation}
The dimensionless constants $a_s$ and $x_s$ can be determined numerically,
and are fixed by the boundary conditions imposed at the horizon.
The behavior of the scalar field along the potential 
is represented schematically in figure \ref{fig17}. The
CP diagram for this geometry is represented in figure \ref{fig18}, 
where it is compared with the standard diagram for an open cosmology.
Let us remark that the global structure of this model is related to the earlier
work of Hawking and Turok \cite{HawkTurok}, where the top half of diagram \ref{fig18}(a) is
glued to an Euclidean gravitational instanton.

Before we investigate the geometry near the singularity, 
let us comment on the issue of cosmological particle production for
this geometry. Generically, from the existence of a horizon, one expects Hawking 
thermal radiation. In fact, in the two--dimensional toy model associated to the
shifted--boost orbifold, we showed that, starting from the asymptotic vacuum in the
contracting region, an observer in the far future
sees a thermal spectrum. To extend this result to a four--dimensional cosmological 
solution one just needs to compute the surface gravity of the cosmological horizon 
with respect to the appropriate Killing vector field. The natural
Killing vector fields to use are the generators of the $SO(1,3)$
isometries of $H_3$. With this in mind, one
arrives at the general result
\begin{equation}
T=\frac{{\mathcal E}}{2\pi\,a(\tau)}\ .
\label{temp}
\end{equation}
This is expected for radiation in a three--dimensional cosmology, 
since, from Boltzmann law $\rho\sim T^4$, we obtain $\rho\sim 1/a^4(\tau)$, 
which follows from the radiation equation of state $\rho=3\,p$.

Like in the two--dimensional toy model of the previous section,
to make sense of this geometry it is fundamental to understand
the naked singularity. From the behavior of the scalar field at 
the singularity $(\psi\rightarrow-\infty)$ and at the horizon 
$(\psi=\psi_0)$, and from the fact that the scalar field is a 
growing function as one moves from the singularity to the horizon, 
it is natural to use $\Lambda$ as a radial variable instead of $x$. 
At the singularity we have $\Lambda=0$  and at the light--cone 
$\Lambda=\Lambda_0\equiv e^{2\psi_0/7}$. Then, near the singularity,
in the limit $\Lambda\ll\Lambda_0$, we obtain the following
ten--dimensional metric
\beq
{\cal E}^2ds^2 = 
\Lambda^{-1/2}\left(\,\mu\,ds^2(dS_3)
+ds^2\left(\bT^6\right)\right)
+\Lambda^{1/2}d\Lambda^2\ .
\notag
\eeq
The dilaton field and the cosmological constant are still given by 
(\ref{eq1000}) and the constant $\mu$ can be determined from $a_s$ in the
expansion (\ref{eq30}). This geometry looks very
similar to the $O8$--plane geometry of section 4.1, with the important difference that the 
orientifold has worldvolume $dS_3\times \bT^6$, with deSitter radius $\sqrt{\mu}$.
In the limit $\Lambda\rightarrow 0$ the curvature of the induced metric on the
orientifold worldvolume vanishes, so that locally the orientifold looks flat. This is 
simply a higher dimensional generalization of the $O8/\overline{O}8$ system, since now 
one has a single $O8$--plane with spatial worldvolume $S^2\times T^6$. Antipodal points
on the sphere will have opposite charges, and therefore the near singularity geometry is 
locally BPS but breaks SUSY globally. These arguments are purely based on the supergravity
description of the system and should be taken only as such. In fact, while 
there is a perturbative string theory definition of the $O8/\overline{O}8$ system at
zero coupling, such definition does not exist for this curved $O8$--plane. If the 
locally flat description of the $O8$--plane is valid in string theory,
one can quantize the charge as before, with the result
\beq
2\pi\hat{l}_s\,{\cal E} = 8\, \hat{g}_s\ .
\notag
\eeq

Naively the solution just described is parameterized by $\hat{g}_s$, ${\cal E}$ and the
position of the horizon $\Lambda_0$. However, we can set
$\Lambda_0= 1$ using the rescaling of the coordinates 
$\Lambda\to c\Lambda$  and 
$\bT^6\to c\,\bT^6$,
which leaves the form of the solution invariant, if we also redefine
$\hat{g}_s\to  c^{-1/4}\,\hat{g}_s\,$,  
${\cal E}\to c^{5/4}\,{\cal E}$. Therefore,
after imposing the quantization of the orientifold charge, the solution
is parametrized only by the string coupling.
The situation is analogous to the $O8/\overline{O}8$ system, where the
geometry is fixed by a single constant of motion which is determined only, 
after charge quantization, by the string coupling.

\subsubsection{Cosmological acceleration}

When the scale factor vanishes, according to the behavior (\ref{solHI}), one
has a non--singular cosmological horizon. Naturally, to evade the
singularity theorems briefly discussed in the introduction, it must be that
the strong energy condition is violated. In fact, the boundary condition
imposed on the scalar field at the horizon was $\dot{\psi}=0$, and therefore
the kinetic energy at the horizon vanishes. At this point, all the field's
energy is in the form of the potential energy $V(\psi_0)$, which acts as a
positive cosmological constant. Hence, in this region the field does not
obey the strong energy condition, explaining why in both regions I there are
no singularities. Moreover, the power expansion (\ref{solHI}) shows that the
scale factor starts with an acceleration, as noted in \cite{CC1}. The fact
that a period of transient acceleration is quite generic, whenever the
kinetic energy vanishes and the potential is positive, was shown in the
context of String/M--theory compactifications in \cite{EmpaGarr}.

The above acceleration occurs near the cosmological horizon. Another
important question, related to the current observed acceleration of the
universe, is whether this geometry exhibits a late time acceleration.
This problem has been considered recently by many people in the
context of time--dependent string compactifications [104]--[115].
Clearly, this can be answered by analyzing the convergence to the asymptotic
solution (\ref{larget}). This is a well posed problem in dynamical systems
and was investigated in great detail by Halliwell \cite{Halliwell} for the cases
of open ($k=-1$), flat ($k=0$) and closed ($k=1$) four--dimensional
cosmologies with an exponential potential of the type here considered. Let
us review here Halliwell's results. Start with a scalar field with canonically 
normalized kinetic energy and consider a family of potentials, 
parametrized by a constant $a>0$, of the form 
\begin{equation*}
V(\psi )=\Lambda \,e^{-a\psi }\,,
\ \ \ \ \ \ \ \ \ \ \ \ \ \ \ \ \ \ \ \ 
(\Lambda>0)\,,
\end{equation*}
together with the FRW geometry 
\begin{equation*}
-N^{2}\left( t\right) \,dt^{2}+e^{2A\left( t\right) }\,ds^{2}\left( \mathcal{
M}_{k}\right) ,
\end{equation*}
where we introduce the lapse function $N\left(t\right)$ to allow for a more
general time coordinate, and where $\mathcal{M}_{k}$ is $\mathbb{M}^{3}$, $
S^{3}$ or $H^{3}$ depending on whether $k=0,1,-1$. The Friedman equation reads 
\begin{equation*}
\frac{\dot{A}^2}{N^2}
+k\,e^{-2A}=\frac{1}{12}\,\frac{\dot{\psi}^{2}}{N^2}\,
+\frac{1}{6}\,V\left( \psi \right),
\end{equation*}
and the usual equations of motion for $A$ and $\psi $ are 
\begin{eqnarray*}
\frac{1}{N}\,\frac{d}{dt}\left(\frac{\dot{A}}{N}\right) 
-k\,e^{-2A}+\frac{1}{4}\,\frac{\dot{\psi}^{2}}{N^2}=0\,,&&
\\
\frac{1}{N}\,\frac{d}{dt}\left( \frac{\dot{\psi}}{N}\right) 
+3N^{-2}\dot{A}\,\dot{\psi}+V^{\prime }\left( \psi \right)=0\,.&&
\end{eqnarray*}
We shall choose a time coordinate $t$ such that $N^{2}=V^{-1}$. It is then
easy to check that the Friedman equation becomes
\begin{equation*}
\dot{A}^{2}-\frac{1}{12}\,\dot{\psi}^{2}
=\frac{1}{6}-\frac{k}{\Lambda }\,e^{a\psi -2A}\,.
\end{equation*}
Therefore the hyperbola $\dot{A}^{2}=\frac{1}{12}\,\dot{\psi}^{2}+\frac{1}{6}$
divides the regions where $k$ is positive or negative. Moreover, the
equations of motion reduce to the following dynamical system
\begin{eqnarray*}
&&
\ddot{A}=\frac{1}{6}-\frac{1}{6}\,\dot{\psi}^{2}-\dot{A}^{2}
+\frac{a}{2}\,\dot{A}\,\dot{\psi}\,, 
\\
&&
\ddot{\psi}=\frac{a}{2}\,\dot{\psi}^{2}-3\dot{A}\,\dot{\psi}+a\,,
\end{eqnarray*}
which has fixed points in the $\dot{\psi}\dot{A}$--plane 
(neglecting the ones obtained by $t\rightarrow-t$, which correspond to a contracting universe) 
\begin{eqnarray}
P_{1} &=&\left(\frac{a\sqrt{2}}{\sqrt{3-a^{2}}}\,,\frac{1}{\sqrt{2(3-a^{2})}}\right) \,,  
\label{fixpoint} \\
P_{2} &=&\left(1,\,\frac{a}{2}\,\right) \,.  
\notag
\end{eqnarray}
Note that $P_{1}$ is always on the $k=0$ parabola.

To check if the corresponding cosmological solution is accelerating, we must
check the positivity of the second derivative of the scale factor with
respect to proper time
\begin{equation*}
\left( \frac{1}{N}\frac{d}{dt}\right) ^{2}e^{A}>0\,.
\end{equation*}
Using the above equations, it is easy to check that this condition is
equivalent to \cite{CCV}
\begin{equation*}
\dot{\psi}^{2}<1\,.
\end{equation*}
Now we can analyze the various trajectories in the $\dot{\psi}\dot{A}$--plane, 
and check whether they correspond to an accelerating
universe. We first note that the attractor solution is $P_{1}$ for $0<a<1$
and is $P_{2}$ otherwise. Moreover, for $a>\sqrt{3}$ the point $P_{1}$ no
longer exists. Generically flux compactifications have $a>\sqrt{3}$. This
includes the string compactification considered in the previous section
where $a=\sqrt{7}$. On the other hand, in hyperbolic compactifications 
\cite{TownWohl}, where the field $\psi$ is related to the size of an internal
compactification hyperbolic manifold of dimension $n\geq 2$, one has
\begin{equation*}
1<a=\sqrt{\frac{n+2}{n}}<\sqrt{3}\,.
\end{equation*}
Therefore, \textit{in both cases, the attractor solution has }$k=-1$
\textit{\ and is given by }$P_{2}$\textit{.}

\begin{figure}[t]
\begin{center}
\begin{tabular}{c}
\epsfysize=10cm\epsfbox{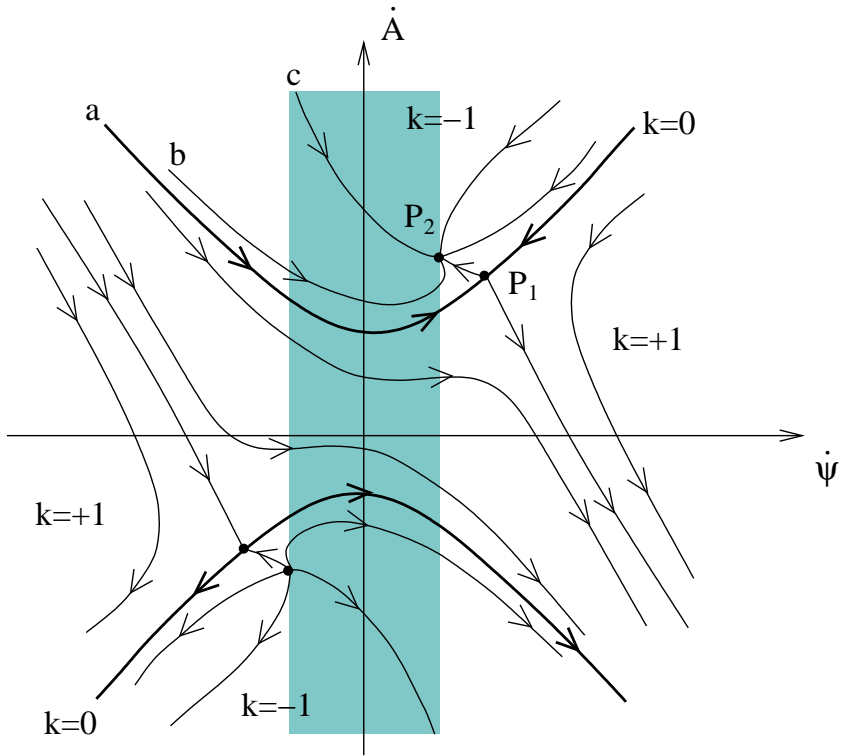}
\end{tabular}
\end{center}
\caption{Trajectories in the $\dot{\psi}\dot{A}$--plane for $1<a<\sqrt{4/3}$. When
$\sqrt{4/3}<a<\sqrt{3}$ the diagram is similar but the trajectories spiral around the 
stable attractors. These are the two regimes associated with hyperbolic compactifications.
Inside the shaded region the universe is accelerating, whereas outside it
decelerates.}
\label{fig19}
\end{figure}

\begin{figure}[t]
\begin{center}
\begin{tabular}{c}
\epsfysize=10cm\epsfbox{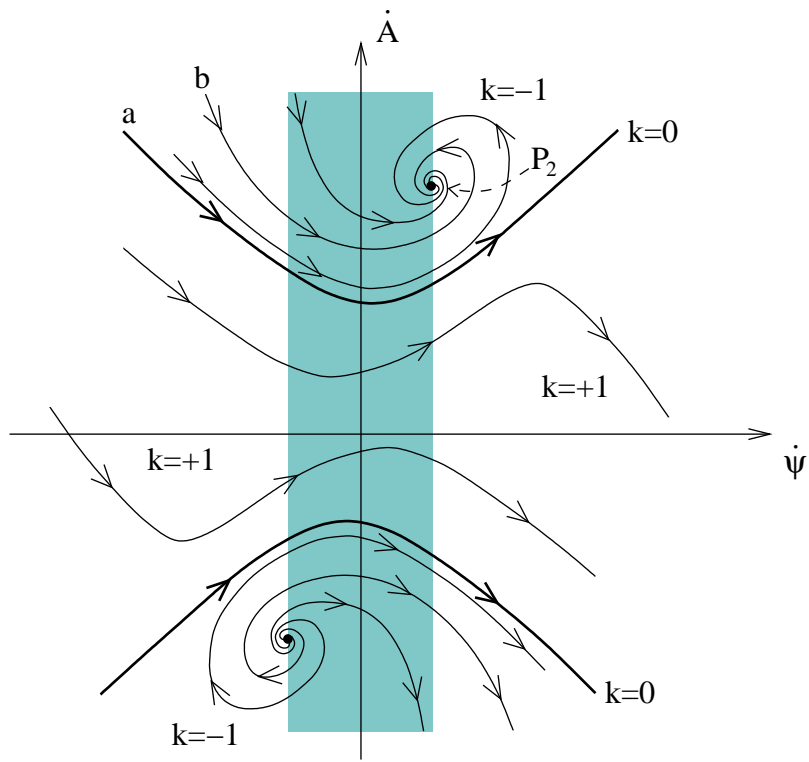}
\end{tabular}
\end{center}
\caption{Trajectories in the $\dot{\psi}\dot{A}$--plane for $a>\sqrt{3}$, corresponding
to flux compactifications. The late time behavior of the acceleration is always 
oscillatory.}
\label{fig20}
\end{figure}

Let us first consider the fixed point $P_{1}$, by concentrating, for the
moment, on a flat universe with $k=0$. The class of solutions for a flat
universe and arbitrary constant $a$ were found explicitly in \cite{Townsend}.
For $0<a<\sqrt{3}$, the point $P_{1}$
is always an attractor, \textit{if we restrict to the }$k=0$ \textit{hyperbola}. 
We must then distinguish two cases. For $a< 1$, and therefore not for
hyperbolic or flux compactifications, the fixed point has $\dot{\psi}^{2}< 1$. 
Hence, the asymptotic solution is \textit{accelerating}. The particular 
case with $a=1$ has a solution which accelerates and a solution which decelerates, 
depending if one starts from $\dot{\psi}=\mp\infty$. In particular, the accelerating 
solution does not have a future event horizon \cite{Townsend}. For $1<a<\sqrt{3}$,
which includes hyperbolic compactifications, the asymptotic solution is 
always decelerating (figure \ref{fig19}). For example, let us consider,
following \cite{TownWohl}, the trajectory (a) in figure \ref{fig19}, starting
with large negative $\dot{\psi}$ (the field rolling up the potential). Then,
as $\dot{\psi}^{2}$ becomes less then $1$, we enter into a period of 
\textit{transient acceleration}, followed again by a period of deceleration.
Finally, for $a>\sqrt{3}$, the $k=0$ trajectory (a) of figure \ref{fig20},
associated to flux compactifications,  has a
runaway behavior, always with a period of transient acceleration.

Let us move to the case of $k=-1$, with $a>1$, where there is always an
attractor at $P_{2}$. Note that $P_{2}$ is on the boundary of the region 
$\dot{\psi}^{2}<1$ of acceleration, because at late times curvature dominates and
one has linear expansion. Consequently, these geometries do not have a 
future event horizon. The behavior of the
trajectories converging to $P_{1}$ changes at $a=\sqrt{4/3}$ (see \cite{Halliwell}
for details). For $a<\sqrt{4/3}$ the trajectories are as in figure \ref{fig19},
whereas for $a>\sqrt{4/3}$ the trajectories spiral to $P_{2}$, like in
figure \ref{fig20}. Whenever $a<\sqrt{4/3}$, we can arrange initial conditions to
have a period of transient acceleration similar to the $k=0$ case, as in the
trajectory (b) in figure \ref{fig19}. Other initial conditions will
give a cosmology with a late time acceleration, as for the
trajectory (c) in the same figure. When $a>\sqrt{4/3}$ (so, in particular, for all
flux compactifications and for hyperbolic compactifications with $n<6$), 
\textit{independently of initial conditions}, we have a cyclic
behavior, with the acceleration oscillating around zero, with decreasing
magnitude. 

To summarize, for open cosmologies, \textit{both} compactifications 
can give at late times periods of positive acceleration. This result is in contrast with the
case of flat universes, where a positive acceleration is always
transient for \textit{both} type of compactifications.

\section{Conclusion}
In these lectures we have extensively discussed the physics of
time--dependent orbifolds of string theory, focusing on the key examples of
orbifolds of three--dimensional Minkowski space, together with their
implications for the study of the cosmological singularity. Although the
free propagation of particles is well understood, the physics of
interactions presents new challenges which are intimately tied with the
physics of quantum gravity at trans--Planckian energies. At first, the
problem seems insurmountable, of the same order of complexity as the
analysis of black--hole singularities. On the other hand, as we have
discussed in the section of eikonal resummation, there are hints that one
can have a better control over these theories, due to their orbifold
structure and to some partial facts known about gravity at high energies in
flat space--time. Certainly these models are a laboratory where we can try to push
our knowledge of quantum gravity to its limits. Such an example is the 
series of string dualities that led us to 
conjecture a remarkably simple quantization condition for the gravitational
constant in the presence of closed time--like curves. This condition is
related to quantization of charge, and should rely only on basic properties
of quantum gravity, together with the requirement of unitarity of the
orbifold theory. Therefore, even though we are probing physics at
trans--Planckian energies, we are in a more controlled situation than with
black--holes, and more conclusive statements can be made.

\vspace{1 cm}
{\bf Acknowledgments:} We are grateful to Costas Kounnas for many discussions and
collaboration on some of the material here presented. We would also like to thank 
Carlos Herdeiro, Costas Bachas, Fernando Quevedo, Frank Ferrari, 
Gianmassimo Tasinato, Ivonne Zavala, Joe Polchinski, Paul Steinhardt, Neil Turok, 
Paul Townsend, Rodolfo Russo and Pedro Vieira for discussions and correspondence. 
We are both grateful to the organizers of the
RTN Winter School in Torino for a great meeting and for their kind hospitality.
M.S.C. thanks Amsterdam University and L.C. thanks Oporto University for hospitality
during the course of these work. We are also grateful to LPTENS and DAMTP for their
hospitality. L.C. is supported by a Marie Curie Fellowship under the European Commission's
Improving Human Potential programme (HPMF--CT--2002--02016). This work
is partially funded by CERN under contract POCTI/FNU/49507/2002--FEDER and
by FCT under contract POCTI/FNU/38004/2001--FEDER.

\end{document}